\documentclass[12pt]{article}
\usepackage[margin=.9in]{geometry}
\usepackage{graphicx}
\usepackage[font=footnotesize]{subcaption}
\usepackage{putex}
\usepackage{amsmath,amsfonts,amssymb,amsthm}
\usepackage{tensor}
\usepackage{booktabs}
\usepackage{bm}
\usepackage{bbm}
\usepackage{braket}
\usepackage{array}
\newcolumntype{R}[1]{>{\raggedleft\let\newline\\\arraybackslash\hspace{0pt}}m{#1}}
\usepackage{tikz}
\usetikzlibrary{decorations.markings,bbox,arrows.meta}
\usepackage[color=orange!20]{todonotes}
\usepackage{hyperref}
\usepackage{dsfont}
\usepackage[font=small]{caption}
\usepackage{cite}

\tikzset{
  mid arrow/.style={postaction={decorate,decoration={
        markings,
        mark=at position 0.5 with {\arrow[xshift=4pt]{Latex[length=8pt,#1]}}
      }}},
  reverse mid arrow/.style={postaction={decorate,decoration={
        markings,
        mark=at position 0.5 with {\arrow[xshift=-4pt,xscale=-1]{Latex[length=8pt,#1]}}
      }}}
}

\newcommand{\sclattice}[2]{\tikz[baseline=-.2cm,xscale=1.5]{
	\draw[ultra thick,gray] (-.2,0) -- (4.2,0); 
	\node at (-.5,0) {$\cdots$};
	\node at (4.5,0) {$\cdots$};
	\node[fill=gray,circle,minimum size=5,inner sep=0] at (0,0) {};
	\node[fill=gray,circle,minimum size=5,inner sep=0] at (1,0) {};
	\node[fill=gray,circle,minimum size=5,inner sep=0] at (2,0) {};
	\node[fill=gray,circle,minimum size=5,inner sep=0] at (3,0) {};
	\node[fill=gray,circle,minimum size=5,inner sep=0] at (4,0) {};
	\node[draw=gray,thick,fill=white,circle,inner sep=2,minimum size=20] at (.5,0) {\footnotesize $\bm{#1}$};
	\node[draw=gray,thick,fill=white,circle,inner sep=2,minimum size=20] at (1.5,0) {\footnotesize $\bm{#2}$};
	\node[draw=gray,thick,fill=white,circle,inner sep=2,minimum size=20] at (2.5,0) {\footnotesize $\bm{#1}$};
	\node[draw=gray,thick,fill=white,circle,inner sep=2,minimum size=20] at (3.5,0) {\footnotesize $\bm{#2}$};
}}

\numberwithin{equation}{section}

\DeclareMathOperator{\grSU}{SU}
\DeclareMathOperator{\grSO}{SO}
\DeclareMathOperator{\grSpin}{Spin}
\DeclareMathOperator{\grUSp}{USp}
\DeclareMathOperator{\Tr}{Tr}
\DeclareMathOperator{\rk}{rk}
\DeclareMathOperator{\Lie}{Lie}
\DeclareMathOperator{\su}{\mathfrak{su}}

\def\F{\hat{\mathcal{F}}}
\def\C{\hat{\mathcal{C}}}
\def\V{\hat{\mathcal{V}}}
\def\U{\hat{\mathcal{U}}}

\institution{PU}{Joseph Henry Laboratories, Princeton University, Princeton, NJ 08544, USA}
\institution{PCTS}{Princeton Center for Theoretical Science, Princeton University, Princeton, NJ 08544, USA}

\title{More About the Lattice Hamiltonian for Adjoint QCD$_2$}

\authors{Ross Dempsey,\worksat{\PU} Silviu S.~Pufu,\worksat{\PU, \PCTS} Benjamin T. S\o gaard\worksat{\PU} 
\\[10pt] and
Igor R.~Klebanov\worksat{\PU, \PCTS}}

\abstract{
In our earlier work~\cite{Dempsey:2023fvm}, we introduced a lattice Hamiltonian for Adjoint QCD$_2$ using staggered Majorana fermions. We found the gauge invariant space of states explicitly for 
the gauge group $\grSU(2)$ and used them for numerical calculations of observables, such as the spectrum and the expectation value of the fermion bilinear. 
In this paper, we carry out a more in-depth study of our lattice model, extending it to any compact and simply-connected gauge group $G$. We show how to find the gauge invariant space of states and use it to study various observables.
We also use the lattice model to calculate the mixed 't Hooft anomalies of Adjoint QCD$_2$ for arbitrary $G$. We show that the matrix elements of the lattice Hamiltonian can be expressed in terms of the Wigner 6$j$-symbols of $G$. For $G = \grSU(3)$, we perform exact diagonalization for lattices of up to six sites and study the low-lying spectrum, the fermion bilinear condensate, and the string tension. We also show how to write the lattice strong coupling expansion for ground state energies and operator expectation values in terms of the Wigner 6$j$-symbols. For $\grSU(3)$ we carry this out explicitly and find good agreement with the exact diagonalizations, and for $\grSU(4)$ we give expansions that can be compared with future numerical studies.}

\begin{document}

\preprint{PUPT-2654}

\maketitle

\tableofcontents

\section{Introduction}

Adjoint QCD$_2$ is a $(1+1)$-dimensional gauge theory with a single multiplet of Majorana fermions in the adjoint representation of the gauge group $G$~\cite{Dalley:1992yy}. When $G = \grSU(N_c)$, its action is
\begin{equation}\label{eq:suN_action}
    S = \int d^2x\,\tr_\text{fund}\left(-\frac{1}{2g^2}F_{\mu\nu}F^{\mu\nu} + i \bar\psi \gamma^\mu D_\mu \psi - m\bar\psi\psi\right)\,.
\end{equation}
This theory is of interest as a model of non-perturbative gauge dynamics: it appears to be the simplest 2D model that has an infinite number of Regge trajectories in the large $N$ limit~\cite{Bhanot:1993xp}
and exhibits a thermal deconfinement transition ~\cite{Kutasov:1993gq,Semenoff:1996xg}. 
The model has several surprising qualitative features, including a mass gap even when the adjoint fermion is massless~\cite{Kutasov:1993gq,Bhanot:1993xp,Delmastro:2021otj},
a vanishing string tension 
when $m = 0$~\cite{Gross:1995bp,Gross:1997mx,Semenoff:1996xg,Komargodski:2020mxz,Dempsey:2024ofo},\footnote{This vanishing applies to the model (\ref{eq:suN_action}), which does not contain the double-trace four-fermion operator ~\cite{Cherman:2019hbq}.}
and $(1,1)$ supersymmetry at masses $m = \pm m_\text{SUSY}\equiv \pm\sqrt{\frac{g^2 N_c}{2\pi}}$~\cite{Kutasov:1994xq,Popov:2022vud,Dempsey:2024ofo}.

The earliest numerical method for studying Adjoint QCD$_2$ was discretized lightcone quantization~\cite{Pauli:1985ps,Eller:1986nt,Hornbostel:1988fb}. This method was used shortly after the introduction of Adjoint QCD$_2$ to study the spectrum of the $\grSU(N_c)$ in the large-$N_c$ limit~\cite{Bhanot:1993xp,Demeterfi:1993rs,Gross:1997mx}, and it has since been generalized to finite $N_c$~\cite{Antonuccio:1998uz,Dempsey:2022uie}. One result of these calculations is that the mass gap for $m=0$ scales almost exactly like $g_\text{YM}\sqrt{N_c}$, with very small but nonzero corrections at subleading powers of $N_c$~\cite{Dempsey:2022uie,Trittmann:2024jkf}. Additionally, discretized lightcone quantization gives numerical evidence for the supersymmetric spectrum at $m = m_\text{SUSY}$~\cite{Gross:1997mx,Dempsey:2021xpf,Dempsey:2022uie}. Other methods based on lightcone quantization have also produced good numerical results~\cite{Katz:2013qua,Trittmann:2015oka,Trittmann:2023dar}.

One limitation of the lightcone quantization approach is that, in its present formulation, it constructs the Hilbert space from a single Fock vacuum~\cite{Eller:1986nt,Hornbostel:1988fb}. However, in adjoint QCD$_2$ there are generically many different ``universes'' (i.e., sectors of the Hilbert space) distinguished by the expectation values of generators of the one-form center symmetry~\cite{Komargodski:2020mxz,Witten:1978ka,Smilga:1994hc,Lenz:1994du,Cherman:2019hbq}. It is not yet known how to access these different universes using discretized lightcone quantization, and in practice the results seem to correspond only to a single topologically trivial universe~\cite{Dempsey:2023fvm}. This makes it challenging to address questions regarding confinement. Physically, the different universes of adjoint QCD$_2$ are distinguished by having different kinds of background chromoelectric flux~\cite{Witten:1978ka}, and the string tensions are most directly computed as the energy densities of those flux tubes. Thus, calculations based on lightcone quantization have so far not been able to directly access the string tensions, although indirect calculations using heavy probe quark quarks~\cite{Dempsey:2021xpf} have provided useful information.

In view of these limitations of the lightcone approach, in~\cite{Dempsey:2023fvm} we introduced an alternative non-perturbative approach to Adjoint QCD$_2$ for $G = \grSU(N_c)$: we formulated it as a Hamiltonian lattice gauge theory with staggered fermions~\cite{Kogut:1974ag,Banks:1975gq}. For $G = \grSU(2)$, we gave an explicit gauge-invariant expression for the Majorana fermion operators and used it to carry out explicit lattice calculations that agree with the results of lightcone quantization and extend them to a topologically nontrivial universe~\cite{Dempsey:2023fvm}.
In this paper, we generalize the results of~\cite{Dempsey:2023fvm} to an arbitrary (compact, simply-connected) gauge group $G$. We find that for any $G$, the matrix elements of the lattice Hamiltonian can be expressed in terms of the Wigner $6j$-symbols of $G$. These symbols are difficult to compute efficiently, and so in this paper we only perform exact diagonalization of the lattice Hamiltonian for $G = \grSU(3)$, and only compute the lattice strong coupling expansion for $G = \grSU(3)$ and $G = \grSU(4)$. In addition, we use our lattice model to derive the mixed 't Hooft anomalies of Adjoint QCD$_2$ for any $G$, generalizing the results of~\cite{Cherman:2019hbq} for the $\grSU(N_c)$ theory. 

The rest of this paper is organized as follows. In Section~\ref{sec:suN}, we discuss the lattice model for $G = \grSU(N_c)$. In particular, we show how to build a gauge-invariant Hilbert space and how to explicitly implement the fermion operators on the lattice, which gives a means of computing the lattice Hamiltonian. In Section~\ref{sec:strong_coupling} we set up the lattice strong coupling expansion, and in Section~\ref{sec:numerics} we carry out exact diagonalization of the lattice Hamiltonian for $\grSU(3)$. Our numerical results are in good agreement with the light-cone methods~\cite{Dempsey:2022uie} and the recent Euclidean Monte Carlo calculations~\cite{Bergner:2024ttq}. We also match our lattice results with the small circle continuum calculations both for the periodic~\cite{Dempsey:2024ofo} and antiperiodic~\cite{Lenz:1994du} conditions for the fermions. In Section~\ref{sec:arbitrary_group}, we show how the results of Section~\ref{sec:suN} generalize to an arbitrary gauge group, and we compute the symmetries and anomalies of adjoint QCD$_2$ with an arbitrary gauge group. Technical details are relegated to the appendices.

\section{Lattice model for $\grSU(N_c)$}\label{sec:suN}

Here we will review the Hamiltonian lattice formulation of Adjoint QCD$_2$ introduced in~\cite{Dempsey:2023fvm}, and discuss how it can be concretely implemented for any $\grSU(N_c)$ gauge group (generalizing the construction given in~\cite{Dempsey:2023fvm} for $\grSU(2)$). In Section~\ref{sec:suN_staggered}, we review the staggering prescription for placing fermions on lattice sites, and show how the fermions decompose into representations of the gauge symmetry. In Section~\ref{sec:suN_hilbert}, we show how to construct the Hilbert space of gauge-invariant states. In Section~\ref{sec:suN_symmetry}, we show how the symmetries of the continuum theory are implemented on the lattice and compute their anomalies, again following~\cite{Dempsey:2023fvm}. In Section~\ref{sec:suN_fermions} we generalize the gauge-invariant formulation given for the $\grSU(2)$ theory in~\cite{Dempsey:2023fvm} to a general $\grSU(N_c)$ gauge group, and in Section~\ref{sec:suN_fermions} we show how to compute the Hamiltonian matrix elements in this setup.

\subsection{Staggered fermions}\label{sec:suN_staggered}

We use the Kogut-Susskind prescription for distributing the components of the two-component spinor $\psi$ onto our lattice~\cite{Kogut:1974ag}. We place the upper component onto even sites and the lower component onto odd sites~\cite{Banks:1975gq}, so that the total number of sites $N$ is always even. We denote the lattice spacing by $a$. We number the lattice sites by $n = 0, \ldots, N -1$, and we always identify $n\sim n + N$ so that the theory is quantized on a spatial circle of length $L = Na$.

Thus, each site $n$ has a single multiplet $\chi_n^A$ of Majorana fermions, where $A = 1,\ldots,N_c^2 - 1$ is an index for the adjoint representation of $\su(N_c)$. These fermions satisfy the algebra
\begin{equation}\label{eq:clifford}
    \left\lbrace \chi_m^A, \chi_n^B\right\rbrace = \delta_{mn}\delta^{AB}\,.
\end{equation}

\begin{figure}[t]
    \centering
    \begin{tikzpicture}[xscale=2.5,yscale=1.2]
		\draw[ultra thick,dashed] (-1,0) -- (-.5,0);
		\draw[ultra thick] (-.5,0) -- (4.5,0);
		\draw[ultra thick,dashed] (4.5,0) -- (5,0);
		
		\foreach \x in {0,...,3} {
			\draw[ultra thick,-latex] (\x, 0) -- ({\x + .6},0);
			\node at ({\x + .5}, 0.3) {$U^{AB}_\x$};
		}
		\foreach \x in {0,2,4} {
			\node[green!50!black,fill,circle,minimum size=5,inner sep=0,label=90:{\color{green!50!black} $\chi^A_\x$}] at (\x, 0) {};
		}
		\foreach \x in {1,3} {
			\node[blue,fill,circle,minimum size=5,inner sep=0,label=90:{\color{blue} $\chi_\x^A$}] at (\x, 0) {};
		}
		\foreach \x in {0,...,3} {
			\node at (\x+0.15, -0.4) {$L^A_\x$};
			\node at (\x+0.85, -0.4) {$R^A_\x$};
		}
	\end{tikzpicture}
    \caption{}
    \label{fig:lattice_schematic}
\end{figure}

The lattice analog of the gauge field is a bit more subtle. For the spatial component, we should replace the infinitesimal gauge connection $A_1(x)$ with an operator that connects finitely-separated lattice points. Schematically, that is, we need to keep track of the operators $\exp\left(i\int_{na}^{(n+1)a} A_1(x)\,dx\right)$. These are elements of $\grSU(N_c)$, and we denote them by $U_n$. The conjugate variables to $U_n$ are $\su(N_c)$-valued left- and right-acting electric fields, $L_n^A$ and $R_n^A$, obeying the algebra~\cite{Dempsey:2023fvm}
\begin{equation}\label{eq:LRU_algebra}
    \left\lbrack L^A_n, U_m\right\rbrack = \delta_{nm} T^A U_n\,, \qquad \left\lbrack R^A_n, U_m\right\rbrack = \delta_{nm} U_n T^A\,.
\end{equation}
Here we are taking $U_n$ to be in the fundamental representation so that we can multiply it with the fundamental $\su(N_c)$ generators $T^A$ on the right-hand sides. These generators are normalized such that
\begin{equation}\label{eq:suN_generator_orthogonality}
    \tr\left(T^A T^B\right) = \frac{1}{2}\delta^{AB}\,.
\end{equation}

It is also sometimes convenient to work with the link operators in the adjoint representation, which we denote by $U_n^{AB}$. We can compute this from $U_n$ in the fundamental representation by acting with the adjoint action on a generator and using \eqref{eq:suN_generator_orthogonality} on the result:
\begin{equation}
    U_n^{AB} = 2\tr\left(T^A U_n T^B U_n^{-1}\right)\,.
\end{equation}
Using this expression along with \eqref{eq:LRU_algebra}, we can work out
\begin{equation}\label{eq:LRU_algebra_adjoint}
    [L_m^A, U_n^{BC}] = -i\delta_{mn} f^{ABD} U_n^{DC}\,,\qquad [R_m^A, U_n^{BC}] = -i\delta_{mn} f^{ADC} U_n^{BD}\,,
\end{equation}
where $f^{ABC}$ is defined by $[T^A, T^B] = if^{ABC} T^C$.

In the Hamiltonian formulation, we eliminate the time component of the gauge field, at the cost of an explicit Gauss law constraint. This constraint is~\cite{Dempsey:2023fvm}
\begin{equation}\label{eq:gauss}
    L_n^A - R_{n-1}^A = Q_n^A\,, \qquad n=0,1,\ldots,N-1\,,
\end{equation}
where
\begin{equation}\label{eq:charge}
    Q_n^A = -\frac{i}{2}f^{ABC}\chi_n^B \chi_n^C\,.
\end{equation}

With these ingredients, we can write down our lattice Hamiltonian (more details can be found in~\cite{Dempsey:2023fvm}). With periodic boundary conditions for the fermions, the Hamiltonian is
\begin{equation}\label{eq:hamiltonian}
    H = \sum_{n=0}^{N-1} \left\lbrack \frac{g^2 a}{2}L^A_n L^A_n -\frac{i}{2}\left(a^{-1} + (-1)^n m\right)\chi_n^A U_n^{AB} \chi_{n+1}^B\right\rbrack\,.
\end{equation}
The first term corresponds to the gauge-kinetic term, while the second term corresponds to the fermion kinetic and mass terms. 
The mass term couples the Majorana fermions on adjacent sites~\cite{Seiberg:2023cdc}.
The presence of $U_n^{AB}$ is essential for gauge invariance. Indeed, let $g_n\in \grSU(N_c)$ be a choice of a group element for every site, and let $g_n^{AB}$ be the adjoint representation matrix elements of $g_n$. Then \eqref{eq:hamiltonian} is invariant under
\begin{equation}\label{eq:gauge_transform}
    \chi_n^A \mapsto g_n^{AB} \chi_n^B\,, \qquad L_n \mapsto g_n^{AB} L_n^B\,, \qquad U_n^{AB} \mapsto g_n^{AC} U_n^{CD} g_{n+1}^{BD}\,.
\end{equation}

With antiperiodic boundary conditions, we have a very similar Hamiltonian in which the sign on one link is altered:
\begin{equation}\label{eq:hamiltonian_ap}
   \begin{split}
   	 H^\text{(AP)} &= \frac{g^2 a}{2} \sum_{n=0}^{N-1} L^A_n L^A_n -\frac{i}{2}\sum_{n=0}^{N-2}\left(a^{-1} + (-1)^n m\right)\chi_n^A U_n^{AB} \chi_{n+1}^B\\
     &\qquad+ \frac{i}{2}\left(a^{-1} + (-1)^n m\right)\chi_{N-1}^A U_{N-1}^{AB} \chi_0^B\,.
   \end{split}
\end{equation}

\subsection{Gauge-invariant Hilbert space}\label{sec:suN_hilbert}

Prior to imposing gauge invariance, the Hamiltonian \eqref{eq:hamiltonian} acts on $\mathcal{H}_\text{full} = \mathcal{H}_F\otimes \mathcal{H}_B$, where $\mathcal{H}_F$ is the Hilbert space of the fermions living on sites and $\mathcal{H}_B$ is the Hilbert space of the gauge fields living on links. The $N(N_c^2-1)$ Majorana fermion operators on the sites can be combined into $\frac{N(N_c^2-1)}{2}$ Dirac fermions, and so $\dim \mathcal{H}_F = 2^{\frac{N(N_c^2-1)}{2}}$. The bosonic Hilbert space is infinite-dimensional, because on each link we have a quantum mechanical particle on the $\grSU(N_c)$ manifold with Hilbert space $L^2(\grSU(N_c))$.

Let us now discuss how to find the physical Hilbert space $\mathcal{H}_\text{phys} \subset \mathcal{H}_\text{full}$ that satisfies the Gauss law \eqref{eq:gauss}. We first need to understand how the states of $\mathcal{H}_F$ and $\mathcal{H}_B$ transform under a local gauge transformation \eqref{eq:gauge_transform}. In~\cite{Dempsey:2023fvm}, it is shown that the states of $\mathcal{H}_F$ transform in $2^{\frac{N(N_c-1)}{2}}$ copies of the representation $\bm{R}^{\otimes N}$ of the gauge symmetry $\grSU(N_c)^{\otimes N}$, where $\bm{R}$ is the representation with Dynkin labels $[11\cdots 1]$.\footnote{We can also verify that $\bm{R}$ is the on-site representation by calculating the quadratic Casimir using the charges \eqref{eq:charge}. Using the algebra \eqref{eq:clifford} and the fact that $f^{ABC}f^{BCD} = N_c \delta^{AD}$~\cite{Haber:2019sgz}, we find
\begin{equation}
    Q_n^A Q_n^A = -\frac{1}{4}f^{ABC} f^{ADE} \chi_n^B \chi_n^C \chi_n^D \chi_n^E = \frac{N_c(N_c^2 - 1)}{8}\mathbbm{1}\,.
\end{equation}
And indeed, $C_2(\bm{R}) = \frac{N_c(N_c^2 - 1)}{8}$, as one can check easily for $\grSU(2)$ or $\grSU(3)$.} Indeed, the dimension of $\bm{R}$ is $2^{\frac{N_c^2 - N_c}{2}}$, and so this gives the full dimension of $\mathcal{H}_F$. For $\grSU(2)$, $\bm{R} = \bm{2}$ is the fundamental representation; for $\grSU(3)$, $\bm{R} = \bm{8}$ is the adjoint representation; for $N_c>3$, $\bm{R}$ is a rather high-dimensional representation that does not coincide with any other special representation (for instance, for $\grSU(4)$ we have $\bm{R} = \bm{64}$). We can thus represent $\mathcal{H}_F$ by $\bm{R}^{\otimes N}$, with states of the form
\begin{equation}
    \bigotimes_{n=0}^{N-1} \ket{\bm{R}; m_n}\,,
\end{equation}
where $m_n$ is an index for the representation $\bm{R}$, tensored with states of $\mathbb{C}^{N(N_c-1)/2}$. We can roughly think of the fermionic Hilbert space as constructed from $\bm{R}^{\oplus 2^{(N_c-1)/2}}$ at each site, so that the factor of $\mathbb{C}^{N(N_c-1)/2}$ keeps track of the multiplicity of $\bm{R}$ on the sites. Of course, this cannot be the complete story since $2^{(N_c-1)/2}$ is not always an integer; we will give a more precise discussion in Section~\ref{sec:suN_fermions}.

Under a gauge transformation parametrized by $g_n\in \grSU(N_c)$, the factor $\mathbb{C}^{N(N_c - 1)/2}$ is inert, while the $\bm{R}$ kets transform according to the matrices $D_{\bm{R}}(g_n)$ of the $\bm{R}$ representation:
\begin{equation}
    \bigotimes_{n=0}^{N-1} \ket{\bm{R}; m_n} \mapsto \bigotimes_{n=0}^{N-1} \ket{\bm{R}; m'_n} D_{\bm{R}}(g_n)_{m_n m'_n}\,.
\end{equation}

To decompose the bosonic Hilbert space, we can use the Peter-Weyl theorem, which says that $L^2(\grSU(N_c))$ is spanned by the matrix elements of irreducible representations of $\grSU(N_c)$ (or likewise for any compact topological group). Thus, a basis for $\mathcal{H}_B$ is given by states of the form
\begin{equation}
    \bigotimes_{n=0}^{N - 1} \ket{\bm{r}_n;\mathfrak{m}_{nL},\mathfrak{m}_{nR}}\,.
\end{equation}
Here $\bm{r}_n$ is an irreducible representation of $\grSU(N_c)$ and $\mathfrak{m}_{nL},\mathfrak{m}_{nR} = 1,\ldots,\dim \bm{r}_n$ are indices for its matrices. For a gauge transformation parametrized by $g_n\in \grSU(N_c)$, this state transforms as
\begin{equation}\label{eq:boson_gauge_transformation}
    \bigotimes_{n=0}^{N - 1} \ket{\bm{r}_n;\mathfrak{m}_{nL},\mathfrak{m}_{nR}}\mapsto \bigotimes_{n=0}^{N - 1} \ket{\bm{r}_n;\mathfrak{m}_{nL}',\mathfrak{m}_{nR}'}D_{\bm{r}_n}(g_n)_{\mathfrak{m}_{nL}\mathfrak{m}'_{nL}}D_{\bm{r}_n}(g_{n+1})_{\mathfrak{m}_{nR}'\mathfrak{m}_{nR}}\,.
\end{equation}

To build a gauge-invariant state, we must contract the indices $\mathfrak{m}_{(n-1)R}$, $m_n$, and $\mathfrak{m}_{nL}$ with an invariant symbol in the representations $(\bm{r}_{n-1}, \bm{R}, \bm{r}_n)$. These are the Clebsch-Gordan coefficients $C^{\bm{r}_{n-1}\,\bm{R}\,\bm{r}_n;e_n}_{\mathfrak{m}_{n-1,R}\,m_n\,\mathfrak{m}_{nL}}$, where $e_n$ labels different invariants and runs from 1 to the multiplicity of $\bm{r}_n$ in the tensor product $\bm{r}_{n-1}\otimes \bm{R}$. (In the $\grSU(2)$ theory considered in~\cite{Dempsey:2023fvm}, the multiplicity never exceeds 1 and so this additional index is not needed.)

Following this procedure, we find gauge-invariant contractions of the form
\begin{equation}\label{eq:state_contraction}
\begin{split}
    &\ket{(\bm{r}_0,e_0),\cdots,(\bm{r}_{N-1},e_{N-1})} =  \\
    &\qquad\sum_{m_n,\mathfrak{m}_{nL},\mathfrak{m}_{nR}} \left\lbrack \left(\bigotimes_{n=0}^{N-1} \ket{\bm{R},m_n}\right)\otimes\left(\bigotimes_{n=0}^{N-1} \frac{C^{\bm{r}_{n-1}\,\bm{R}\,\bm{r}_n;e_n}_{\mathfrak{m}_{n-1,R}\,m_n\,\mathfrak{m}_{nL}}}{\sqrt{\dim \bm{r}_n}}\ket{\bm{r}_n;\mathfrak{m}_{nL},\mathfrak{m}_{nR}}\right)\right\rbrack\,.
\end{split}
\end{equation}
The states of $\mathcal{H}_\text{phys}$ are spanned by tensor products of \eqref{eq:state_contraction} with states of $\mathbb{C}^{N(N_c-1)/2}$ (which keeps track of the different copies of $\bm R^{\otimes N}$ is used in the construction of $\mathcal{H}_F$ above). Note that the Clebsch-Gordan symbols are normalized according to \eqref{eq:clebsch_orthogonality}, so these states are normalized.

We can represent the contraction \eqref{eq:state_contraction} with a ``birdtracks'' diagram, as in Figure~\ref{fig:birdtrack_state}. See Appendix~\ref{app:group_theory} for details of the notation.

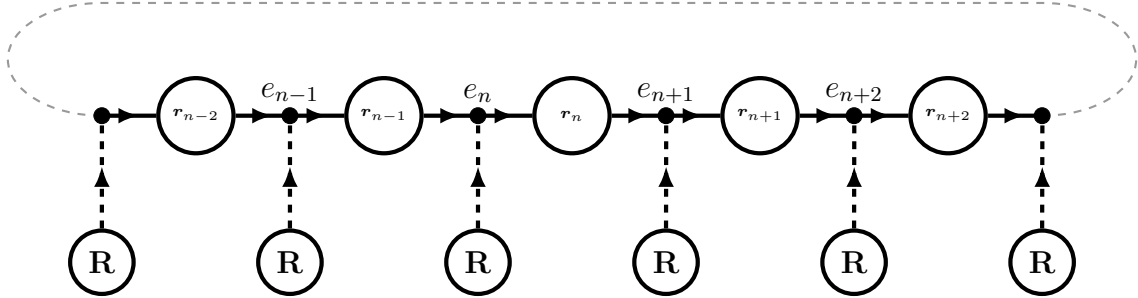
\begin{figure}[t]
    \centering
    \begin{tikzpicture}[xscale=2.5,yscale=1.5]
        \foreach \x/\sub in {1/n-2,2/n-1,3/n,4/n+1,5/n+2} {
        		\node[ultra thick,draw,circle,minimum size=1cm] (\x) at ({\x+.5},0) {\tiny $\bm{r}_{\sub}$};
            \draw[ultra thick,mid arrow] (\x,0) -- (\x);
            \draw[ultra thick,mid arrow] (\x) -- ({\x+1}, 0);
        };
        \draw[gray!80,thick,dashed] (6,0) arc(-90:90:.5) -- (1,1) arc(90:270:.5);
        \foreach \x in {1,...,6} {
            \node[draw,circle,fill,minimum size=6,inner sep=0] at (\x,0) (\x) {};
            \draw[ultra thick,dashed,mid arrow] (\x,-1) node[below,draw,solid,circle] {$\mathbf{R}$} -- (\x);
        };
        \node at (2,.2) {$e_{n-1}$};
        \node at (3,.2) {$e_n$};
        \node at (4,.2) {$e_{n+1}$};
        \node at (5,.2) {$e_{n+2}$};
    \end{tikzpicture}
    \caption{A birdtracks diagram for a gauge-invariant state as in \eqref{eq:gauge_invariant_state}. Circles indicate the kets in \eqref{eq:state_contraction}, lines indicate representations of $\grSU(N_c)$, and vertices are invariant tensors of $\grSU(N_c)$. At site $n$, we have a Clebsch-Gordan coefficient with multiplicity label $e_n$ that combines representations $\bm{r}_{n-1}$, $\bm{R}$, and $\bm{r}_{n}$. Note that the normalization factor $\left(\prod_{n=0}^{N-1} \dim \bm{r}_n\right)^{-1/2}$ is not included here.}
    \label{fig:birdtrack_state}
\end{figure}

Note in particular that gauge-invariance requires that the tensor product of $\bm{R}$ with $\bm{r}_n$ contains $\bm{r}_{n+1}$. When $N_c$ is even, $\bm{R}$ has $N_c$-ality of $\frac{N_c}{2}$, and so the $N_c$-ality alternates between even and odd links. When $N_c$ is odd, $\bm{R}$ has $N_c$-ality 0, and so the $N_c$-ality is the same on each link.

\subsection{Symmetries and anomalies}\label{sec:suN_symmetry}

The symmetries of $\grSU(N_c)$ adjoint QCD$_2$ are discussed in~\cite{Cherman:2019hbq}. When $N_c \ge 3$, the symmetry group is
\begin{equation}
    \left\lbrack \mathbb{Z}^{[1]}_{N_c} \rtimes (\mathbb{Z}_2)_C\right\rbrack \times (\mathbb{Z}_2)_F\,.
\end{equation}
The factors here are as follows. The fermion parity symmetry $(\mathbb{Z}_2)_F$, generated by $\F$, acts on the fermions via
\begin{equation}
    \F \psi \F = -\psi\,.
\end{equation}
The charge conjugation symmetry $(\mathbb{Z}_2)_C$, generated by $\C$, acts on the adjoint-valued gauge field and fermion by
\begin{equation}\label{eq:suN_charge_conjugation_continuum}
    \C A^\mu_{ij} \C = -A^\mu_{ji}\,, \qquad \C \psi_{ij} \C = -\psi_{ji}\,.
\end{equation}
The one-form center symmetry $\mathbb{Z}^{[1]}_{N_c}$ acts on Wilson lines. Letting $\U(x)$ be a generator of $\mathbb{Z}^{[1]}_{N_c}$, the charge of a Wilson line under $\U(x)$ is $e^{\frac{2\pi i}{N_c}p(\bm{r})}$ where $p(\bm{r})$ is the $N_c$-ality of $\bm{r}$. The full symmetry involves a semidirect product of the center symmetry and the charge conjugation symmetry because charge conjugation changes a representation with $N_c$-ality $m$ to its conjugate with $N_c$-ality $-m$.

When $m = 0$, the symmetry group is extended by a direct product with the chiral symmetry $(\mathbb{Z}_2)_\chi$. Its generator $\V$ acts upon the fermions by
\begin{equation}
    \V \psi \V = \gamma^5\psi\,.
\end{equation}
In~\cite{Cherman:2019hbq}, it is shown that the quantum theory generically has mixed 't Hooft anomalies between the chiral symmetry and the other symmetries. When the fermions obey periodic boundary conditions on the spatial circle, the anomalous phases are given by\footnote{In~\cite{Cherman:2019hbq}, charge conjugation is defined by $\C_\text{theirs} = \C_\text{ours}\F$. However, they find the same anomalies \eqref{eq:suN_anomalies} that we calculate using $\C_\text{ours}$. This slight discrepancy has no physical consequence, as we are always free to make such a redefinition in what we mean by charge conjugation, so we will not explore it further here.}
\begin{equation}\label{eq:suN_anomalies}
    \begin{split}
        \F \V &= (-1)^{N_c - 1} \V \F\,, \\
        \C \V &= (-1)^{\left\lfloor \frac{N_c-1}{2}\right\rfloor}\V\C\,,\\
        \U(x)\V &= (-1)^{N_c - 1}\V \U(x)\,.
    \end{split}
\end{equation}
When the fermions obey antiperiodic boundary conditions, $\V$ commutes with $\F$ and $\C$, and the anomaly between the chiral symmetry and the center symmetry is unchanged.

In~\cite{Dempsey:2023fvm}, we showed that the lattice Hamiltonian formulation reproduces all these symmetries and anomalies. Here we show this again in a way more suited for generalizing to an arbitrary gauge group, as we do in Section~\ref{sec:arbitrary_group}.
First, we identify the symmetry generators in the lattice model. We will start with the zero-form symmetries. The simplest of these is the fermion parity symmetry, which should act on the fermions by
\begin{equation}\label{eq:fermion_parity}
    \F \chi^A_n \F = -\chi^A_n\,.
\end{equation}
We can express $\F$ in terms of the fermion operators as
\begin{equation}\label{eq:fermion_parity_def}
    \F = \left(2i\right)^{N(N_c^2-1)/2}\prod_{n=0}^{N-1}\prod_{A=1}^{N_c^2 - 1}\chi_n^A\,.
\end{equation}
The action of charge conjugation can be written like in \eqref{eq:suN_charge_conjugation_continuum} using the fundamental representation, but instead we will express it directly in the adjoint representation. By defining a charge conjugation matrix $\mathsf{C}^{AB}$ via
\begin{equation}
    -T^A_{ji} = \mathsf{C}^{AB} T^B_{ij}\,,
\end{equation}
we can write
\begin{equation}
    \C \chi^A_n \C = \mathsf{C}^{AB} \chi^B_n\,, \qquad \C L^A_n \C = \mathsf{C}^{AB} L^B_n\,,\qquad \C U_n^{AB} \C= \mathsf{C}^{AC}\mathsf{C}^{BD} U_n^{CD}\,.
\end{equation}
The expression for $\C$ in terms of the fermions is given in~\cite{Dempsey:2023fvm}, but we will not need it here.

Finally, when the fermions obey periodic boundary conditions, chiral symmetry acts as a one-site translation:
\begin{equation}
    \V \chi^A_n \V = \chi^A_{n+1}\,, \qquad \V L^A_n \V = L^A_{n+1}\,, \qquad \V U^{AB}_n \V = U^{AB}_{n+1}\,.
\end{equation}
We can write it in terms of the fermions as
\begin{equation}\label{eq:v_implementation}
    \V = \F^{N_c^2 - 2}\left(\prod_{A=1}^{N_c^2 - 1}\left(\chi_0^A + \chi_1^A\right)\right)\left(\prod_{A=1}^{N_c^2 - 1}\left(\chi_1^A + \chi_2^A\right)\right)\ldots \left(\prod_{A=1}^{N_c^2 - 1}\left(\chi_{N-2}^A + \chi_{N-1}^A\right)\right)\,.
\end{equation}
When the fermions obey antiperiodic boundary conditions, we need to modify $\V$ so that it does not move the link with the flipped sign in \eqref{eq:hamiltonian_ap}. This is accomplished by defining
\begin{align}\label{eq:move_defect}
	\V^{\text{(AP)}} = \V\F_0\,, \qquad \F_0 = (2i)^{(N_c^2-1)/2} \mathcal F^{N_c^2-1} \prod_{A=1}^{N_c^2-1}\chi_0^A\,.
\end{align}

Now we will compute the projective factors in the algebra of these zero-form symmetries. Consider some operator $\mathcal{O}$ which acts on the fermions on each site by
\begin{equation}
    \mathcal{O} \chi^A_n \mathcal{O}^{-1} = {M^A}_B \chi^B_n\,,
\end{equation}
with ${M^A}_B$ an orthogonal matrix (so $\det M = \pm 1$). The canonical anti-commutation relations of $\chi^A_n$ imply
\begin{align}
	\mathcal O \left( \prod_{A=1}^{\dim G} \chi_n^A\right) \mathcal O^{-1} = (\det M) \prod_{A=1}^{\dim G} \chi_n^A\,.
\end{align}
This means that $\mathcal{O}\mathcal{F} = (\det M)^N \mathcal{F} \mathcal{O} = \mathcal{F}\mathcal{O}$ (since $N$ is an even integer). Taking $\mathcal{O} = \C$, this tells us that there is no anomaly between fermion parity and charge conjugation. Also, we have
\begin{equation}
    \V^{\text{(AP)}} \mathcal{O} = (\det M)^{N-1} (\det M) \mathcal{O} \V^{\text{(AP)}} = \mathcal{O} \V^{\text{(AP)}}\,,
\end{equation}
so with antiperiodic boundary conditions there are no anomalies involving the zero-form symmetries. But in the case of periodic boundary conditions, we find
\begin{equation}\label{eq:app_anomaly_general}
    \mathcal{V}\mathcal{O} = (\det M)^{N-1} \mathcal{O}\mathcal{V} = (\det M) \mathcal{O}\mathcal{V}\,.
\end{equation}
Thus, in this case we can compute the anomalies between chiral symmetry and the other zero-form symmetries by computing a determinant.

Since fermion parity flips the sign of each of the $N_c^2 - 1$ fermions on a site, we immediately find
\begin{equation}
    \F \V = (-1)^{N_c^2 - 1}\V \F = (-1)^{N_c - 1}\V \F\,,
\end{equation}
reproducing the first line of \eqref{eq:suN_anomalies}.

To compute the determinant of the action of charge conjugation, we could think about the fundamental generators of $\grSU(N_c)$. We can work with a basis in which there are $N_c - 1$ generators of the form $\diag(0,\ldots,1,-1,\ldots,0)$, $\binom{N_c}{2}$ real symmetric generators, and $\binom{N_c}{2}$ imaginary antisymmetric generators. When we take the negative transpose, the $N_c - 1$ traceless diagonal generators and the $\binom{N_c}{2}$ imaginary antisymmetric generators all have their signs flipped, and so
\begin{equation}
    \C \V = (-1)^{\binom{N_c}{2} + (N_c - 1)}\V \C\,.
\end{equation}
One can check that this agrees with the second line of \eqref{eq:suN_anomalies}.

We can also compute this determinant by working in the Cartan-Weyl basis for the adjoint representation of $\grSU(N_c)$. In this basis there are $N_c - 1$ generators of a Cartan subalgebra, $\binom{N_c}{2}$ positive roots (which are simultaneous eigenvectors of the adjoint action of the Cartan generators), and a corresponding set of $\binom{N_c}{2}$ negative roots. The action of charge conjugation is to reverse the order of the Cartan generators, which reverses the Dynkin labels of all representations (e.g. the fundamental $[10\cdots 0]$ is exchanged with the antifundamental $[0\cdots 01]$). This corresponds to a symmetry of the Dynkin diagram for $\grSU(N_c)$, in which the $N_c - 1$ simple roots (one per Cartan generator) are represented by vertices of a graph:
\begin{center}
    \begin{tikzpicture}
        \foreach \i/\label in {1/1,2/2,3/3,5/N_c-3,6/N_c-2,7/N_c-1} \node[draw,thick,circle,minimum size=7,inner sep=0,label=90:{\tiny $\label$}] (\i) at (\i, 0) {};
        \draw (1) -- (2) -- (3) (5) -- (6) -- (7);
        \node at (4,0) {$\cdots$};
        \draw[latex-latex,gray,dashed] (1,-.1) arc(-180:0:3 and 1.5);
        \draw[latex-latex,gray,dashed] (2,-.1) arc(-180:0:2 and 1);
        \draw[latex-latex,gray,dashed] (3,-.1) arc(-180:0:1 and .5);
    \end{tikzpicture}
\end{center}
This permutation of the simple roots leads to some permutation of the positive roots, but the negative roots undergo the same permutation, and so these do not contribute to the determinant. Thus, we only need to keep track of the Cartan generators, and from the diagram above we see that there are $\left\lfloor \frac{N_c - 1}{2}\right\rfloor$ exchanges in their permutation. This gives the second line of \eqref{eq:suN_anomalies}.

Now let's turn to the one-form center symmetry. Let $k\in \mathbb{Z}_{N_c} \subset \grSU(N_c)$ be an element of the center symmetry; then the corresponding one-form center symmetry generator at site $n$ acts on the gauge-invariant lattice states by
\begin{equation}\label{eq:center_symmetry_generator}
    \U_{k,n}\ket{(\bm{r}_0,e_0),\cdots,(\bm{r}_{N-1},e_{N-1})} = (-1)^{n (N_c - 1)}\exp\left(\frac{2\pi i}{N_c} k p(\bm{r}_n)\right)\ket{(\bm{r}_0,e_0),\cdots,(\bm{r}_{N-1},e_{N-1})}\,,
\end{equation}
where $p(\bm{r})$ gives the $N_c$-ality of $\bm{r}$. Indeed, if we act on our lattice state with a spatial Wilson line in, say, the fundamental representation, then the representations on link $n$ will be some $\bm{r'}_n \in \bm{r}_n\otimes \mathbf{fund}$, and in particular $p(\bm{r'}_n) = p(\bm{r}_n) + 1 \pmod{N_c}$. Thus, this Wilson line will carry charge $\exp\left(\frac{2\pi i}{N_c}\right)$ under $\U_{1,n}$. Additionally, we can check that this is a topological operator on physical states, i.e., $\U_{k,n} = \U_{k,n+1}$. We know that when $N_c$ is odd, $p(\bm{r}_{n+1}) = p(\bm{r}_n)$, so no compensating sign is required; when $N_c$ is even, $p(\bm{r}_{n+1}) = p(\bm{r}_n) + \frac{N_c}{2}\pmod{N_c}$, and so the alternating sign in \eqref{eq:center_symmetry_generator} again makes the operator topological.

From \eqref{eq:center_symmetry_generator}, we see plainly that regardless of the boundary condition for the fermions,
\begin{equation}
    \U_{k,n} \V = (-1)^{N_c - 1}\V \U_{k,n}\,.
\end{equation}
This reproduces the third line of \eqref{eq:suN_anomalies}.

\subsection{Fermion operators and spin-charge separation}\label{sec:suN_fermions}
In Section~\ref{sec:suN_hilbert}, we discussed how $\mathcal{H}_F$ decomposes under the $\grSU(N_c)^{\otimes N}$ gauge symmetry into $2^{N(N_c-1)/2}$ copies of $(\bm R,\bm R,\cdots,\bm R)$. In particular, we can choose a basis where the gauge charges explicitly realize this factorization by acting only on the $\bm R$ local to the site:
\begin{equation}\label{eq:decoupled_Q}
	Q^A_n = D_{\bm{R},n}^A \otimes \mathbbm{1}_{\mathbb C^{N(N_c-1)/2}}\,,
\end{equation}
where $D_{\bm{R},n}^A$ is the $\bm{R}$-representation matrix of the $A$th generator of $\grSU(N_c)$ on site $n$. We will now give an explicit construction of the fermion operators $\chi^A_n$ such that the charge $Q^A_n = -\frac{i}{2}f^{ABC}\chi^B_n\chi^C_n$ is given by \eqref{eq:decoupled_Q}.

The form of the fermion operators is highly constrained, since they need to act covariantly on the $\bm R^{\otimes N}$ factor of the Hilbert space. If we assume that $\chi^A_n$ acts only on the $\bm R$ factor at site $n$, then it must have the form
\begin{equation}\label{eq:chi_ansatz_init}
	\chi^A_n = \sum_{j=1}^{N_c - 1} C^A_{n,j}\mathcal O_{n,j}\,,
\end{equation}
where $C^A_{n,j}$ is an invariant symbol on the representations $(\mathbf{adj},\bm{R},\bm{\bar{R}})$ (with the $\bm{R}$ indices suppressed) and $\mathcal O_{n,j}$ are some operators action on $\mathbb C^{N(N_c-1)/2}$. The representation matrices $D_{\bm{R}}^A$ are one of the invariant symbols on the representations $(\mathbf{adj},\bm{R},\bm{\bar{R}})$, but for $N_c>2$ there are several such symbols. For instance, for $\grSU(3)$ the representation matrices of $\bm{R} = \bm{8}$ are given by the structure constant $f^{ABC}$, but we also have the symmetric three-point invariant $d^{ABC}$. In general, there are $N_c - 1$ such invariant symbols~\cite{king_wybourne}, explaining the range of the sum \eqref{eq:chi_ansatz_init}. Considering the fermonic nature of $\chi_n^A$ and the dimensionality of $\mathbb C^{N(N_c-1)/2}$, it is natural to conjecture that the operators $\mathcal O_{n,j}$ should be $N_c-1$ Majorana fermions living on each lattice site. To that end, we define Majorana operators $\lambda_{n,j}$ that satisfy the Clifford algebra
\begin{equation}\label{eq:sigma_alg}
	\left\lbrace \lambda_{n,j}, \lambda_{m,k}\right\rbrace = 2\delta_{nm}\delta_{jk}\,.
\end{equation}
Note that $\lambda_{n,j}$ do not part take in gauge transformations as they are realized on $\mathbb C^{N(N_c-1)/2}$. Thus, we have the following ansatz for the fermions $\chi^A_n$ constructed out of the $N_c -1$ invariants and  $N_c - 1$ Majorana fermions on each site:
\begin{equation}\label{eq:chi_ansatz}
    \chi^A_n = \sum_{j=1}^{N_c - 1} C^A_{n,j}\lambda_{n,j}\,.
\end{equation}

Let us see if this ansatz can satisfy the properties we desire. First, we need the fermions to obey
\begin{equation}
    \left\lbrace \chi^A_n\,, \chi^B_m\right\rbrace = \delta^{AB} \delta_{nm}\,.
\end{equation}
Substituting \eqref{eq:chi_ansatz}, we find that this holds provided the following two conditions are obeyed:
\begin{align}
	\sum_{j=1}^{N_c-1}\{C^A_{n,j},C^B_{n,j}\} &= \delta^{AB}\mathds{1}\,,\label{eq:cond1}\\
	[C^A_{n,j},C^B_{n,k}]+[C^B_{n,j},C^A_{n,k}]&=0\, .\label{eq:cond2}
\end{align}
Additionally, in order for $Q^A_n$ to take the form \eqref{eq:decoupled_Q}, we need
\begin{equation}\label{eq:cond3}
    f^{ABC} C^B_{n,j} C^C_{n,k} - f^{ABC} C^B_{n,k} C^C_{n,j} = 0\,.
\end{equation}

The conditions \eqref{eq:cond2} and \eqref{eq:cond3} are not sensitive to the choice of basis for the $C^A_{n,j}$ invariants, while the condition \eqref{eq:cond1} requires choosing an appropriate basis. We have checked explicitly for $\grSU(2)$, $\grSU(3)$, and $\grSU(4)$ that a basis satisfying \eqref{eq:cond1} exists and that \eqref{eq:cond2} and \eqref{eq:cond3} are satisfied. We find it very likely that these conditions can be made to hold for any $N_c$ (or indeed for any gauge group, as we discuss in Section~\ref{sec:arbitrary_group}), but we do not yet have a proof of this. Here we will show how these conditions are satisfied for $\grSU(2)$ and $\grSU(3)$. The general case is discussed in Appendix~\ref{app:c_properties}.

Let us give a couple of examples. For $N_c=2$, $\bm R = \bm{2}$. Up to normalization there is only a single $(\bm{3}, \bm{2}, \bm{\bar{2}})$ invariant given by the Pauli matrices $\sigma^A$. This makes the conditions \eqref{eq:cond2} and \eqref{eq:cond3} trivial. Using the property $\sigma^A\sigma^B=\delta^{AB}+i\epsilon^{ABC}\sigma^C$, we see that \eqref{eq:cond1} fixes the normalization to
\begin{equation}
    C^A_1 = \frac{1}{\sqrt{2}}\sigma^A\,.
\end{equation}
Thus, the Majorana fermions can be represented as 
\begin{align}
	\chi^A_n = \frac{1}{\sqrt 2}\sigma^A_n\lambda_n\,.
\end{align}
This is the construction given in~\cite{Dempsey:2023fvm} for $\grSU(2)$ (and there the $\lambda_n$ were written using Pauli matrices via a Jordan-Wigner transformation).

When $N_c=3$, the representation $\bm R$ is the adjoint. There are two invariants of the representations $(\bm{8}, \bm{8}, \bm{8})$ given by the $f$- and $d$-symbols; defining these symbols in terms of the fundamental generators by
\begin{equation}
    [T^A, T^B] = if^{ABC} T^C\,, \qquad \left\lbrace T^A, T^B\right\rbrace = \frac{1}{N_c}\delta^{AB}\mathbbm{1} + d^{ABC} T^C\,,
\end{equation}
we can build matrices
\begin{align}
	{(F^A)^B}_C=-if_{ABC}\, , \qquad {(D^A)^B}_C=d_{ABC}\,.
\end{align}
In~\cite{Haber:2019sgz}, it is shown that
\begin{equation}
    \left\lbrace F^A, F^B\right\rbrace + 3\left\lbrace D^A, D^B\right\rbrace = 2\delta^{AB}\mathbbm{1}\,.
\end{equation}
This means we can satisfy \eqref{eq:cond1} by taking
\begin{equation}
    C^A_1 = \frac{1}{\sqrt{2}} F^A\,, \qquad C^A_2 = \sqrt{\frac{3}{2}} D^A\,.
\end{equation}
The conditions \eqref{eq:cond2} and \eqref{eq:cond3} then follow from properties of the $f$- and $d$-symbols found in~\cite{Haber:2019sgz}. Thus, the Majorana fermions in the $\grSU(3)$ theory can be realized as
\begin{align}
	\chi^A_n = \frac{1}{\sqrt 2} F^A_n\lambda_{n,1}+\sqrt{\frac{3}{2}} D^A_n\lambda_{n,2}\,.
\end{align}

We have also explicitly constructed the $\chi^A_n$ operators for the next simplest case of $N_c=4$. Since this construction involves the invariants on the irreps $(\bm{15},\bm{64},\bm{64})$ of $\grSU(4)$, which are difficult to express, we will not reproduce the construction here.

The realization \eqref{eq:chi_ansatz} of the Majorana fermions exhibits a form of ``spin-charge separation''~\cite{Tomonaga:1950zz,Luttinger:1963zz,Haldane:1981zza}. The first factor $C^A_{n,j}$ carries all the information about how $\chi^A_n$ transforms under the gauge symmetry, and so the charge operators \eqref{eq:decoupled_Q} act trivially on the second factor. That is, the first factor encodes the chromoelectric charge. The second factor, $\lambda_{n,j}$, encodes the fermionic nature of $\chi^A_n$. For instance, the fermion parity operator \eqref{eq:fermion_parity_def} can be expressed solely in terms of the $\lambda_{n,j}$ operators as
\begin{align}
	\hat{\mathcal F} = i^{N(N_c-1)/2}\prod_{n=0}^{N-1}\prod_{j=1}^{N_c-1}\lambda_{n,j}\,.
\end{align}

It is not clear whether this factorization property has any physical consequences for the dynamics of adjoint QCD$_2$; we leave this question to future work. For the purposes of this paper, the factorization has practical implications: we can consider the action of $C^A_{n,j}$s on the gauge invariant states \eqref{eq:state_contraction} separately from the action of the $\lambda_{n,j}$s. We will make use of this fact in the following section.

\subsection{Hamiltonian matrix elements}\label{sec:suN_hamiltonian}

We will now determine how the Hamiltonian operator \eqref{eq:hamiltonian} acts upon the gauge-invariant states. First, for the sake of being explicit, let us use a Jordan-Wigner transformation to express the Hilbert space of the $N_c-1$ Majorana fermions on each site as the Hilbert space of $N_c - 1$ qubits shared between pairs of sites. We will denote the states of the qubits shared between sites $2k$ and $2k+1$ by
\begin{equation}\label{eq:qubit_state}
    \ket{s_{k,1}\cdots s_{k,N_c-1}}\,, \qquad s_{k,j} = \pm 1\,.
\end{equation}
The Majorana fermions $\lambda_{n,j}$ are then expressed as
\begin{equation}\label{eq:sigma_explicit}
\begin{split}
    \lambda_{2k,j} &= \left(\bigotimes_{k'=0}^{k-1} \bigotimes_{j'=1}^{N_c-1} (\sigma_3)_{k',j'}\right)\otimes (\sigma_3)_{k,1} \otimes\cdots \otimes (\sigma_3)_{k,j-1}\otimes (\sigma_1)_{k,j} \\
    \lambda_{2k+1,j} &= \left(\bigotimes_{k'=0}^{k-1} \bigotimes_{j=1}^{N_c-1} (\sigma_3)_{k',j}\right)\otimes (\sigma_3)_{k,1} \otimes\cdots \otimes (\sigma_3)_{k,m-1}\otimes (\sigma_2)_{k,m}\,,
\end{split}
\end{equation}
where $(\sigma_i)_{k,j}$ is the $i$th Pauli matrix acting upon the $j$th qubit at the $k$th pair of sites.

We will compute the matrix elements in a basis constructed of tensor products of the states \eqref{eq:qubit_state} with contractions of the form \eqref{eq:state_contraction}. Explicitly, we will show how to compute matrix elements of the Hamiltonian between the two states
\begin{equation}\label{eq:gauge_invariant_state}
\begin{split}
    \ket{\psi} &= \left(\bigotimes_{k=0}^{N/2 - 1} \ket{s_{k,1}\cdots s_{k,N_c-1}}\right) \otimes \ket{(\bm{r}_0,e_0),\cdots,(\bm{r}_{n-1},e_{n-1})}\,,\\
    \ket{\psi'} &= \left(\bigotimes_{k=0}^{N/2 - 1} \ket{s'_{k,1}\cdots s'_{k,N_c-1}}\right) \otimes \ket{(\bm{r'}_0,e'_0),\cdots,(\bm{r'}_{n-1},e'_{n-1})}\,.
\end{split}
\end{equation}

The gauge-kinetic term is simple. As explained in~\cite{Dempsey:2023fvm}, we have
\begin{equation}\label{eq:h_0_eigenvalue}
    \Braket{\psi'|\sum_n L^A_n L^A_n|\psi} = \left(\sum_n C_2(\bm{r}_n)\right)\braket{\psi'|\psi}\,.
\end{equation}
In particular, the gauge-kinetic term is diagonal in our basis.

The fermion kinetic and mass terms are much more difficult to evaluate. They each are written as a sum of terms of the form $\chi_n^A U_n^{AB} \chi_{n+1}^B$, so we will show how to evaluate the matrix element
\begin{equation}
    M_n = \Braket{\psi'|\chi_n^A U_n^{AB} \chi_{n+1}^B|\psi}\,.
\end{equation}
We will assume an expression \eqref{eq:chi_ansatz} for the fermions. First, we note that the orthogonality of the basis states for the links,
\begin{equation}\label{eq:link_orthogonality}
\begin{split}
	\Braket{\bm{r'}; \mathfrak{m}_L', \mathfrak{m}_R'|\bm{r}; \mathfrak{m}_L, \mathfrak{m}_R} &= \delta_{\bm{r},\bm{r'}} \delta_{\mathfrak{m}_L,\mathfrak{m}_L'}\delta_{\mathfrak{m}_R,\mathfrak{m}_R'} \\[.5em]
	\begin{tikzpicture}[xscale=2,yscale=1.3,baseline=.5cm]
		\node[ultra thick,draw,circle,minimum size=1cm] (top) at (.5,1) {$\bm{r}$};
        \draw[ultra thick,mid arrow] (0,1) -- (top);
        \draw[ultra thick,mid arrow] (top) -- (1,1);
		\node[ultra thick,draw,circle,minimum size=1cm] (bottom) at (.5,0) {$\bm{r'}$};
        \draw[ultra thick,mid arrow] (1,0) -- (bottom);
        \draw[ultra thick,mid arrow] (bottom) -- (0,0);
    \end{tikzpicture} &= \delta_{\bm{r},\bm{r'}} \times \begin{tikzpicture}[baseline=.5cm,xscale=2]
    		\draw[ultra thick,mid arrow] (0,1) arc(90:-90:.25 and .5);
    		\draw[ultra thick,mid arrow] (1,0) arc(270:90:.25 and .5);
    	\end{tikzpicture}
\end{split}
\end{equation}
means that for any link $m\neq n$ we must have $\bm{r}_m = \bm{r'}_m$ in order to have a nonzero matrix element. Provided this holds, we show in Appendix~\ref{app:hamiltonian} that 
\begin{equation}\label{eq:matrix_element_6j}
	\begin{split}
    M_n &= \frac{1}{\dim \bm{r}_{n+1}}\sum_{j,k=1}^{\rk G} q_{n,jk} D_{jk}^{e_n,e_{n+1};e'_n,e'_{n+1}}(\bm{r}_{n-1},\bm{r}_{n+1};\bm{r}_n,\bm{r'}_n)\,,\qquad\text{where}\\    
    q_{n,jk} &= \braket{s'|\lambda_{n,j} \lambda_{n,k}|s}\,,\\
    D_{jk}^{e_n,e_{n+1};e'_n,e'_{n+1}}(\bm{r}_{n-1},&\bm{r}_{n+1};\bm{r}_n,\bm{r'}_n) = \frac{1}{\dim \bm{r'}_n}\sum_{l} \left\lbrack \begin{tikzpicture}[xscale=1.7,baseline=-1cm]
        \foreach \x in {1,2} {
            \node[draw,circle,fill,minimum size=6,inner sep=0] at ({2*\x+1},0) (\x) {};
            \node[draw,circle,fill,minimum size=6,inner sep=0] at ({2*\x+1},-2) (a\x) {};
            \draw[ultra thick,dashed,mid arrow] (a\x) -- ({2*\x+1},-1);
            \draw[ultra thick,dashed,mid arrow] ({2*\x+1},-1) -- (\x);
        };
        \draw[ultra thick,mid arrow] (3,-2) arc(270:90:.5 and 1) node[left,pos=0.5] {$\bm{r}_{n-1}$};
        \draw[ultra thick,mid arrow] (5,0) arc(90:-90:.5 and 1) node[right,pos=0.5] {$\bm{r}_{n+1}$};
        \node at (2.9,.4) {$e_n$};
        \node at (5.1,.4) {$e_{n+1}$};
        \node at (2.9,-2.4) {$e'_n$};
        \node at (5.1,-2.4) {$e'_{n+1}$};
        \draw[ultra thick,dotted,blue,mid arrow] (3,-1) node[black,fill,minimum size=6,inner sep=0,label=180:{\color{black}{$j$}}] {} -- (3.5,-1) node[black,fill,circle,minimum size=6,inner sep=0,label=0:{\color{black}{$l$}}] {};
        \draw[ultra thick,mid arrow] (3,0) arc(90:0:.5 and 1) node[midway,right] {$\bm{r}_n$};
        \draw[ultra thick,mid arrow] (3.5,-1) arc(0:-90:.5 and 1) node[midway,right] {$\bm{r'}_n$};
        \draw[ultra thick,dotted,blue,mid arrow] (4.5,-1) node[black,fill,circle,minimum size=6,inner sep=0,label=180:{\color{black}{$l$}}] {} -- (5,-1) node[black,fill,minimum size=6,inner sep=0,label=0:{\color{black}{$k$}}] {};
        \node at (4,-1) {$\times$};
        \draw[ultra thick,mid arrow] (4.5,-1) arc(180:90:.5 and 1) node[midway,left] {$\bm{r}_n$};
        \draw[ultra thick,mid arrow] (5,-2) arc(-90:-180:.5 and 1) node[midway,left] {$\bm{r'}_n$};
    \end{tikzpicture}
    \right\rbrack\,.
    \end{split}
\end{equation}
The values $D_{jk}^{e_n,e_{n+1};e'_n,e'_{n+1}}(\bm{r}_{n-1},\bm{r}_{n+1};\bm{r}_n,\bm{r'}_n)$ are all numbers, which depend on the neighboring link representations $\bm{r}_{n-1}$ and $\bm{r}_{n+1}$, the representation $\bm{r}_n$ on the $n$th link in $\ket{\psi}$ and the representation $\bm{r'}_n$ on the $n$th link in $\ket{\psi'}$, multiplicity labels $e_n$, $e_{n+1}$ and $e'_n$, $e'_{n+1}$ in $\ket{\psi}$ and $\ket{\psi'}$ respectively, and the multiplicity labels $j$ and $k$ appearing in the Hamiltonian when expanded using \eqref{eq:chi_ansatz}. They can be calculated by evaluating the contractions of Clebsch-Gordan symbols indicated by the diagrams, as explained in Appendix~\ref{app:group_theory}. The values $q_{n,jk}$ can be computed using the explicit realization \eqref{eq:sigma_explicit} of the Majorana fermions $\lambda_{n,j}$ and $\lambda_{n,k}$.

The contractions of four three-point invariants appearing in these diagrams are called Wigner 6$j$-symbols. For $\grSU(2)$ the 6$j$-symbols are known in closed form, but for $N_c\ge 3$ they are difficult to compute. In Section~\ref{sec:numerics} we discuss methods for explicitly computing 6$j$-symbols to build the lattice Hamiltonian for the $\grSU(3)$ theory.

The multiplicity of the invariant indexed by $\ell$ is 1 when $\bm{r}_n \neq \bm{r'}_n$, and when $\bm{r}_n = \bm{r'}_n$ it is the number of nonzero Dynkin labels of $\bm{r}_n$.\footnote{Both of these statements follow from counting Littelmann paths~\cite{littelmann}, and noting that all nonzero weights in the adjoint representation (i.e., roots) appear with multiplicity 1. See also~\cite{king_wybourne} for a proof of the second statement.}

\section{Strong coupling expansion}\label{sec:strong_coupling}

Here we discuss results that can be obtained by working in the lattice strong coupling limit $ga\to\infty$. We will work at $m = 0$, where the Hamiltonian can be expressed as
\begin{equation}\label{eq:strong_coupling_hamiltonian}
    H = H_0 + x V\,, \qquad \text{with} \quad H_0 = \frac{g}{2\sqrt{x}}\sum_{n=0}^{N-1} L^A_n L^A_n\,, \qquad V = -\frac{ig}{2\sqrt{x}}\sum_{n=0}^{N-1} \chi^A_n U^{AB}_n \chi^B_{n+1}\, ,
\end{equation}
with $x= \frac{1}{(ga)^2}$. The leading term $H_0$ is diagonal in our basis of gauge-invariant states of the form \eqref{eq:state_contraction}. Thus, we can use perturbation theory to expand various quantities in inverse powers of $ga$. This process is discussed in~\cite{Dempsey:2023fvm} for the $\grSU(2)$ theory. In Section~\ref{sec:strong_coupling_vac}, we enumerate the strong coupling vacua for several groups, and conjecture a pattern for all $\grSU(N_c)$ groups. In Section~\ref{sec:strong_coupling_bw} we set up the perturbation theory in the general case, and in Section~\ref{sec:strong_coupling_su3} we carry out the first few orders of the expansion for the $\grSU(3)$ and $\grSU(4)$ theories.

\subsection{Strong coupling vacua}\label{sec:strong_coupling_vac}

To leading order in the strong coupling expansion, we need to determine the eigenvalues of $H_0$, which is diagonal in our basis (see \eqref{eq:h_0_eigenvalue}). However, since the Gauss law requires that $\bm{r}_{n+1} \in \bm{r}_n\otimes \bm{R}$, it is still a nontrivial problem to determine the states with minimal energy at leading order.

For example, for the $\grSU(2)$ theory, the minimal energy configuration in the trivial universe is
\begin{equation}
	\bm{r}_n = \begin{cases} \bm{1} & n\text{ even},\\\bm{2} & n\text{ odd}. \end{cases}
\end{equation}
In general, since $\bm{1}\in \bm{R}\otimes\bm{R}$, we can always use a two-fold periodic pattern of representations.\footnote{In principle, the two-fold periodicity is not required; if we have several pairs of representations with the same energy \eqref{eq:sc_min_energy}, then we could string them together in a lattice state. The simplest case where this could happen is in the trivial universe of the $\grSU(3)$ theory. However, in a gauge-invariant state we can have the sequence $\bm{1},\bm{8},\bm{8},\bm{1}$, but \emph{not} the sequence $\bm{8},\bm{1},\bm{1},\bm{8}$ (since $\bm{1}\not\in \bm{1}\otimes\bm{R} = \bm{1}\otimes\bm{8}$), so we cannot build a state on a periodic chain that uses both of the minimal pairs. In fact, in all cases we have looked at, the structure of the tensor product with $\bm{R}$ implies that all the strong coupling ground states have a two-fold periodic pattern of representations on links.} Our problem is then reduced to determining the patterns $(\bm{r}_0, \bm{r}_1)$ that minimize
\begin{align}\label{eq:sc_min_energy}
    E^{(0)} = \min_{\bm r_1\subset \bm r_0\otimes \bm R} \frac{g^2 L}{4}\left(C_2(\bm r_0)+C_2(\bm r_1)\right)\, .
\end{align}
Generically, for a given universe of a given theory, there are several different patterns that produce the same minimal energy. A number of examples of strong coupling ground states can be found in Table~\ref{tab:strong_coupling_vacua}.

\begingroup
\renewcommand{\arraystretch}{1.7}
\begin{table}[p]
\centering
\begin{tabular}{m{1.8cm}m{2cm}m{9.4cm}m{1.5cm}}
\toprule
$G$ & Universe & Ground state(s) & $E^{(0)}/g^2 L$ \\
\midrule
SU(2) & 0 & \sclattice{1}{2} & $\frac{3}{16}$\\
 & 1 & \sclattice{2}{1} & $\frac{3}{16}$\\
SU(3) & 0 & \sclattice{1}{8} & $\frac{3}{4}$\\
 &  & \sclattice{8}{1} & \\
 & 1 & \sclattice{3}{3} & $\frac{2}{3}$\\
 & 2 & \sclattice{\overline{3}}{\overline{3}} & $\frac{2}{3}$\\
SU(4) & 0 & \sclattice{15}{6} & $\frac{13}{8}$\\
 & 1 & \sclattice{20}{\overline{4}} & $\frac{27}{16}$\\
 & & \sclattice{4}{\overline{20}} & \\
  & 2 & \sclattice{6}{15} & $\frac{13}{8}$\\
 & 3 & \sclattice{\overline{20}}{4} & $\frac{27}{16}$\\
 & & \sclattice{\overline{4}}{20} & \\
 $\grSU(5)$ & 0 & \sclattice{75}{24} & $\frac{13}{4}$ \\
 & & \sclattice{24}{75} & \\
 & 1 & \sclattice{45}{45} & $\frac{16}{5}$ \\
 & 2 & \sclattice{40}{40} & $\frac{33}{10}$ \\
 & & \sclattice{10}{175} & \\
 & & \sclattice{175}{10} & \\
 & 3 & \sclattice{\overline{40}}{\overline{40}} & $\frac{33}{10}$ \\
 & & \sclattice{\overline{10}}{\overline{175}} & \\
 & & \sclattice{\overline{175}}{\overline{10}} & \\
 & 4 & \sclattice{\overline{45}}{\overline{45}} & $\frac{16}{5}$ \\
\bottomrule
\end{tabular}
\caption{Examples of strong coupling ground states. In all of these cases (and conjecturally in general), there is no outer multiplicity to account for in the tensor product with $\bm{R}$.}
\label{tab:strong_coupling_vacua}
\end{table}
\endgroup

By looking at the Dynkin labels of the representations appearing in Table~\ref{tab:strong_coupling_vacua}, we can conjecture a pattern for the strong coupling ground states for $\grSU(N_c)$. The highest-weight vectors $\vec{w}_0$ and $\vec{w}_1$ of $\bm{r}_0$ and $\bm{r}_1$ appear to always satisfy
\begin{equation}
	\vec{w}_0 + C(\vec{w}_1) = \rho = (1,1,\ldots,1)\,,
\end{equation}
where $C$ denotes charge conjugation (explicitly, $C(\langle w_1,w_2,\ldots,w_{N_c-1}\rangle) = \langle w_{N_c-1},\ldots, w_2, w_1\rangle$. Additionally, the vectors $\vec{w}_0$ and $\vec{w}_1$ are constructed by repeating the patterns $1,0$ or $0,1$ many times, and then taking a subsequence (in technical terms, $\vec{w}_0$ and $\vec{w}_1$ are subsequences of sequences in the regular language $(10|01)^*$). One can prove that all pairs of representations satisfying these properties give rise to states with energies
\begin{equation}\label{eq:strong_coupling_leading_energy_nc}
	\frac{E^{(0)}(N_c, p)}{g^2L} = \frac{1}{192}\begin{cases} 5N_c^3 - 2N_c & N_c\text{ even} \\ 5N_c^3 - 2N_c - 3/N_c & N_c\text{ odd}\end{cases} + \frac{1}{4N_c} f_{N_c}(p)\,,
\end{equation}
where
\begin{equation}\label{eq:sc_energy_split}
	f_{N_c}(p) = \begin{cases}
		(N_c/4)^2 - \min(|p - N_c/4|^2, |p - 3N_c/4|^2) & N_c\text{ even} \\
		(\lfloor (N_c+3)/4\rfloor - 1/2)^2 - (p - N_c/2)^2 & N_c\text{ odd}, \quad N_c/4 < p < 3N_c/4 \\
		(\lfloor (N_c+1)/4\rfloor)^2 - \min(p, N_c - p)^2 & N_c\text{ odd}, \quad\text{otherwise}.
	\end{cases}
\end{equation}
The number of these states is given by
\begin{equation}\label{eq:sc_leading_degeneracy}
	N_0(N_c, p) = 2^{N(N_c-1)/2} \begin{cases}
		\binom{m}{\min(p,2m - p)} & N_c = 2m,\\
		\binom{2\lfloor m/2\rfloor + 1}{p-\lfloor (m+1)/2\rfloor} & N_c = 2m + 1, \quad N_c/4 < p < 3N_c/4 \\
		\binom{2\lfloor (m+1)/2\rfloor}{\max(\lfloor (m + 1)/2\rfloor - p, p-\lfloor 3m/2\rfloor - 1} & N_c = 2m + 1, \quad\text{otherwise}.
		\end{cases}
\end{equation}
The prefactor comes from the $N_c - 1$ Majorana fermions on each site, which do not affect the leading-order energy. Although we have not proved that this is the global minimum leading-order energy, nor that these are all the states with that energy, we have found that this is true in the many cases we have checked.

Provided this holds in general, we can use \eqref{eq:strong_coupling_leading_energy_nc} and \eqref{eq:sc_energy_split} to read off the leading-order fundamental string tension in the massless theory for any $N_c$:
\begin{equation}
	\quad\lim_{ga\to\infty} \frac{E^{(0)}(N_c, p = 1) - E^{(0)}(N_c, p = 0)}{g^2 L} = -\frac{1}{4N_c} + \begin{cases}
		1/8 & N_c\text{ even},\\
		0 & N_c\text{ odd}.
	\end{cases}
\end{equation}

\subsection{Perturbation theory}\label{sec:strong_coupling_bw}

In general, to solve the eigenvalue problem $(H_0 + xV)\ket{\psi} = (E_0 + E_1 + E_2 + \ldots)\ket{\psi}$ perturbatively in $x$, it is more efficient to use Brillouin-Wigner perturbation theory (especially in light of the large degeneracy \eqref{eq:sc_leading_degeneracy}). See~\cite{Dempsey:2023fvm} for an example of this approach applied to the lattice Hamiltonian for $\grSU(2)$ adjoint QCD$_2$. However, here we will only work to second-order in $x$ for the eigenvalue, and at this order there is no difference between the Brillouin-Wigner approach and the more familiar Rayleigh-Schr\"odinger theory, so in the following we will use the latter.

Our first step is to diagonalize $V$ on the subspace of degenerate ground states of $H_0$. Since the action of $V$ can only change the representation on one link, $V$ cannot connect the ground states with different patterns of link representations, so it is block diagonal. On a block with pattern $(\bm{r}_0, \bm{r}_1)$, we can follow the discussion in Section~\ref{sec:suN_hamiltonian} to find the projection of the Hamiltonian to the degenerate ground-state subspace (i.e., the $2^{N(N_c-1)/2}$-dimensional space on which the Majorana fermions $\lambda_{n,j}$ act):
\begin{equation}\label{eq:h1}
\begin{split}
	H^{(1)}(\bm{r}_0, \bm{r}_1) = -\frac{i}{2a}\sum_{k=0}^{\frac{N}{2}-1}\sum_{j,j'=1}^{N_c-1} \Big(&D_{jj'}^{11;11}(\bm{r}_1, \bm{r}_1; \bm{r}_0, \bm{r}_0)\times  \lambda_{2k,j} \lambda_{2k+1,j'} \\
	+{}&D_{jj'}^{11;11}(\bm{r}_0, \bm{r}_0; \bm{r}_1, \bm{r}_1) \times \lambda_{2k+1,j} \lambda_{2k+2,j'}\Big)\,.
\end{split}
\end{equation}
This projection of the Hamiltonian takes the form of a Majorana chain with nearest-neighbor interactions. The first-order corrections to the energy levels are given by the eigenvalues of $H^{(1)}$, and at subsequent orders we work with the eigenstates of $H^{(1)}$. In the following, we will focus on the lowest-energy states in each universe, so we only need to compute the ground state $\ket{\psi_0}$ of $H^{(1)}$ and its eigenvalue $E_1$. The solution of Majorana chains of the form \eqref{eq:h1} is reviewed in Appendix~\ref{app:majorana_chain}.

Once we determine $\ket{\psi_0}$, the result of second-order perturbation theory is the next-to-leading energy correction
\begin{equation}
	E^{(2)} = g x^{3/2}\sum_{\ket{\psi'}} \frac{\braket{\psi_0|V|\psi'}\braket{\psi'|V|\psi_0}}{E^{(0)} - E^{(0)'}}\,,
\end{equation}
where $E^{(0)'}$ is the leading-order energy of $\ket{\psi'}$, and we exclude any $\ket{\psi'}$ that are in the ground state subspace from the sum. Again, since the action of $V$ can only change the representation on one link, we can write this as a sum over the link acted upon and the representation it was changed to (along with multiplicity labels $e_L$ and $e_R$ for the sites surrounding that link). We find
{\small
\begin{align}\label{eq:sc_epsilon_2_first}
	&E^{(2)} = \frac{gx^{3/2}}{2}\sum_{n=0}^{\frac{N}{2} - 1} \Bigg(\sum_{\substack{\bm{r'}_{2n}\\e_L,e_R}}\sum_{j,k,j',k'=1}^{N_c - 1} \frac{D_{jk}^{1,1;e_L,e_R}(\bm{r}_1,\bm{r}_1; \bm{r}_0, \bm{r'}_{2n}) D_{j'k'}^{e_L,e_R;1,1}(\bm{r}_1,\bm{r}_1; \bm{r'}_{2n}, \bm{r}_0)\langle \lambda_{2n,j'}\lambda_{2n+1,k'} \lambda_{2n,j} \lambda_{2n+1,k}\rangle}{C_2(\bm{r'}_{2n}) - C_2(\bm{r}_{0})}\nonumber \\
	+{}&\sum_{\substack{\bm{r'}_{2n+1}\\e_L,e_R}} \sum_{j,k,j',k'=1}^{N_c-1} \frac{D_{jk}^{1,1;e_L,e_R}(\bm{r}_0, \bm{r}_0; \bm{r}_1, \bm{r'}_{2n+1}) D_{j'k'}^{e_L,e_R;1,1}(\bm{r}_0, \bm{r}_0; \bm{r'}_{2n+1}, \bm{r}_1) \langle \lambda_{2n+1,j'}\lambda_{2n+2,k'} \lambda_{2n+1,j}\lambda_{2n+2,k}\rangle}{C_2(\bm{r'}_{2n+1}) - C_2(\bm{r}_{1})}	\Bigg)
\end{align}
}%
The expectations of the Majorana fermions $\lambda$ are evaluated with respect to $\ket{\psi_0}$. Using the Clifford algebra, we can rewrite the expectation in the first sum as
\begin{equation}
	\begin{split}
		\langle \lambda_{2n,j'}\lambda_{2n+1,k'} \lambda_{2n,j} \lambda_{2n+1,k}\rangle = -\left\langle\left(\delta_{jj'} + \frac{1}{2} [\lambda_{2n,j'},\lambda_{2n,j}]\right)\left(\delta_{kk'} + \frac{1}{2} [\lambda_{2n+1,k'},\lambda_{2n+1,k}]\right)\right\rangle\,,
	\end{split}
\end{equation}
and likewise for the expectation in the second sum. Using the expression \eqref{eq:matrix_element_6j} for $D_{jk}$, which in this case has only one term in the sum because $\bm{r}\neq\bm{r'}$, we find that the commutator terms vanish when inserted into \eqref{eq:sc_epsilon_2_first}. Thus, $\epsilon_2$ simplifies to
\begin{equation}\label{eq:sc_epsilon_2}
\begin{split}
	E^{(2)} = -\frac{g x^{3/2} N}{4} \Bigg(&\sum_{\substack{\bm{r'}_{0}\\e_L,e_R}}\sum_{j,k=1}^{N_c-1} \frac{D_{jk}^{1,1;e_L,e_R}(\bm{r}_1,\bm{r}_1; \bm{r}_0, \bm{r'}_{0}) D_{jk}^{e_L,e_R;1,1}(\bm{r}_1,\bm{r}_1; \bm{r'}_{0}, \bm{r}_0)}{C_2(\bm{r'}_{0}) - C_2(\bm{r}_{0})} \\
	+{}&\sum_{\substack{\bm{r'}_{1}\\e_L,e_R}} \sum_{j,k=1}^{N_c-1} \frac{D_{jk}^{1,1;e_L,e_R}(\bm{r}_0, \bm{r}_0; \bm{r}_1, \bm{r'}_{1}) D_{jk}^{e_L,e_R;1,1}(\bm{r}_0, \bm{r}_0; \bm{r'}_{1}, \bm{r}_1)}{C_2(\bm{r'}_{1}) - C_2(\bm{r}_{1})}	\Bigg)\,.
\end{split}
\end{equation}
In particular, we see that $E_2$ actually does not depend on the ground state $\ket{\psi_0}$ of $H^{(1)}$, which significantly simplifies the calculation. (In the language of Brillouin-Wigner perturbation theory, we would say that the second-order Hamiltonian is proportional to the identity, which was found in~\cite{Dempsey:2023fvm} for the case of $\grSU(2)$.)

We are also interested in the strong-coupling expansion of the fermion bilinear condensate
\begin{equation}
	\left\langle \tr\left(\bar\psi \psi\right)\right\rangle = \frac{1}{L}\left.\frac{\partial E_0}{\partial m}\right|_{m = 0}\,.
\end{equation}
On the lattice, we can compute this using
\begin{equation}
	\left\langle \tr\left(\bar\psi \psi\right)\right\rangle = \frac{1}{Na}\langle H_\text{mass}\rangle\,, \qquad H_\text{mass} = -\frac{i}{2}\sum_{n=0}^{N-1} (-1)^n \chi^A_n U^{AB}_n \chi^B_{n+1}\,.
\end{equation}
Since $H_\text{mass}$ only differs from $xV$ by the factor of $(-1)^n$, the calculation is similar to the perturbation theory for the energy. We find
\begin{equation}
	\left\langle \tr\left(\bar\psi \psi\right)\right\rangle = \left\langle \tr\left(\bar\psi \psi\right)\right\rangle^{(0)} + \left\langle \tr\left(\bar\psi \psi\right)\right\rangle^{(1)} + \mathcal{O}\left(x^2\right)\,,
\end{equation}
where
\begin{equation}
\begin{split}
	\left\langle \tr\left(\bar\psi \psi\right)\right\rangle^{(0)} &= \frac{1}{Na}\braket{\psi_0|H_\text{mass}|\psi_0}\,,\\
	\left\langle \tr\left(\bar\psi \psi\right)\right\rangle^{(1)} &= -\frac{x^2}{2a} \Bigg(\sum_{\substack{\bm{r'}_{0}\\e_L,e_R}}\sum_{j,k=1}^{N_c-1} \frac{D_{jk}^{1,1;e_L,e_R}(\bm{r}_1,\bm{r}_1; \bm{r}_0, \bm{r'}_{0}) D_{jk}^{e_L,e_R;1,1}(\bm{r}_1,\bm{r}_1; \bm{r'}_{0}, \bm{r}_0)}{C_2(\bm{r'}_{0}) - C_2(\bm{r}_{0})} \\
	&\qquad \qquad -\sum_{\substack{\bm{r'}_{1}\\e_L,e_R}} \sum_{j,k=1}^{N_c-1} \frac{D_{jk}^{1,1;e_L,e_R}(\bm{r}_0, \bm{r}_0; \bm{r}_1, \bm{r'}_{1}) D_{jk}^{e_L,e_R;1,1}(\bm{r}_0, \bm{r}_0; \bm{r'}_{1}, \bm{r}_1)}{C_2(\bm{r'}_{1}) - C_2(\bm{r}_{1})}	\Bigg)\,.
\end{split}
\end{equation}

\subsection{$\grSU(3)$ and $\grSU(4)$}\label{sec:strong_coupling_su3}

Here we will use the perturbation theory developed in Section~\ref{sec:strong_coupling_bw} to derive expansions for the ground state energies in each universe for the $\grSU(3)$ and $\grSU(4)$ theories, as well as for the fermion bilinear condensate in each universe of these theories.

Let's start with $\grSU(3)$. The leading-order ground states are given in Table~\ref{tab:strong_coupling_vacua}. To evaluate the expressions in Section~\ref{sec:strong_coupling_bw}, we need the values of $D_{jk}$ for various representations. These are given in Appendix~\ref{app:su3_6j}. Using these group theory results, we can evaluate, for instance, \eqref{eq:h1} for the $p = 0$ and $p = 1$ universes. These are
\begin{equation}
\begin{split}
	H^{(1)}_{\grSU(3),p=0}(\bm{1},\bm{8}) &= -\frac{i}{2a}\sum_{k=0}^{\frac{N}{2}-1}\left(\frac{5}{2}\lambda_{2k+1,1} \lambda_{2k+2,1} + \frac{3}{2}\lambda_{2k+1,2} \lambda_{2k+2,2}\right)\,,\\
	H^{(1)}_{\grSU(3),p=1}(\bm{3},\bm{3}) &= -\frac{i}{64a}\sum_{n=0}^{N - 1}\Big(25\lambda_{n,1} \lambda_{n+1,1} + 15\sqrt{3}(\lambda_{n,1}\lambda_{n+1,2}-\lambda_{n,2}\lambda_{n+1,1}) \\
	&\phantom{=-\frac{i}{64}\sum_{n=0}^{N - 1}\Big(} - 27\lambda_{n,2} \lambda_{n+1,2}\Big)\,.
\end{split}
\end{equation}
Using the method given in Appendix~\ref{app:majorana_chain}, we find that the ground state energy of $H^{(1)}_{\grSU(3),p=0}(\bm{1},\bm{8})$ is $-N/a$ and the ground state energy of $H^{(1)}_{\grSU(3),p=1}(\bm{3},\bm{3})$ is $-c_1 N/a$ where
\begin{equation}
	c_1 = \frac{1}{64\sqrt{2}\pi}\int_0^{2\pi} dk\,\sqrt{1351-\cos k} \approx 0.812199\,.
\end{equation}
We can also evaluate $E^{(2)}$ using \eqref{eq:sc_epsilon_2} and the results given in Appendix~\ref{app:su3_6j}. We find
\begin{equation}
	E^{(2)}_{p=0} = -\frac{1}{6}Ngx^{3/2}\,, \qquad E^{(2)}_{p=1} = -\frac{201}{512}Ngx^{3/2}\,.
\end{equation}

Putting this together, we find
\begin{equation}
\begin{split}
	\frac{E^{\grSU(3)}_{p=0}}{N} &= g^2a\left(\frac{3}{4} - x - \frac{1}{6}x^2 + \mathcal{O}\left(x^3\right)\right)\,,\\
	\frac{E^{\grSU(3)}_{p=1}}{N} &= g^2a\left(\frac{2}{3} - c_1 x - \frac{201}{512}x^2 + \mathcal{O}\left(x^3\right)\right)\,,
\end{split}
\end{equation}
(Note that the ground state in the $p = 0$ universe is doubly-degenerate to all orders in perturbation theory due to the mixed anomalies discussed in Section~\ref{sec:suN_symmetry}, and $E_{p = 2}^{\grSU(3)} = E_{p = 1}^{\grSU(3)}$). From this, we find a fundamental string tension of
\begin{equation}
	\frac{E^{\grSU(3)}_{p=1} - E^{\grSU(3)}_{p=0}}{g^2L} = -\frac{1}{12} + (1-c_1)x - \frac{347}{1536}x^2 + \mathcal{O}\left(x^3\right)\,.
\end{equation}
The chiral condensate can be computed similarly. In the $p = 1$ universe it vanishes, and in the $p = 0$ universe the two vacua have opposite values,
\begin{equation}
	\left\langle \tr\left(\bar\psi \psi\right)\right\rangle _{p=0}^{\grSU(3)} = \pm\frac{1}{a}\left(1 - \frac{1}{3}x + \mathcal{O}\left(x^2\right)\right)\,.
\end{equation}
As an illustration of the utility of the strong coupling expansion, we can estimate the continuum value of $\langle \tr\left(\bar\psi\psi\right)\rangle^{\grSU(3)}_{p = 0}$ using a Pad\'e approximant. In order to obtain a quantity that is proportional to $g$ in the $x\to\infty$ limit, we replace $1 - \frac{1}{3}x$ with $\left(1 + \frac{2}{3}x\right)^{-1/2}$. This gives a continuum estimate of $\left\langle \tr\left(\bar\psi\psi\right)\right\rangle^{\grSU(3)}_{p = 0} = \pm g\sqrt{\frac{3}{2}} \approx \pm 1.22g$. In Figure~\ref{fig:vev} we show that this agrees well with a numerical lattice calculation.

For the $\grSU(4)$ theory, we can carry out an exactly analogous calculation, but there are many more $6j$-symbols to compute and transitions to consider. They are given in Appendix \ref{app:su3_6j}. To find the first-order energy corrections, we need the ground state energies of
\begin{equation}
\begin{split}
	H^{(1)}_{\grSU(4),p=0}(\bm{15},\bm{6}) = -\frac{i}{2a}\sum_{k=0}^{\frac{N}{2} - 1}\Bigg\lbrack&\vec{\lambda}_{2k}^{T}\begin{pmatrix}
		\frac{25}{12} & 0 & 0 \\
		0 & \frac{2}{5} & \frac{7}{10} \\
		0 & \frac{7}{10} & \frac{49}{40}
	\end{pmatrix}\vec{\lambda}_{2k+1} \\
	&+\vec{\lambda}_{2k+1}^{T}\begin{pmatrix}
		0 & 0 & 0 \\
		0 & \frac{121}{100} & -\frac{77}{100} \\
		0 & -\frac{77}{100} & \frac{49}{100}
	\end{pmatrix}\vec{\lambda}_{2k+2}\Bigg\rbrack,\\
	H^{(1)}_{\grSU(4),p=1}(\bm{20},\bm{\overline{4}}) = -\frac{i}{2a}\sum_{k=0}^{\frac{N}{2} - 1}\Bigg\lbrack&\vec{\lambda}_{2k}^{T}\begin{pmatrix}
		\frac{325}{288} & \frac{181}{288}\sqrt{\frac{5}{2}} & \frac{133}{288}\sqrt{\frac{5}{2}} \\
 		\frac{181}{288}\sqrt{\frac{5}{2}} & \frac{6541}{2880} & -\frac{707}{2880} \\
 		\frac{133}{288}\sqrt{\frac{5}{2}} & -\frac{707}{2880} & \frac{2989}{2880}
	\end{pmatrix}\vec{\lambda}_{2k+1} \\
	&+\vec{\lambda}_{2k+1}^{T}\begin{pmatrix}
		\frac{5}{6} & \frac{2}{3} \sqrt{\frac{2}{5}} & \frac{7}{3 \sqrt{10}} \\
		\frac{2}{3} \sqrt{\frac{2}{5}} & \frac{16}{75} & \frac{28}{75} \\
 		\frac{7}{3 \sqrt{10}} & \frac{28}{75} & \frac{49}{75}
	\end{pmatrix}\vec{\lambda}_{2k+2}\Bigg\rbrack\,.\\
\end{split}
\end{equation}
The ground state energies of these chains are $-c_2N/a$ and $-c_3N/a$ respectively, with
\begin{equation}
	c_2 \approx 1.35161, \qquad c_3 \approx 1.53406.
\end{equation}
The ground state energy of $H^{(1)}_{\grSU(4),p=1}(\bm{4},\bm{\overline{20}})$ is also $-c_3 N/a$. We can also compute the second-order energy corrections using \eqref{eq:sc_epsilon_2} and the results given in Appendix~\ref{app:su3_6j}. Putting everything together, we find
\begin{equation}
\begin{split}
	\frac{E^{\grSU(4)}_{p = 0}}{N} &= g^2 a\left(\frac{13}{8} - c_2 x -\frac{6104881}{6912000}x^2 + \mathcal{O}\left(x^3\right)\right)\,,\\
	\frac{E^{\grSU(4)}_{p = 1}}{N} &= g^2 a\left(\frac{27}{16} - c_3 x -\frac{902873}{1536000}x^2 + \mathcal{O}\left(x^3\right)\right)\,.
\end{split}
\end{equation}

This gives a fundamental string tension of
\begin{equation}
	\frac{E^{\grSU(4)}_{p=1} - E^{\grSU(4)}_{p=0}}{g^2L} = \frac{1}{16} - (c_3-c_2)x + \frac{816781}{2764800}x^2 + \mathcal{O}\left(x^3\right)\,.
\end{equation}
We can also compute the corrections to the chiral condensate in each of the strong-coupling ground states; for the $p = 0$ universe and the two ground states in the $p = 1$ universe, the results are
\begin{equation}
\begin{split}
	\left\langle \tr\left(\bar\psi \psi\right)\right\rangle_{p=0}^{\grSU(4)} &= \frac{1}{a}\left(-c_4 + \frac{1159631}{3456000} x + \mathcal{O}\left(x^2\right)\right)\,,\\
	\left\langle \tr\left(\bar\psi \psi\right)\right\rangle_{p=1}^{\grSU(4)} &= \pm\frac{1}{a}\left(c_5 + \frac{9459}{256000} x + \mathcal{O}\left(x^2\right)\right)\,,\\
\end{split}
\end{equation}
where
\begin{equation}
	c_4 \approx 0.502073\,, \qquad c_5 \approx 0.684369\,.
\end{equation}

\section{Numerical results for $\grSU(3)$}\label{sec:numerics}

Here we use exact diagonalization to compute numerical results for the theory with $G = \grSU(3)$. In Sections \ref{sec:numerics_truncation} and \ref{sec:numerics_hamiltonian}, we discuss how we truncate the Hilbert space to a finite basis, and how we compute the Hamiltonian matrix elements on this basis. In Section~\ref{sec:numerics_massless}, we compute the spectra of the $p = 0$ and $p = 1$ universes along with the fermion bilinear condensate and the string tension when $m = 0$. In Section~\ref{sec:numerics_susy}, we compute the spectrum at the supersymmetric mass $m_\text{SUSY} = g\sqrt{\frac{3}{2\pi}}$.

\subsection{Basis truncation}\label{sec:numerics_truncation}

In order to compute a Hamiltonian matrix from \eqref{eq:hamiltonian}, our first step is to truncate the infinite Hilbert space spanned by states of the form \eqref{eq:gauge_invariant_state}. We will use a truncation scheme motivated by an exact understanding of the low-lying states in a particular large-mass continuum limit: $m\gg g$ with $a = m^{-1}$. Many terms of the Hamiltonian \eqref{eq:hamiltonian} vanish when $a = m^{-1}$, and so an exact analysis is possible. In Appendix D of~\cite{Dempsey:2023fvm}, it is shown for the $\grSU(2)$ theory that in this limit, the low-lying states in universe $p$ have link representations
\begin{equation}
    \bm{r}_n \in \begin{cases} 
        \{ \bm{r} \} & n\text{ even},\\
        \bm{R}\otimes \bm{r} & n\text{ odd}
    \end{cases}
\end{equation}
for some fixed representation $\bm{r}$ having $k_{\bm{r}} = p$ (and with $\bm{R} = \bm{2}$, as is appropriate for $\grSU(2)$). Moreover, these states have energy levels $\frac{g^2 L}{2}C_2(\bm{r})$, as we should expect when the adjoint fermions decouple.

This argument carries over to a general group, and so we will keep only the representations that appear in the first few of these large-mass eigenstates. In particular, we will fix a truncation parameter $c_\text{max}$ and keep the representations appearing in the eigenstates corresponding to the first $c_\text{max}$ distinct Casimir eigenvalues above the lowest one in a given universe.

We will focus on the $\grSU(3)$ theory, so let us explain in detail what happens in this case. The Casimir eigenvalue of the representation with Dynkin label $(m,n)$ is
\begin{equation}
	C_{m,n} = \frac{1}{3}\left(m^2 + n^2 + 3(m + n) + mn\right)\,.
\end{equation}
From this one can show that the representations in the $p = 0$ and $p = 1$ universes with the lowest Casimir eigenvalues are
\begin{equation}\label{eq:su3_casimirs}
	\begin{aligned}
	p &= 0 &\qquad p &= 1 \\
	\hline
	C_{0,0} &= 0 &\qquad C_{1,0} &= 4/3 \\
	C_{1,1} &= 3 &\qquad C_{0,2} &= 10/3 \\
	C_{2,0} = C_{0,2} &= 6 &\qquad C_{2,1} &= 16/3 \\
	C_{2,2} &= 8 &\qquad C_{1,3} &= 25/3
	\end{aligned}
\end{equation}
Thus, for instance, if we take $c_\text{max} = 1$ in the $p = 1$ universe, then we want to be able to represent the large mass states corresponding to representations $(1,0)$ and $(0,2)$. This means that we include these representations along with those in the products with $\bm{R} = (1,1)$:
\begin{equation}
\begin{split}
    (1,0)\otimes (1,1) &= (1,0)\oplus (0,2)\oplus (2,1)\,, \\ (0,2)\otimes (1,1) &= (1,0)\oplus (0,2) \oplus (2,1) \oplus (1,3)\,.
\end{split}
\end{equation}
So, in total, we truncate the space of $\grSU(3)$ representations to $\{(1,0),(0,2),(2,1),(1,3)\}$.

In Table~\ref{tab:state_counts}, we give the number of states in the $p = 0$ and $p = 1$ universes of $\grSU(3)$ for different numbers of sites and truncation parameters. 

\begin{table}[t]
  \centering
  \begin{tabular}{p{1.5cm}R{2cm}R{2cm}R{2cm}R{2cm}R{2cm}R{2cm}}
    \multicolumn{7}{c}{$p = 0$} \\
    \toprule
    $c_\text{max}$ & 0 & 1 & 2 & 3 & 10 & 20 \\
    \midrule
    $N = 2$ & 12 & 44 & 76 & 96 & 352 & 728 \\
    $N = 4$ & 272 & 2,832 & 5,552 & 8,224 & 38,272 & 84,928 \\
    $N = 6$ & 6,336 & 209,216 & 506,752 & 894,336 & 5,427,520 & 12,877,184 \\
    \bottomrule
  \end{tabular}\\[1cm]
  \begin{tabular}{p{1.5cm}R{2cm}R{2cm}R{2cm}R{2cm}R{2cm}R{2cm}}
    \multicolumn{7}{c}{$p = 1$} \\
    \toprule
    $c_\text{max}$ & 0 & 1 & 2 & 3 & 10 & 20 \\
    \midrule
    $N = 2$ & 24 & 40 & 66 & 92 & 236 & 460 \\
    $N = 4$ & 1,088 & 2,400 & 5,064 & 7,760 & 24,112 & 51,504 \\
    $N = 6$ & 50,688 & 161,632 & 470,976 & 817,952 & 3,199,328 & 7,503,520 \\
    \bottomrule
  \end{tabular}
  \caption{The number of states in the $p = 0$ and $p = 1$ universes for $\grSU(3)$ as a function of $N$ and $c_\text{max}$.}
  \label{tab:state_counts}
\end{table}

\subsection{Computing the Hamiltonian}\label{sec:numerics_hamiltonian}

In Section~\ref{sec:suN_hamiltonian}, we showed that the gauge kinetic term is diagonal in our basis while the fermion kinetic and fermion mass terms require the calculation of 6$j$-symbols for the group $G$.

For the calculations in this paper, we computed these 6$j$-symbols by explicitly constructing the Clebsch-Gordan symbols and contracting them as in \eqref{eq:matrix_element_6j}. We have used two publicly available tools in order to construct the Clebsch-Gordan symbols. The first is \texttt{GroupMath}~\cite{Fonseca:2020vke}, which supports the calculation of invariant tensors for any set irreducible representations of any group $G$. This is extremely useful, but in practice the algorithm relies upon computing nullspaces of large matrices in exact arithmetic, which quickly becomes intractable. Thus, for our $\grSU(3)$ numerics, we have used the package \href{https://github.com/QuantumKitHub/SUNRepresentations.jl}{\texttt{SUNRepresentations.jl}} which uses the Gelfand-Tseytlin basis to carry out a more rapid construction of Clebsch-Gordan symbols for the $\grSU(N_c)$ groups.

Unfortunately, neither method can possibly scale well to larger groups, because the Clebsch-Gordan symbols become very large tensors. Indeed, $\dim \bm{R} = 2^{(\dim G-\rk G)/2}$, and so the $(\textbf{adj},\bm{R},\bm{\bar{R}})$ symbols appearing in the Majorana operators for $\grSU(N_c)$ have up to $(N_c^2-1)\times 2^{N_c(N_c-1)}$ components. For $N_c = 3$ this is only 512, but for $N_c = 4$ it is 61,440, and for $N_c = 5$ it is 25,165,824. Worse yet, this is only one of the many sets of Clebsch-Gordan symbols required to evaluate all possible 6$j$-symbols appearing in \eqref{eq:matrix_element_6j}. For this reason, it is of great interest to explore methods for calculating the 6$j$-symbols that do not rely on the explicit construction of Clebsch-Gordan symbols, such as~\cite{butler_wybourne}, but we leave this for future work.

Once we compute the Hamiltonian from the $6j$-symbols and the equations in Section~\ref{sec:suN_hamiltonian}, we diagonalize it using \texttt{SLEPc}~\cite{petsc-user-ref,petsc-efficient,slepc-manual,slepc-toms} to obtain eigenvalues. We use the center symmetry and the fermion parity symmetry to break the Hamiltonian into blocks and diagonalize in each sector.\footnote{One could also use charge conjugation symmetry to further split up the $p = 0$ Hilbert space, but we have not implemented the charge conjugation symmetry here.}

\begin{figure}
    \centering
    \includegraphics[width=.8\linewidth]{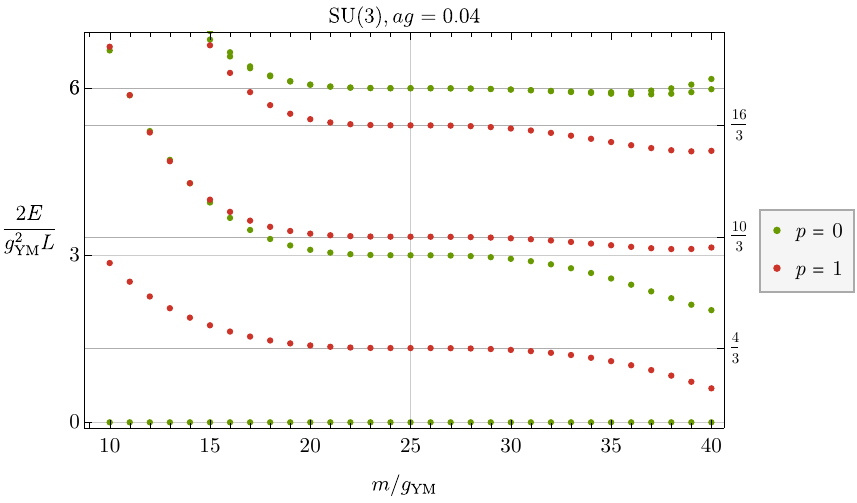}
    \caption{The energy levels for the $\grSU(3)$ theory with $ag = 1/25$ for fermion mass in the vicinity of $m/g = 25$.}
    \label{fig:large_mass}
\end{figure}

One simple consistency check is to compute the spectrum in the large-mass limit described in Section~\ref{sec:numerics_truncation} and see that the energy differences are indeed given by
\begin{equation}
    \Delta E_m = \frac{g^2 L}{2} C_2(\bm{r}_m)
\end{equation}
for the lowest few Casimir eigenvalues $C_2(\bm{r}_m)$. In Figure~\ref{fig:large_mass}, we plot the energy differences above the $p = 0$ vacuum for the first few states in the $p = 0$ and $p = 1$ universes at $m/g = (ag)^{-1} = 25$. We see that both sets of energy levels are as expected from \eqref{eq:su3_casimirs}.

\subsection{Massless theory}\label{sec:numerics_massless}

\begin{figure}
    \centering%
    \begin{subfigure}[t]{\linewidth}%
    \centering
    \includegraphics[width=.9\linewidth]{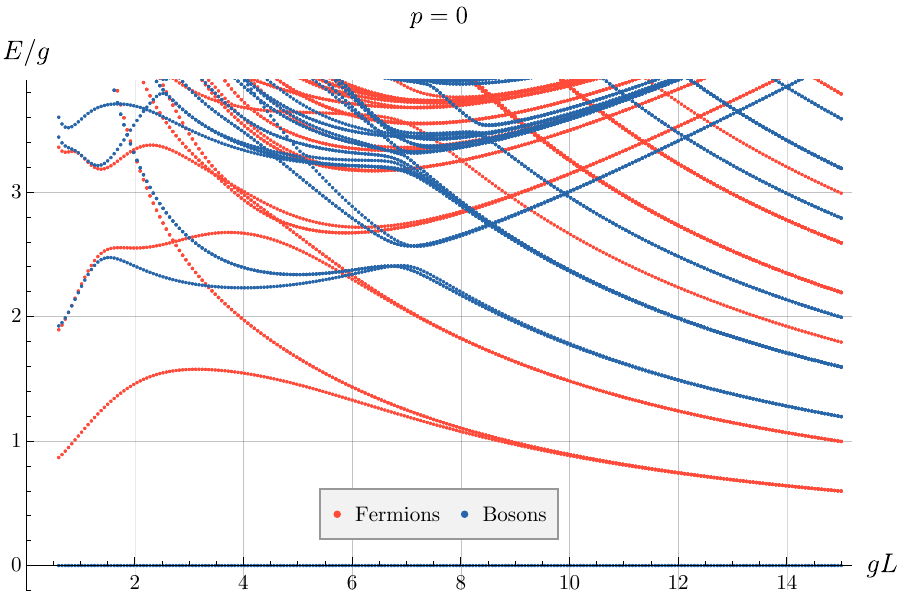}
    \caption{}
    \label{fig:massless_spectrum_p0}
    \end{subfigure}\\[1em]
    \begin{subfigure}[t]{\linewidth}%
    \centering
    \includegraphics[width=.9\linewidth]{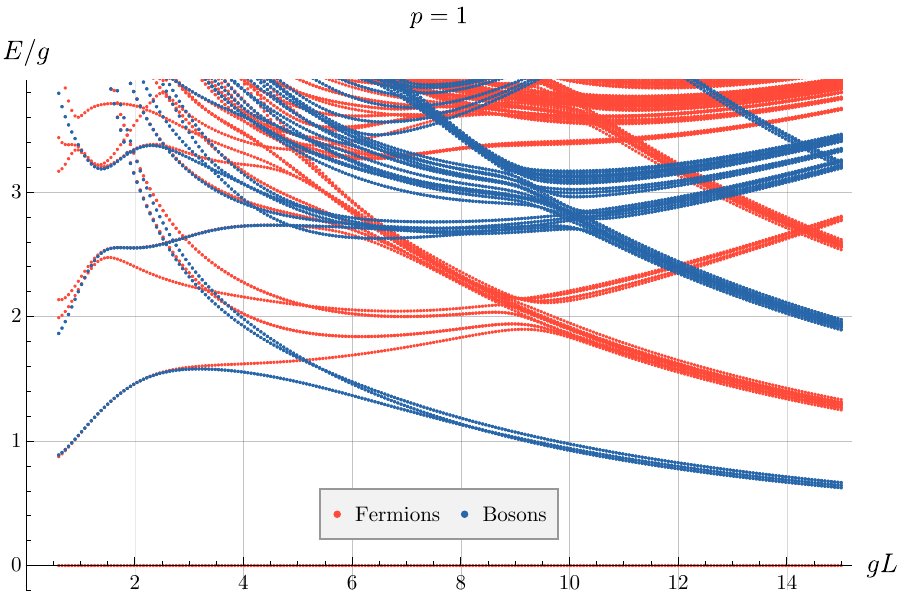}
    \caption{}
    \label{fig:massless_spectrum_p1}
    \end{subfigure}%
    \caption{The spectra in the $p = 0$ and $p = 1$ universes of $\grSU(3)$ massless adjoint QCD$_2$ from a lattice with $N = 6$ sites as a function of the circle length $L$. The truncation parameter is $c_\text{max} = 3$.}
    \label{fig:massless_spectra}
\end{figure}

In Figure~\ref{fig:massless_spectra}, we give the spectrum of the Hamiltonian \eqref{eq:hamiltonian} for $N = 6$ sites at $m = 0$ as a function of $gL$, with truncation parameter $c_\text{max} = 3$ (empirically, the spectrum is very well-converged already at $c_\text{max} = 3$, except in the $gL\ll 1$ limit). All states in the $p = 0$ universe are doubly-degenerate at $m = 0$, which follows from the mixed 't Hooft anomaly between the chiral symmetry and the charge conjugation symmetry~\cite{Dempsey:2023fvm}.

When $gL \ll 1$, the spectrum can be understood in terms of the effective theory of light modes on a small circle~\cite{Dempsey:2024ofo}. In the continuum limit, the leading-order spectrum at $m = 0$ has energy levels equally spaced by\footnote{On the lattice this spacing is slightly modified, as described in~\cite{Dempsey:2024ofo}. For $N = 6$ sites it is altered by less than 5\%\ from the continuum value.} $g\sqrt{\frac{3}{2\pi}}$, with various numbers of boson and fermion states at each level as described in Figure 2 of~\cite{Dempsey:2024ofo}. Our lattice results are consistent with this zeroth-order expectation. We study the $gL\ll 1$ regime in more detail in Section~\ref{sec:numerics_small_circle}.

When $ga\gg 1$, we can in principle use a lattice strong-coupling expansion to describe the spectrum, as described in Section~\ref{sec:strong_coupling}. However, we have not yet carried out the strong-coupling expansion for excited states, so we cannot compare with the energy gaps in Figure~\ref{fig:massless_spectra}.

\begin{figure}
    \centering
    \includegraphics[width=.6\linewidth]{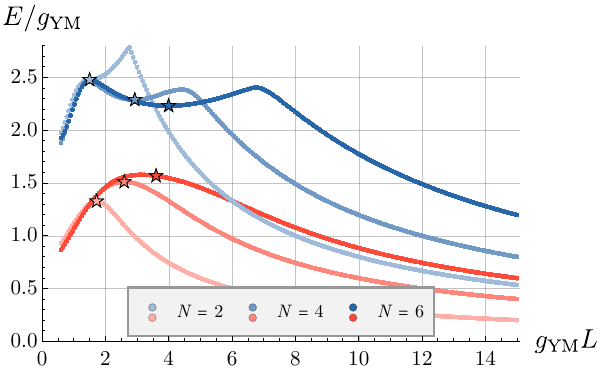}
    \caption{We estimate the masses of the lowest fermionic and bosonic excitations in the $p = 0$ universe by finding the extrema of the plateaus in their energy levels as a function of $gL$ and extrapolating to $N\to\infty$. The extrapolated values are $M_F \approx 1.69g$ and $M_B\approx 2.11g$.}
    \label{fig:plateaus}
\end{figure}

We can obtain information about the continuum spectrum of the $\grSU(3)$ theory on a line from the intermediate region in $gL$ in Figure~\ref{fig:massless_spectra}. By comparing the curves for $N = 2,4,6$, we can see the development of plateaus in the energies which we expect would extend to $gL\to\infty$ if we first took $N\to\infty$. In Figure~\ref{fig:plateaus}, we show these plateaus in the lowest fermionic and bosonic excitations in the $p = 0$ universe. By extrapolating the maximum of the fermion plateau or the minimum of the boson plateau to $N\to\infty$, we find the following rough estimates of the first two bound state energies:
\begin{equation}
    M_F \approx 1.69g\,, \qquad M_B \approx 2.11g\,.
\end{equation}
The values obtained from DLCQ are~\cite{Dempsey:2022uie}
\begin{equation}
    M_F = 1.65g\,, \qquad M_B = 2.27g\,,
\end{equation}
and so our lattice results are roughly consistent. The fermion which is degenerate with the lowest boson excitation in the small circle limit likely becomes the lowest $C$-odd fermion, which according to DLCQ has $M = 2.87g$~\cite{Dempsey:2022uie}, but we would need a larger lattice to reliably extract the continuum mass of this state. The lowest $C$-odd boson has an energy above the two-particle continuum of the lowest fermion, so this would be even more difficult to extract from our lattice data.

\begin{figure}[t]
    \centering
    \begin{subfigure}[t]{0.45\linewidth}
        \centering
        \includegraphics[width=\linewidth]{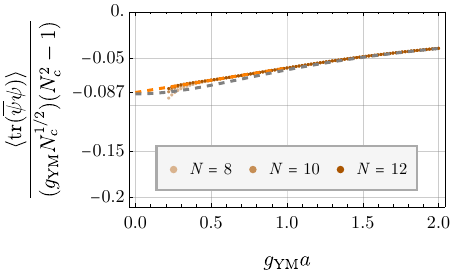}
        \caption{$\grSU(2)$}
        \label{fig:vev_su2}
    \end{subfigure}
    \hspace{.05\linewidth}
    \begin{subfigure}[t]{0.45\linewidth}
        \centering
        \includegraphics[width=\linewidth]{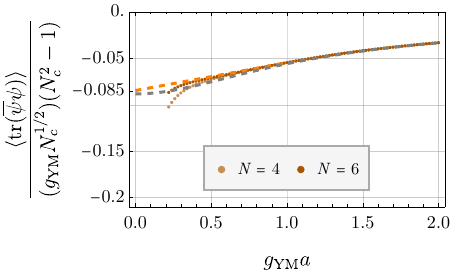}
        \caption{$\grSU(3)$}
        \label{fig:vev_su3}
    \end{subfigure}
    \caption{The VEV of the chiral condensate $\Tr_\text{fund}(\bar\psi\psi)$ obtained from lattice results for the $\grSU(2)$ and $\grSU(3)$ theories. The dashed orange lines are linear extrapolations of the lattice data to $g_\text{YM}a\to 0$, and the dashed gray lines are a Pad\'e approximant to lattice strong coupling expansions up to $\mathcal{O}(x)$ (from \cite{Dempsey:2023fvm} in the case of $\grSU(2)$, and from Section~\ref{sec:strong_coupling_su3} in the case of $\grSU(3)$).}
    \label{fig:vev}
\end{figure}

We can also study the vacuum expectation value of the chiral condensate $\Tr_\text{fund}(\bar\psi \psi)$. We can extract this from the lattice using
\begin{equation}
    \left\langle \Tr_\text{fund}(\bar\psi \psi)\right\rangle = \frac{1}{L}\left\langle H_\text{mass}\right\rangle\,.
\end{equation}
This expression is ambiguous at the massless point in the $p = 0$ universe\footnote{In the $p = 1$ and $p = 2$ universes, the chiral condensate VEV vanishes exactly as a consequence of the mixed 't Hooft anomaly between the chiral symmetry and the charge conjugation symmetry.} because we have two degenerate ground states, which have opposite values of $\langle\Tr_\text{fund}(\bar\psi \psi)\rangle$. We split them by letting $m/g = +\epsilon$ so that we isolate the ground state with the negative chiral condensate. In Figure~\ref{fig:vev}, we plot the chiral condensate in the $p = 0$ universe obtained from the lattice and extrapolate to the continuum value of
\begin{equation}
    \langle\Tr_\text{fund}(\bar\psi \psi)\rangle \approx -0.085(N_c^2-1)\sqrt{g^2 N_c}\,.
\end{equation}
Furthermore, we compare with results from the $\grSU(2)$ lattice theory to show that the chiral condensate values are roughly equal when rescaled in this way, as demonstrated also with the Euclidean lattice in~\cite{Bergner:2024ttq}.

Finally, we can calculate the difference between ground state energies in the $p = 1$ and $p = 0$ universes, $E_{p = 1} - E_{p = 0}$.
In the continuum treatment of adjoint QCD$_2$ without the four-fermion terms, this quantity vanishes for any size of the circle due to the non-invertible symmetry~\cite{Komargodski:2020mxz,Dempsey:2024ofo}.

\begin{figure}[t]
    \centering
    \includegraphics[width=.5\linewidth]{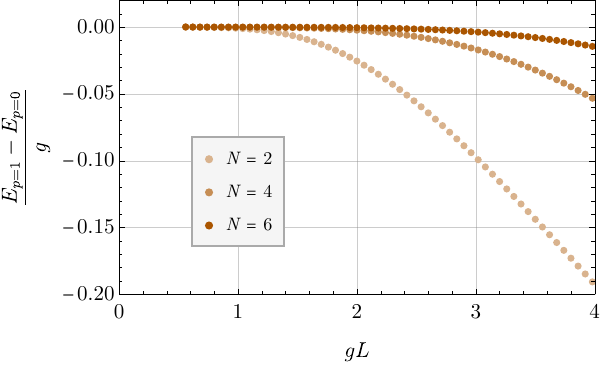}
    \caption{The ground state energy difference $(E_{p=1}-E_{p=0})/g$ in the massless $\grSU(3)$ theory as a function of circle length. This energy difference seems to converge towards zero at any value of $gL$ as we increase $N$.}
    \label{fig:tension}
\end{figure}

In Figure~\ref{fig:tension} we plot $\frac{E_{p = 1} - E_{p = 0}}{g}$ on a finite-size lattice, and we see nascent evidence that it vanishes for all $gL$ when $N\to\infty$. Thus, the non-invertible symmetry appears to be restored in the continuum limit, and we find no observable effects of the four-fermion terms studied in~\cite{Cherman:2019hbq}.

\subsection{Supersymmetric point}\label{sec:numerics_susy}

Adjoint QCD$_2$ is known to exhibit (1,1) supersymmetry at\footnote{For a general group $G$, the number of colors is replaced by the dual Coxeter number $h^\vee(G)$~\cite{Popov:2022vud,Dempsey:2024ofo}.} $m = g\sqrt{\frac{N_c}{2\pi}}$. In this section, we will study the spectrum of the $\grSU(3)$ theory at its supersymmetric point $m = g\sqrt{\frac{3}{2\pi}}$. 

\begin{figure}
    \centering%
    \begin{subfigure}[t]{\linewidth}%
    \centering
    \includegraphics[width=.9\linewidth]{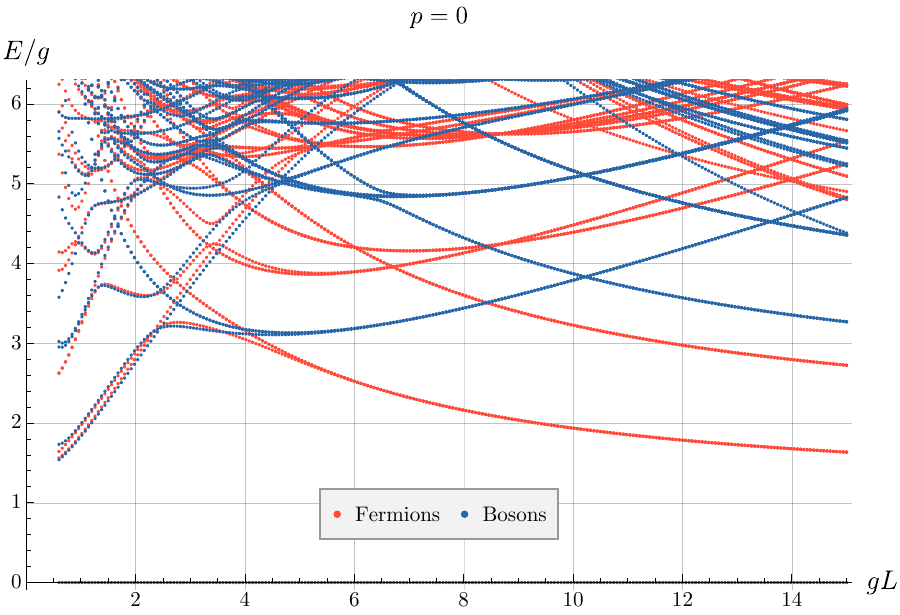}
    \caption{}
    \label{fig:susy_spectrum_p0}
    \end{subfigure}\\[1em]
    \begin{subfigure}[t]{\linewidth}%
    \centering
    \includegraphics[width=.9\linewidth]{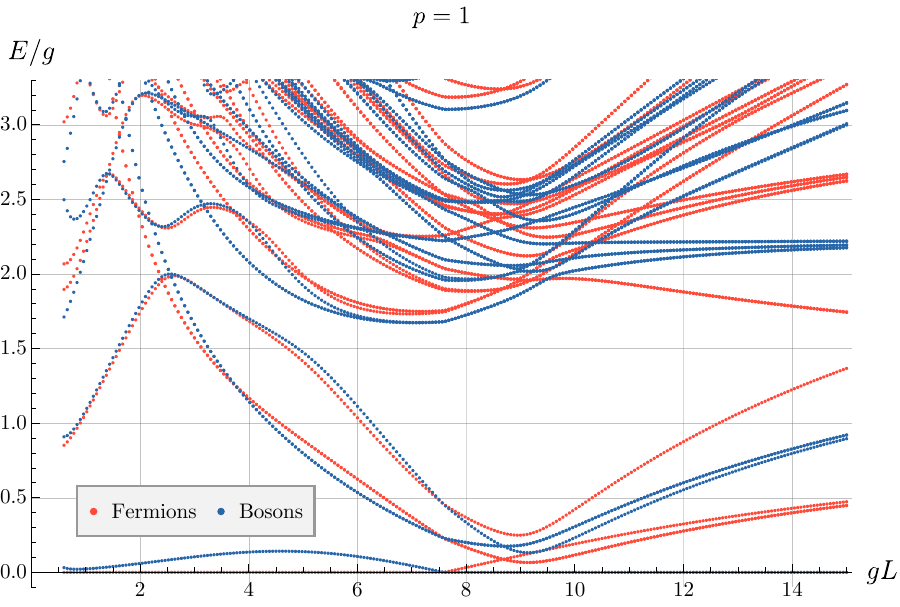}
    \caption{}
    \label{fig:susy_spectrum_p1}
    \end{subfigure}%
    \caption{The spectra in the $p = 0$ and $p = 1$ universes of $\grSU(3)$ adjoint QCD$_2$ at its supersymmetric point $m = g\sqrt{\frac{3}{2\pi}}$ from a lattice with $N = 6$ sites as a function of the circle length $L$.}
    \label{fig:susy_spectra}
\end{figure}

In Figure~\ref{fig:susy_spectra}, we see in both the $p = 0$ and $p = 1$ universes that in the small-circle limit and for intermediate lengths, there is near boson-fermion degeneracy in all states except for the bosonic vacuum of the $p = 0$ universe. This degeneracy is broken in the lattice strong coupling limit $ga\gg 1$, but this regime does not directly correspond to the physical theory.

\begin{figure}[t]
    \centering
    \includegraphics[width=.6\linewidth]{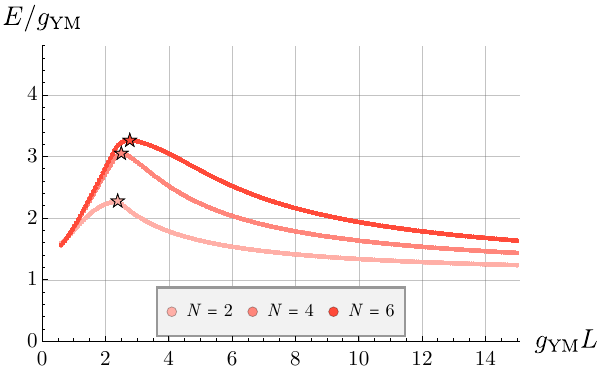}
    \caption{We estimate the mass of the lowest boson-fermion degenerate excitation in the $p = 0$ universe at $m = g\sqrt{\frac{3}{2\pi}}$ by extrapolating the maxima of the lowest fermion energy levels as a function of $gL$ to $N\to\infty$. The extrapolated value is $M_1\approx 3.8g$, roughly consistent with the DLCQ value $M_1\approx 3.5g$.}
    \label{fig:susy_plateau}
\end{figure}

\begin{figure}[t]
    \centering
    \begin{subfigure}[t]{0.45\linewidth}
        \centering
        \includegraphics[width=\linewidth]{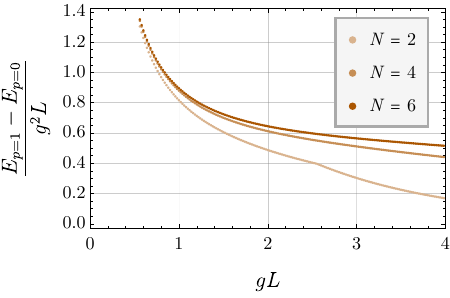}
        \caption{The fundamental string tension in the $\grSU(3)$ theory at $m = m_\text{SUSY}$ as a function of circle length.}
        \label{fig:susy_tension_L}
    \end{subfigure}
    \hspace{.05\linewidth}
    \begin{subfigure}[t]{0.45\linewidth}
        \centering
        \includegraphics[width=\linewidth]{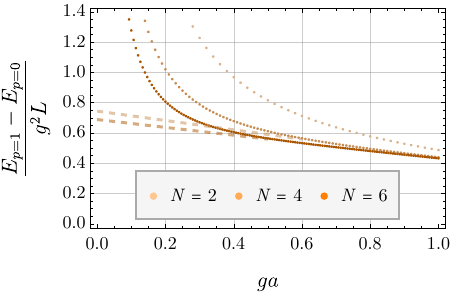}
        \caption{The same data as a function of $(ga)$, with rough extrapolations indicating that $\sigma_1(m = m_\text{SUSY})\approx 0.6g^2$ in the continuum limit.}
        \label{fig:susy_tension_a}
    \end{subfigure}
    \caption{The fundamental string tension $\sigma_1$ extracted from lattice data at $m = m_\text{SUSY}$.}
    \label{fig:susy_tension}
\end{figure}

We can compare the leading-order spectrum in the limit of $gL\to 0$ with Figure 2 of~\cite{Dempsey:2024ofo}. In the $p = 0$ spectrum, the prediction is a unique bosonic vacuum and then two bosons and two fermions each at the next two excited levels. This is reproduced in Figure~\ref{fig:susy_spectrum_p0}. In the $p = 1$ spectrum, the prediction is a boson-fermion degenerate vacuum followed by another boson-fermion degenerate level, and then two bosons and two fermions at the next level. The first two levels are clearly reproduced in Figure~\ref{fig:susy_spectrum_p1}, and there seem to be two bosons and two fermions forming the second excited level, although as described in Section~\ref{sec:numerics_massless} the numerics are not completely converged at small $gL$ due to the relatively small value of $c_\text{max}$ (see Section~\ref{sec:numerics_truncation}).

From the DLCQ results in~\cite{Dempsey:2022uie}, the mass of the lowest (boson-fermion degenerate) excitation in the $p = 0$ universe is found to be $M_1 \approx 3.5g$. We can try to estimate this mass by looking at the plateau in the energy of the lowest fermion excitation as a function of $gL$, and extrapolating to large $N$, as in Section~\ref{sec:numerics_massless}. We find $M_1 \approx 3.8g$, roughly consistent with DLCQ. The plateau maxima for $N = 2,4,6$ are shown in Figure~\ref{fig:susy_plateau}.

We can also compare the ground state energies in the $p = 0$ and $p = 1$ universes to estimate the fundamental string tension
\begin{equation}
	\sigma_1 = \frac{E_{p = 1} - E_{p = 0}}{L}
\end{equation}
at $m = m_\text{SUSY}$. In Figure~\ref{fig:susy_tension}, we plot this energy difference divided by the circle length as functions of both $gL$ and $ga$. To extract the string tension in the continuum limit, we would ideally extrapolate $N\to\infty$ and then take $ga\to 0$. However, with only 6 sites we do not have enough data to perform this extrapolation systematically. By extrapolating the finite-$N$ data in Figure~\ref{fig:susy_tension_a} to $ga\to 0$, we can roughly estimate $\sigma_1(m=m_\text{SUSY}) \approx 0.6g^2$.

\subsection{Antiperiodic fermions}\label{sec:numerics_small_circle}

It is interesting to compare the ground state energies in the $p = 0$ and $p = 1$ universes when the fermions obey antiperiodic boundary conditions. In this case we can also reinterpret our lattice results as corresponding to the theory on a spatial line at temperature $\beta \equiv L^{-1}$. If we take $gL\ll 1$, we can study the high-temperature limit of adjoint QCD$_2$. The difference $E_{p = 1}-E_{p = 0}$ of the ground state energies in different universes is related to the difference in the free energy density with or without the insertion of a fundamental flux tube around the compact direction. When this difference is nonzero, the partition function is sensitive to the insertion of this flux tube, and so the theory is in a confining phase. When the difference becomes zero, as is known to occur in the large-$N_c$ limit and with $m\gg g$~\cite{Semenoff:1996xg}, the theory is in a deconfined phase.

At finite $N_c$, the small-circle limit of adjoint QCD$_2$ with antiperiodic boundary conditions for fermions was first considered in~\cite{Smilga:1994hc,Lenz:1994du}. In~\cite{Lenz:1994du}, the leading-order contribution to the fermion bilinear condensate at $m=0$ and small $gL$ in the $\grSU(2)$ theory is found to be
\begin{align}
	\grSU(2):\qquad \frac{1}{L}\left.\frac{\partial E^{\text{AP}}_{p}}{\partial m}\right|_{m=0} = (-1)^{p+1}\frac{4\pi^{3/2}}{gL^2}\exp\left(-\frac{\pi^{3/2}}{gL}\right)\,.
\end{align}
Integrating this, we find that the energy difference between the two universes behaves at small $m/g$ like
\begin{equation}\label{eq:su2_abc_gap}
    \grSU(2):\qquad \frac{E^{\text{AP}}_{p = 1} - E^{\text{AP}}_{p = 0}}{g} = \left(8\pi^{3/2} \frac{m}{g} + \mathcal{O}\left((m/g)^3\right)\right)(gL)^{-1}\exp\left(-\frac{\pi^{3/2}}{gL}\right)\,.
\end{equation}
For $\grSU(3)$,~\cite{Lenz:1994du} shows that the quadratic condensate vanishes and that the quartic condensate is suppressed as $\exp\left(-\sqrt{\frac{8}{3}}\frac{\pi^{3/2}}{gL}\right)$ at the massless point. From this, we can infer that the energy difference at small $m/g$ is
\begin{equation}\label{eq:su3_abc_gap}
   \grSU(3):\qquad  \frac{E^{\text{AP}}_{p = 1} - E^{\text{AP}}_{p = 0}}{g} = \left( A \left(\frac{m}{g}\right)^2 + \mathcal{O}\left((m/g)^4\right)\right)(gL)^{-2}\exp\left(-\sqrt{\frac{8}{3}}\frac{\pi^{3/2}}{gL}\right)
\end{equation}
for some numerical constant $A$. (Note that the $p = 1$ and $p = 2$ universes are degenerate because they are interchanged by parity or by charge conjugation). The exponential suppression and the power $m$ in the energy splitting can also be understood as an instanton effect~\cite{Smilga:1996dn}. In particular, the exponential suppression is given by the instanton action interpolating between neighboring universes and the power of $m$ is related to the number of would-be fermionic zero modes about the instanton configuration. This perspective is explained further in Appendix~\ref{app:instanton}.

\begin{figure}[t]
    \centering
    \begin{subfigure}[t]{0.45\linewidth}
        \centering
        \includegraphics[width=\linewidth]{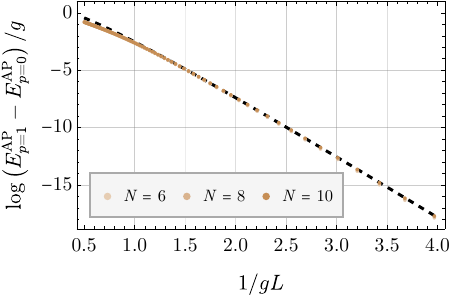}
        \caption{The difference between the ground state energies of the $p = 0$ and $p = 1$ universes for the $\grSU(2)$ theory at $m/g = 0.5$ as a function of the thermal circle length $gL$, compared with a fit of the form $A\cdot (gL)^{-1} \exp\left(-\pi^{3/2}/gL\right)$.}
        \label{fig:su2_abc_gap}
    \end{subfigure}
    \hspace{.05\linewidth}
    \begin{subfigure}[t]{0.45\linewidth}
        \centering
        \includegraphics[width=\linewidth]{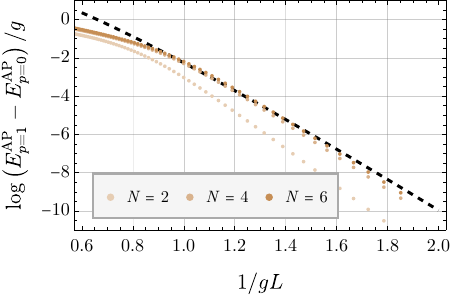}
        \caption{The difference between the ground state energies of the $p = 0$ and $p = 1$ universes for the $\grSU(3)$ theory at $m/g = 0.5$ as a function of the thermal circle length $gL$, compared with a fit of the form $A\cdot (gL)^{-2} \exp\left(-\sqrt{8/3}\pi^{3/2}/gL\right)$.}
        \label{fig:su3_abc_gap}
    \end{subfigure}
    \caption{}
    \label{fig:thermal_energy_gap}
\end{figure}

On the lattice, we can first check that we reproduce this exponential behavior at small $gL$. Indeed, in Figure~\ref{fig:thermal_energy_gap}, we plot the energy difference between the $p = 0$ and $p = 1$ universes for $\grSU(2)$ and $\grSU(3)$ at $m/g = 0.5$. We fit functions of the form \eqref{eq:su2_abc_gap} and \eqref{eq:su3_abc_gap} respectively (with only the overall coefficient determined by the fit). For $\grSU(2)$ we can easily work with 10 sites and take the representation cutoff very high, and the convergence to the analytic result is rapid; for $\grSU(3)$, we have only used 6 sites and set $c_\text{max} = 20$, which already requires a $\sim 10^7$-dimensional Hilbert space (see Table~\ref{tab:state_counts}). Nevertheless, we see good agreement with the $(gL)^{-2}\exp\left(-\sqrt{\frac{8}{3}}\frac{\pi^{3/2}}{gL}\right)$ scaling.

We can also use the numerical data to fit the dependence of the coefficient on $m/g$ for relatively small masses (we work with $m/g\in [0,1]$). For the $\grSU(2)$ theory, we can fit the dependence on $m/g$ at several values of $N$ and then extrapolate $N\to\infty$. This gives
\begin{equation}
\grSU(2):\qquad \frac{E^{\text{AP}}_{p = 1} - E^{\text{AP}}_{p = 0}}{g} \approx 8\pi^{3/2}\left( 0.97\frac{m}{g} + 0.44\frac{m^3}{g^3}\right)(gL)^{-1}\exp\left(-\frac{\pi^{3/2}}{gL}\right)\,.
\end{equation}
For the $\grSU(3)$ theory, we do not have as many values of $N$ to work with. We do clearly see that the prefactor scales like $(m/g)^2$ at small mass for any fixed $N$, and from the $N = 6$ data we can estimate a coefficient of roughly $3.3\times 10^3$.

\begin{figure}[t]
	\centering
	\includegraphics[width=0.5\linewidth]{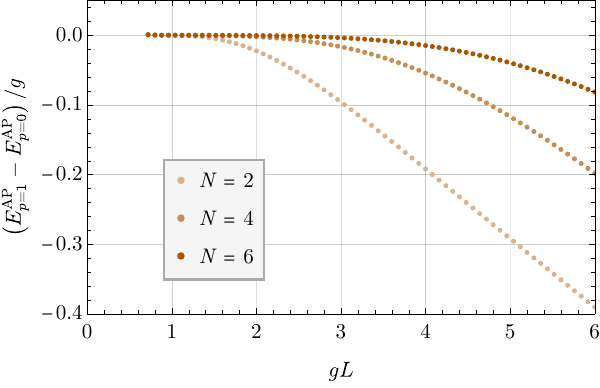}
	\caption{The difference between the ground state energies of the $p = 0$ and $p = 1$ universes for the $\grSU(3)$ theory at $m = 0$ as a function of the thermal circle length $gL$.
The region where this quantity approximately vanishes expands as $N$ is increased.}
	\label{fig:antiperiodic_massless_gap}
\end{figure}

When $m = 0$, if the non-invertible symmetry is present on a circle with anti-periodic boundary conditions, then   
the $p = 0$ and $p = 1$ vacua of the $\grSU(3)$ theory are degenerate for any circle length. While this degeneracy is certainly not present for a lattice with a finite $N$, we see evidence of it emerging in the continuum limit. 
In Figure~\ref{fig:antiperiodic_massless_gap}, we plot the energy difference between the $p = 0$ and $p = 1$ vacua as a function of $gL$ when $m = 0$. For sufficiently small $gL$, we find an energy difference very close to zero. At any $gL$ it appears to converge towards 0 as $N$ is increased.

These numerical results suggest that the four-fermion term $\kappa \int dx (\tr \bar \psi \psi)^2$, whose effect would be to break the degeneracy between the $p=1$ and $p=0$ sectors
\cite{Cherman:2019hbq,Komargodski:2020mxz}, is not induced by the lattice discretization effects. Let us suggest a reason why. In the lattice Hamiltonian, the four-fermion term would enter as $\sim   \frac{\kappa}{a}
\sum_{n=0}^{N-1} (\tr \chi_n U_n \chi_{n+1} U_n^{-1})^2$. In our model, four-fermion terms only arise when eliminating the gauge field by solving the Gauss law constraint. However, instead of $1/a$ they are multiplied by $g^2 a$. Therefore, it seems impossible for the nearest neighbor four-fermion terms multiplied by $1/a$ to arise as a result of the discretization. In other words, our lattice Hamiltonian has the naive continuuum limit without the four-fermion terms. An analogous argument can be made for the Schwinger model, where the $(\bar \psi \psi)^2 $ term has nontrivial effects if added by hand \cite{
Cherman:2022ecu,ArguelloCruz:2024xzi}. While the usual lattice Hamiltonian contains similar terms upon solving the Gauss law constraints, they are multiplied by $g^2 a$ instead of $1/a$. So, again one can argue that the usual lattice Hamiltonian does not induce the continuum four-fermion term, and this is confirmed by a multitude of numerical results.    

\section{Lattice model for other gauge groups}\label{sec:arbitrary_group}

Essentially everything in Section~\ref{sec:suN} holds for any compact, simply-connected gauge group $G$. In Table~\ref{tab:algebras}, we collect some useful properties of the classical Lie algebras $\{A_n, B_n, C_n, D_n\}$ and the exceptional algebras $\{F_4, G_2, E_6, E_7, E_8\}$. 

Throughout this section, we take $G$ to be a compact and simply-connected Lie group with Lie algebra $\mathfrak{g}$. In Section~\ref{sec:arbitrary_group_generalizations}, we restate some key results in a way that applies to any such $G$, which also serves as a brief recapitulation of the key ingredients in our lattice model. In Section~\ref{sec:arbitrary_group_symmetries}, we compute the symmetries of adjoint QCD$_2$ for any such $G$, and use the lattice model to compute the anomalies.

\begingroup
\renewcommand{\arraystretch}{1.5}
\begin{table}[t]
\centering
\begin{tabular}{m{2.5cm}m{1.3cm}m{1.3cm}m{2.4cm}m{1.4cm}m{1.7cm}m{3cm}}
\hline
$G$ & $\Lie(G)$ & $Z(G)$ & Dimension & Rank & $h^\vee$ & Dynkin Diagram \\
\hline
$\grSU(N_c)$ & $A_{N_c-1}$ & $\mathbb{Z}_{N_c}$ & $N_c^2 - 1$ & $N_c - 1$ & $N_c$ & \begin{tikzpicture}[scale=.4]
    \foreach \i in {1,2,3,5} \node[draw,thick,circle,minimum size=7,inner sep=0] (\i) at (\i, 0) {};
    \draw (1) -- (2) -- (3);
    \draw[dashed] (3) -- (5);
\end{tikzpicture} \\
$\grSpin(2M+1)$ & $B_M$ & $\mathbb{Z}_2$ & \(M(2M+1)\) & \(M\) & \(2M-1\) & \begin{tikzpicture}[scale=.4]
    \foreach \i in {1,2,3,5} \node[draw,thick,circle,minimum size=7,inner sep=0] (\i) at (\i, 0) {};
    \node[fill,circle,minimum size=7,inner sep=0] (6) at (6, 0) {};
    \draw (1) -- (2) -- (3);
    \draw[dashed] (3) -- (5);
    \draw[bend left=30] (5) ..controls (5.5,.3).. (6);
    \draw[bend right=30] (5) ..controls (5.5,-.3).. (6);
\end{tikzpicture} \\
$\grUSp(2M)$ & $C_M$ & $\mathbb{Z}_2$ & \(M(2M+1)\) & \(M\) & \(M+1\) & \begin{tikzpicture}[scale=.4]
    \foreach \i in {1,2,3,5} \node[fill,circle,minimum size=7,inner sep=0] (\i) at (\i, 0) {};
    \node[draw,thick,circle,minimum size=7,inner sep=0] (6) at (6, 0) {};
    \draw (1) -- (2) -- (3);
    \draw[dashed] (3) -- (5);
    \draw[bend left=30] (5) ..controls (5.5,.3).. (6);
    \draw[bend right=30] (5) ..controls (5.5,-.3).. (6);
\end{tikzpicture} \\
$\grSpin(4M)$ & $D_{2M}$ & $\mathbb{Z}_2\times\mathbb{Z}_2$ & \(2M(4M-1)\) & \(2M\) & \(4M-2\) & \begin{tikzpicture}[scale=.4]
    \foreach \i in {1,2,3,5} \node[draw,thick,circle,minimum size=7,inner sep=0] (\i) at (\i, 0) {};
    \node[draw,thick,circle,minimum size=7,inner sep=0] (6+) at (6, .5) {};
    \node[draw,thick,circle,minimum size=7,inner sep=0] (6-) at (6, -.5) {};
    \draw (1) -- (2) -- (3);
    \draw[dashed] (3) -- (5);
    \draw[bend left=30] (5) -- (6+);
    \draw[bend right=30] (5) -- (6-);
\end{tikzpicture} \\
$\grSpin(4M+2)$ & $D_{2M+1}$ & $\mathbb{Z}_4$ &  $(2M+1)\times$\linebreak$(4M+1)$ & \(2M+1\) & \(4M\) & \begin{tikzpicture}[scale=.4]
    \foreach \i in {1,2,3,5} \node[draw,thick,circle,minimum size=7,inner sep=0] (\i) at (\i, 0) {};
    \node[draw,thick,circle,minimum size=7,inner sep=0] (6+) at (6, .5) {};
    \node[draw,thick,circle,minimum size=7,inner sep=0] (6-) at (6, -.5) {};
    \draw (1) -- (2) -- (3);
    \draw[dashed] (3) -- (5);
    \draw[bend left=30] (5) -- (6+);
    \draw[bend right=30] (5) -- (6-);
\end{tikzpicture} \\
\(E_6\) & $E_6$ & $\mathbb{Z}_3$ & 78 & 6 & 12 & \begin{tikzpicture}[scale=.4,xscale=-1]
    \foreach \i in {1,2,3,4,5} \node[draw,thick,circle,minimum size=7,inner sep=0] (\i) at (\i, 0) {};
    \node[draw,thick,circle,minimum size=7,inner sep=0] (6) at (3,1) {};
    \draw (1) -- (2) -- (3) -- (4) -- (5);
    \draw (3) -- (6);
\end{tikzpicture} \\
\(E_7\) & $E_7$ & $\mathbb{Z}_2$ & 133 & 7 & 18 & \begin{tikzpicture}[scale=.4,xscale=-1]
    \foreach \i in {1,2,3,4,5,6} \node[draw,thick,circle,minimum size=7,inner sep=0] (\i) at (\i, 0) {};
    \node[draw,thick,circle,minimum size=7,inner sep=0] (7) at (4,1) {};
    \draw (1) -- (2) -- (3) -- (4) -- (5) -- (6);
    \draw (4) -- (7);
\end{tikzpicture} \\
\(E_8\) & $E_8$ & -- & 248 & 8 & 30 & \begin{tikzpicture}[scale=.4,xscale=-1]
    \foreach \i in {1,2,3,4,5,6,7} \node[draw,thick,circle,minimum size=7,inner sep=0] (\i) at (\i, 0) {};
    \node[draw,thick,circle,minimum size=7,inner sep=0] (8) at (5,1) {};
    \draw (1) -- (2) -- (3) -- (4) -- (5) -- (6) -- (7);
    \draw (5) -- (8);
\end{tikzpicture} \\
\(F_4\) & $F_4$ & -- & 52 & 4 & 9 & \begin{tikzpicture}[scale=.4]
    \foreach \i in {1,2} \node[draw,thick,circle,minimum size=7,inner sep=0] (\i) at (\i, 0) {};
    \foreach \i in {3,4} \node[fill,circle,minimum size=7,inner sep=0] (\i) at (\i, 0) {};
    \draw (1) -- (2) (3) -- (4);
    \draw (2) ..controls (2.5,.3).. (3);
    \draw (2) ..controls (2.5,-.3).. (3);
\end{tikzpicture} \\
\(G_2\) & $G_2$ & -- & 14 & 2 & 4 & \begin{tikzpicture}[scale=.4]
    \foreach \i in {1} \node[draw,thick,circle,minimum size=7,inner sep=0] (\i) at (\i, 0) {};
    \foreach \i in {2} \node[fill,circle,minimum size=7,inner sep=0] (\i) at (\i, 0) {};
    \draw (1) -- (2);
    \draw (1) ..controls (1.5,.3).. (2);
    \draw (1) ..controls (1.5,-.3).. (2);
\end{tikzpicture} \\
\hline
\end{tabular}
\caption{Some data associated to compact, simply-connected Lie groups and their Lie algebras.}
\label{tab:algebras}
\end{table}
\endgroup

\subsection{Generalizations}\label{sec:arbitrary_group_generalizations}

First, let us write the action of adjoint QCD$_2$ for an arbitrary group. In \eqref{eq:suN_action}, we used the fundamental trace to write the action, but this is not well-defined in general. Instead, we will use the trace in the adjoint and write
\begin{equation}\label{eq:G_action}
    S = \frac{1}{2h^\vee} \int d^2x\,\tr_\text{adj}\left(-\frac{1}{2g^2}F_{\mu\nu}F^{\mu\nu} + i\bar\psi\gamma^\mu D_\mu \psi - m\bar\psi \psi\right)\,.
\end{equation}
Here $h^\vee$ is the dual Coxeter number of $G$. For $\grSU(N_c)$, $\tr_\text{fund} = \frac{1}{2h^\vee}\tr_\text{adj}$, so this definition reduces to that of \eqref{eq:suN_action}.

The lattice Hamiltonian still takes the form \eqref{eq:hamiltonian} (or \eqref{eq:hamiltonian_ap} with antiperiodic boundary conditions for the fermions). The representation $\bm{R}$ has its highest weight equal to the Weyl vector $\rho$ of $\mathfrak{g}$, given by half the sum of the positive roots, and its dimension is
\begin{equation}
    \dim \bm{R} = 2^{\frac{\dim G-\rk G}{2}}\,.
\end{equation}
In addition to the fermions transforming in the $\bm{R}$ representation, there are $\rk G$ Majorana fermions on each site, which we label $\lambda_{n,j}$ with $j = 1,\ldots,\rk G$.

There are always $\rk G$ invariant symbols for the representations $(\mathbf{adj}, \bm{R}, \bm{\bar R})$. Thus, we can represent the operators $\chi^A_n$ by
\begin{equation}
    \chi^A_n = \sum_{j=1}^{\rk G} C^A_{n,j} \lambda_{n,j}\,.
\end{equation}
The conditions for this ansatz to obey the properties required of $\chi^A_n$ are discussed in Appendix~\ref{app:c_properties}.

The matrix elements of the Hamiltonian on gauge-invariant states are given by $6j$-symbols of $G$, exactly as in \eqref{eq:matrix_element_6j}. The $6j$-symbols can in principle be calculated by contracting Clebsch-Gordan symbols, but for groups of even modest rank this is computationally intractable; for instance, for $E_8$ we have $\dim\bm{R} = 2^{120}\approx 1.3\times 10^{36}$. 

\subsection{Symmetries and anomalies}\label{sec:arbitrary_group_symmetries}

Let us enumerate the invertible symmetries of adjoint QCD$_2$ for an arbitrary gauge group $G$. There are always at least two invertible zero-form $\mathbb{Z}_2$ symmetries to consider: fermion parity $(\mathbb{Z}_2)_F$ generated by $\F$ and, when $m = 0$, chiral symmetry $(\mathbb{Z}_2)_\chi$ generated by $\V$. They act only on the fermions, via
\begin{equation}
    \hat{\mathcal{F}}\psi\hat{\mathcal{F}} = -\psi\,, \qquad \hat{\mathcal{V}}\psi\hat{\mathcal{V}} = \gamma^5\psi\,.
\end{equation}
When the group $G$ has a nontrivial outer automorphism group, we have symmetry generators corresponding to each of the outer automorphisms. From Table~\ref{tab:algebras}, we see that $\grSU(N_c)$ for $N_c\ge 3$, $\grSpin(2M)$, and $E_6$ have such an outer automorphism that exchanges pairs of roots in the Dynkin diagram.\footnote{For $\grSU(N_c)$ this is implemented by charge conjugation symmetry, and for $\grSpin(2M)$ this is implemented by conjugation with a reflection.} In the special case of $\grSpin(8)$, the outer automorphism group is enhanced to $S_3$. In these cases, we define $\hat{\mathcal{C}}_\sigma$ to act on the gauge field and the fermion via the action of the outer automorphism $\sigma$.

In addition to these $\mathbb{Z}_2$ zero-form symmetries, adjoint QCD$_2$ has a one-form center symmetry that acts upon Wilson loops. The most rich case is $\grSU(N_c)$, for which the center is $\mathbb{Z}_{N_c}$; other simply-connected groups have centers that do not depend on their rank, as shown in Table~\ref{tab:algebras}. When the charge conjugation group is $\mathbb{Z}_2$, it always acts on the center symmetry group by inversion, and so for an element $\mathcal{U}(x)$ of the center symmetry $Z(G)^{[1]}$ we have
\begin{equation}
    \hat{\mathcal{U}}(x) \hat{\mathcal{C}} = \hat{\mathcal{C}}\hat{\mathcal{U}}(x)^{-1}\,.
\end{equation}
When every element of the center symmetry group has order 2, this action is trivial. Special care is required in the case of $\grSpin(8)$, when the outer automorphism group is $S_3$. In this case, $S_3$ acts on the center symmetry group $\mathbb{Z}_2\times\mathbb{Z}_2$ by permuting the three non-identity elements.

Putting all this together, the invertible symmetries of adjoint QCD$_2$ for $m\neq 0$ are as follows:
\begin{equation}\label{eq:symmetries}
    \begin{cases}
        \left\lbrack \mathbb{Z}_{N_c}^{[1]}\rtimes (\mathbb{Z}_2)_C\right\rbrack \times (\mathbb{Z}_2)_F & G = \grSU(N_c\ge 3) \\
        \left\lbrack (\mathbb{Z}_{2}\times\mathbb{Z}_2)^{[1]}\rtimes (S_3)_C\right\rbrack \times (\mathbb{Z}_2)_F & G = \grSpin(8) \\
        (\mathbb{Z}_{2}\times\mathbb{Z}_2)^{[1]}\times (\mathbb{Z}_2)_C \times (\mathbb{Z}_2)_F & G = \grSpin(4M)\text{ with }M\ge 3 \\
        \left\lbrack \mathbb{Z}_4^{[1]}\rtimes (\mathbb{Z}_2)_C\right\rbrack \times (\mathbb{Z}_2)_F & G = \grSpin(4M+2) \\
        \left\lbrack \mathbb{Z}_3^{[1]}\rtimes (\mathbb{Z}_2)_C\right\rbrack \times (\mathbb{Z}_2)_F & G = E_6 \\
        \mathbb{Z}_2^{[1]} \times (\mathbb{Z}_2)_F & G \in \{\grSU(2), \grSpin(2M+1), \grUSp(M), E_7\} \\
        (\mathbb{Z}_2)_F & G \in \{E_8,F_4,G_2\}
    \end{cases}
\end{equation}
When $m = 0$, the classical symmetry group is extended by a direct product with $(\mathbb{Z}_2)_\chi$.

To compute the anomalies, we follow the method in Section~\ref{sec:suN_symmetry}. We can express $\F$ and $\V$ in terms of the fermions as
\begin{equation}
    \begin{split}
        \F &= \prod_{n=0}^{N-1} \prod_{A=1}^{\dim G} \chi^A_n\,,\\
        \V &= \F^{\dim G-1} \left(\prod_{A=1}^{\dim G} \left(\chi_0^A + \chi_1^A\right)\right) \left(\prod_{A=1}^{\dim G} \left(\chi_1^A + \chi_2^A\right)\right) \cdots \left(\prod_{A=1}^{\dim G} \left(\chi_{N-2}^A + \chi_{N-1}^A\right)\right)\,.
    \end{split}
\end{equation}
We then find
\begin{equation}
    \F \V = (-1)^{\dim G}\V \F\,, \qquad \C_\sigma\V = (-1)^\sigma \V \C_\sigma\,,
\end{equation}
where by $(-1)^\sigma$ we mean the sign of the permutation of the Dynkin diagram indicated by $\sigma$.

For the one-form center symmetry, we need some notation. Let $p_{\bm{r}} \in \widetilde{Z(G)}$ be the conjugacy class of a representation $\bm{r}$. For $k\in Z(G)$, we then define
\begin{equation}
    \U_{k,n}\ket{(\bm{r}_0,e_0),\cdots,(\bm{r}_{N-1},e_{N-1})} = p_{\bm{R}}(k)^n p_{\bm{r}_n}(k)\ket{(\bm{r}_0,e_0),\cdots,(\bm{r}_{N-1},e_{N-1})}.
\end{equation}
Indeed, since $\bm{r}_{n+1} \in \bm{r}_n \otimes \bm{R}$ we have $p_{\bm{r}_{n+1}}(k) = p_{\bm{r}_{n}}(k) p_{\bm{R}}(k)$, so this is a topological operator. We then find
\begin{equation}
    \U_{k,n} \V = p_{\bm{R}}(k) \V \U_{k,n}\,.
\end{equation}

Explicitly, for the groups $\grSU(2m)$, $\grSpin(8m+q)$ with $q=1,3,5,6,7$, $\grUSp(2(4m+q))$ with $q=1,2$, and $E_7$, $\U_{k,n}$ has a projective sign when $k$ is a generator of $Z(G)$. For $\grSpin(8m+4)$ one of the $(\mathbb{Z}_2\times\mathbb{Z}_2)^{[1]}$ generators has a projective sign and the other does not. For all other groups, there is no anomaly between the center symmetry and the chiral symmetry.

\section{Discussion}

In this paper, we generalized the lattice Hamiltonian formulation of Adjoint QCD$_2$ given in~\cite{Dempsey:2023fvm} to an arbitrary gauge group $G$. The construction relies upon an interesting factorization of the fermionic Hilbert space. We find that the matrix elements of the lattice Hamiltonian as well as the lattice strong coupling expansion can be expressed in terms of the Wigner $6j$-symbols of $G$.

There are a number of natural directions for future work on this topic. One is to extend the explicit calculations to groups beyond $\grSU(3)$. We are limited by two difficulties. One is the calculation of the Hamiltonian matrix elements; the values of the requisite $6j$-symbols are not known explicitly for $G\neq \grSU(2)$, and we do not yet have an efficient method for computing them in other cases. However, even if this problem were solved, the size of the Hilbert space grows extremely quickly with the rank of the gauge group, and so exact diagonalization would quickly become infeasible anyway. For this reason, it would be of great interest to implement our model (perhaps on an open chain) using matrix product states. The matrix product state ansatz allows for efficient calculation of the low-lying spectrum of one-dimensional Hamiltonians even when the full Hilbert space is extremely large, and has provided many precision results for abelian gauge theories in $(1+1)$ dimensions~\cite{Byrnes:2002nv,Banuls:2013jaa,Banuls:2016lkq,Funcke:2023lli,Dempsey:2023gib,Itou:2024psm}. Using it for our lattice model of Adjoint QCD$_2$ will likely require tensor network algorithms that explicitly respect non-abelian symmetries, which is an area of active research~\cite{Weichselbaum:2012,Weichselbaum:2019ofd}.

One very interesting application of a tensor network formulation of our model would be to study its real-time dynamics. A quantum simulation of the hadrons of Adjoint QCD$_2$ would provide a unique window into the phenomenology of this rich model. Such simulations could also be carried out on analog quantum simulators or digital quantum hardware, both of which have been used to simulate abelian lattice gauge theories in $(1+1)$ dimensions~\cite{Kokail:2018eiw,Farrell:2024fit}.

Furthermore, it would be particularly interesting to study the $G = \grSU(4)$ or even $\grSU(5)$ theories, for which the non-invertible symmetries of the continuum theory 
\cite{Komargodski:2020mxz,Dempsey:2024ofo}
predict more degenerate vacua than are required by the anomalies of invertible symmetries. There are quantitative predictions for the ratios of the fermion bilinear condensates in different vacua at $m = 0$~\cite{Komargodski:2020mxz}, which we hope to compare with the lattice model. The non-invertible symmetries are broken by the lattice, which we have seen in this work from the fact that $E_{p=1}- E_{p=0}$ for $\grSU(3)$ at $m = 0$ is not identically zero for any lattice spacing. However, we have provided some evidence that the non-invertible symmetry is restored in the continuum limit, suggesting that the four-fermion terms added in~\cite{Cherman:2019hbq} are not induced in our lattice model.
 Effective calculations for the higher-rank groups would likely require progress both on efficient calculation of Hamiltonian matrix elements and on a tensor network formulation, but we hope to perform some initial studies of these theories in the future.

\section*{Acknowledgments}

We are grateful to Aleksey Cherman, Zohar Komargodski and Grisha Tarnopolsky for useful discussions.  This work was supported in part by the Simons Foundation Grant No.~917464 (Simons Collaboration on Confinement and QCD Strings), and by the US National Science Foundation under Grants No.~PHY-2111977 and PHY-2209997. SSP, BS, and RD are supported in part by the U.S. Department of Energy under Award No.~DE-SC0007968. RD was also supported in part by an NSF Graduate Research Fellowship and a Princeton University Charlotte Elizabeth Procter Fellowship. We thank the SwissMAP Research Station in Les Diablerets, where some of this work was carried out, for the warm hospitality. 

\appendix

\section{Group theory calculations}\label{app:group_theory}

Here we collect some group theory calculations needed in the main text. We will make extensive use of birdtracks notation. Generic representations are shown using black lines with arrows; we use dashed lines for $\bm{R}$, and dotted blue lines for the adjoint representation. Clebsch-Gordan coefficients are represented by circular vertices with two incoming irreps and one outgoing irrep; their conjugates are denoted with one ingoing irrep and two outgoing ones. They are normalized by
\begin{equation}\label{eq:clebsch_orthogonality}
    C^{\bm{r}\,\bm{r'}\,\bm{r''};e}_{ijk}\left(C^{\bm{r}\,\bm{r'}\,\bm{r''};e'}_{ijk}\right)^* = \begin{tikzpicture}[baseline=-.1cm,scale=.8]
    \node[fill,circle,minimum size=5,inner sep=0,label=-90:{$e'$}] at (0,-1) {};
    \node[fill,circle,minimum size=5,inner sep=0,label=90:{$e$}] at (0,1) {};
    \draw[ultra thick,mid arrow] (0,-1) -- node[left] {$\bm{r}$} (0,1);
    \draw[ultra thick,mid arrow] (0,-1) arc(-90:-270:1) node[pos=0.5,left] {$\bm{r'}$};
    \draw[ultra thick,mid arrow] (0,1) arc(90:-90:1) node[pos=0.5,right] {$\bm{r''}$};
\end{tikzpicture} = (\dim \bm{r''})\delta_{ee'}\,,
\end{equation}
We use square vertices to indicate the special three-point invariants appearing in \eqref{eq:chi_ansatz}.

Many of the calculations in this section will involve Wigner 6$j$-symbols. These symbols are recoupling coefficients, also known in other contexts as associators or $F$-symbols. They tell us how to relate a basis for the tensor product $(\bm{r}_1\otimes \bm{r}_2)\otimes \bm{r}_3$ with one for $\bm{r}_1\otimes(\bm{r}_2\otimes \bm{r}_3)$. In each of these cases we would use two Clebsch-Gordan symbols to perform the tensor products, so the 6$j$-symbol is a contraction of four Clebsch-Gordan symbols. We will denote them by drawing the set of four contracted symbols. For instance, one of the 6$j$-symbols in \eqref{eq:matrix_element_6j} can be expanded as
\begin{equation}
	\begin{tikzpicture}[xscale=1.7,baseline=-1cm]
        \foreach \x in {1} {
            \node[draw,circle,fill,minimum size=6,inner sep=0] at ({2*\x+1},0) (\x) {};
            \node[draw,circle,fill,minimum size=6,inner sep=0] at ({2*\x+1},-2) (a\x) {};
            \draw[ultra thick,dashed,mid arrow] (a\x) -- ({2*\x+1},-1);
            \draw[ultra thick,dashed,mid arrow] ({2*\x+1},-1) -- (\x);
        };
        \draw[ultra thick,mid arrow] (3,-2) arc(270:90:.5 and 1) node[left,pos=0.5] {$\bm{r}_{n-1}$};
        \node at (2.9,.4) {$e_n$};
        \node at (2.9,-2.4) {$e'_n$};
        \draw[ultra thick,dotted,blue,mid arrow] (3,-1) node[black,fill,minimum size=6,inner sep=0,label=180:{\color{black}{$j$}}] {} -- (3.5,-1) node[black,fill,circle,minimum size=6,inner sep=0,label=0:{\color{black}{$l$}}] {};
        \draw[ultra thick,mid arrow] (3,0) arc(90:0:.5 and 1) node[midway,right] {$\bm{r}_n$};
        \draw[ultra thick,mid arrow] (3.5,-1) arc(0:-90:.5 and 1) node[midway,right] {$\bm{r'}_n$};
    \end{tikzpicture} = \left(C^A_j\right)_{\alpha\beta} C^{\bm{r}_{n-1},\bm{R},\bm{r}_n;e_n}_{a\alpha b}C^{\bm{r}_n,\bm{R},\bm{r'}_n;l}_{bAc}\left(C^{\bm{r}_{n-1},\bm{R},\bm{r'}_n;e'_n}_{a\beta c}\right)^*\,.
\end{equation}
Here the $C^A_j$ symbol on the right is one of the special $(\mathbf{adj},\bm R,\bm{\bar{R}})$ invariants discussed in Section~\ref{sec:suN_fermions}, and the others are ordinary Clebsch-Gordan symbols in an arbitrary basis. The Greek indices are for $\bm{R}$, the capital indices are for the adjoint, and the lowercase indices are for the other irreps appearing.

\subsection{Hamiltonian matrix elements}\label{app:hamiltonian}

We show here how to compute
\begin{equation}
	M_n = \braket{\psi' | \chi^A_n U_n^{AB} \chi_{n+1}^B | \psi}
\end{equation}
for states of the form \eqref{eq:gauge_invariant_state}. We assume $\bm{r}_m = \bm{r'}_m$ when $m\neq n$; otherwise the matrix element vanishes.

Using \eqref{eq:link_orthogonality}, we can reduce most of $M_n$ to contractions of the form \eqref{eq:clebsch_orthogonality}, which then cancel most of the normalization factors in \eqref{eq:state_contraction}. We are left with
\begin{equation}\label{eq:matrix_element_1}
	\begin{split}
    M_n = \frac{1}{\dim \bm{r}_{n+1}}\sum_{j,k=1}^{\rk G} q_{n,jk} &D_{jk}^{e_n,e_{n+1};e'_n,e'_{n+1}}(\bm{r}_{n-1},\bm{r}_{n+1};\bm{r}_n,\bm{r'}_n)\,, \qquad \text{where} \\
    D_{jk}^{e_n,e_{n+1};e'_n,e'_{n+1}}(\bm{r}_{n-1},\bm{r}_{n+1};\bm{r}_n,\bm{r'}_n) &= \frac{1}{\sqrt{\dim \bm{r}_n \dim\bm{r'}_{n}}}\times \\
    &\quad\begin{tikzpicture}[baseline=-1.4cm,xscale=2.8,yscale=1.4]
    		\node[ultra thick,draw,circle,minimum size=.8cm] (top) at (3.5,0) {\tiny $\bm{r}_n$};
        \draw[ultra thick,mid arrow] (3,0) -- (top);
        \draw[ultra thick,mid arrow] (top) -- (4,0);
        \node[ultra thick,draw,circle,minimum size=.8cm] (bottom) at (3.5,-2) {\tiny $\bm{r'}_n$};
        \draw[ultra thick,mid arrow] (4,-2) -- (bottom);
        \draw[ultra thick,mid arrow] (bottom) -- (3,-2);
        \foreach \x in {3,4} {
            \node[draw,circle,fill,minimum size=6,inner sep=0] at (\x,0) (\x) {};
            \node[draw,circle,fill,minimum size=6,inner sep=0] at (\x,-2) (a\x) {};
            \draw[ultra thick,dashed,mid arrow] (\x,-2) -- (\x,-1);
            \draw[ultra thick,dashed,mid arrow] (\x,-1) -- (\x,0);
        };
        \draw[ultra thick,mid arrow] (3,-2) arc(270:90:.5 and 1) node[left,pos=0.5] {$\bm{r}_{n-1}$};
        \draw[ultra thick,mid arrow] (4,0) arc(90:-90:.5 and 1) node[right,pos=0.5] {$\bm{r}_{n+1}$};
        \node at (2.9,.3) {$e_n$};
        \node at (4.1,.3) {$e_{n+1}$};
        \node at (2.9,-2.3) {$e'_n$};
        \node at (4.1,-2.3) {$e'_{n+1}$};        
    		\node[ultra thick,draw,circle,minimum size=.8cm] (mid) at (3.5,-1) {\tiny $U$};
        \draw[ultra thick,dotted,blue,mid arrow] (3,-1) node[black,fill,minimum size=6,inner sep=0] {} -- (mid);
        \draw[ultra thick,dotted,blue,mid arrow] (mid) -- (4,-1) node[black,fill,minimum size=6,inner sep=0] {};
    \end{tikzpicture}\,.
    \end{split}
\end{equation}
The prefactor $q_{n,jk}$ comes from the qubit sector, and is given by
\begin{equation}
    q_{n,jk} = \braket{s'|\lambda_{n,j}\lambda_{n,k}|s}
\end{equation}
where $\ket{s} \equiv \otimes_{n=0}^{N/2-1} \ket{s_{n,1}\cdots s_{n,N_c-1}}$ and likewise for $\ket{s'}$.

We can simplify \eqref{eq:matrix_element_1} by combining the link operator $U$ with the state on the $n$th link of $\ket{\psi}$, by first fusing the adjoint representation with $\bm{r}_n$ and summing over all irreps $\bm\lambda\in \bm{r}_n\otimes \textbf{adj}$. This gives
\begin{equation}
	\begin{tikzpicture}[baseline=-.6cm,xscale=2.8,yscale=1.4]
    		\node[ultra thick,draw,circle,minimum size=.8cm] (top) at (3.5,0) {\tiny $\bm{r}_n$};
        \draw[ultra thick,mid arrow] (3,0) -- (top);
        \draw[ultra thick,mid arrow] (top) -- (4,0);
        \node[ultra thick,draw,circle,minimum size=.8cm] (mid) at (3.5,-1) {\tiny $U$};
        \draw[ultra thick,dotted,blue,mid arrow] (3,-1) -- (mid);
        \draw[ultra thick,dotted,blue,mid arrow] (mid) -- (4,-1);
    \end{tikzpicture} = \sum_{\bm\sigma\in\bm{r}_n\otimes\textbf{adj}} \sum_{\ell} \sqrt{\frac{\dim \bm{r}_n}{\dim \bm{\sigma}}} \begin{tikzpicture}[xscale=3.5,yscale=1.4,baseline=0cm]
    		\node[ultra thick,draw,circle,minimum size=.8cm] (top) at (3.5,0) {$\bm{\sigma}$};
        \draw[ultra thick,mid arrow] (3,0) node[fill,circle,minimum size=5,inner sep=0,label=-90:{$l$}] {} -- node[above] {$\bm\sigma$} (top);
        \draw[ultra thick,mid arrow] (top) -- node[above] {$\bm\sigma$} (4,0) node[fill,circle,minimum size=5,inner sep=0,label=-90:{$l$}] {};
        \draw[ultra thick,dotted,blue,mid arrow] (2.75,-.5) -- (3,0);
        \draw[ultra thick,dotted,blue,mid arrow] (4,0) -- (4.25,-.5);
        \draw[ultra thick,mid arrow] (2.75,.5) -- (3,0);
        \draw[ultra thick,mid arrow] (4,0) -- (4.25,.5);
    \end{tikzpicture}\,.
\end{equation}
By using this equation in \eqref{eq:matrix_element_1}, and then using \eqref{eq:link_orthogonality}, we find that as long as $\bm{r'}_n\in \bm{r}_n\otimes\textbf{adj}$ we have
\begin{equation}
    D_{jk}^{e_n,e_{n+1};e'_n,e'_{n+1}}(\bm{r}_{n-1},\bm{r}_{n+1};\bm{r}_n,\bm{r'}_n) = \frac{1}{\dim \bm{r'}_n}\sum_{l} \left\lbrack \begin{tikzpicture}[xscale=1.7,baseline=-1cm]
        \foreach \x in {1,2} {
            \node[draw,circle,fill,minimum size=6,inner sep=0] at ({2*\x+1},0) (\x) {};
            \node[draw,circle,fill,minimum size=6,inner sep=0] at ({2*\x+1},-2) (a\x) {};
            \draw[ultra thick,dashed,mid arrow] (a\x) -- ({2*\x+1},-1);
            \draw[ultra thick,dashed,mid arrow] ({2*\x+1},-1) -- (\x);
        };
        \draw[ultra thick,mid arrow] (3,-2) arc(270:90:.5 and 1) node[left,pos=0.5] {$\bm{r}_{n-1}$};
        \draw[ultra thick,mid arrow] (5,0) arc(90:-90:.5 and 1) node[right,pos=0.5] {$\bm{r}_{n+1}$};
        \node at (2.9,.4) {$e_n$};
        \node at (5.1,.4) {$e_{n+1}$};
        \node at (2.9,-2.4) {$e'_n$};
        \node at (5.1,-2.4) {$e'_{n+1}$};
        \draw[ultra thick,dotted,blue,mid arrow] (3,-1) node[black,fill,minimum size=6,inner sep=0,label=180:{\color{black}{$j$}}] {} -- (3.5,-1) node[black,fill,circle,minimum size=6,inner sep=0,label=0:{\color{black}{$l$}}] {};
        \draw[ultra thick,mid arrow] (3,0) arc(90:0:.5 and 1) node[midway,right] {$\bm{r}_n$};
        \draw[ultra thick,mid arrow] (3.5,-1) arc(0:-90:.5 and 1) node[midway,right] {$\bm{r'}_n$};
        \draw[ultra thick,dotted,blue,mid arrow] (4.5,-1) node[black,fill,circle,minimum size=6,inner sep=0,label=180:{\color{black}{$l$}}] {} -- (5,-1) node[black,fill,minimum size=6,inner sep=0,label=0:{\color{black}{$k$}}] {};
        \node at (4,-1) {$\times$};
        \draw[ultra thick,mid arrow] (4.5,-1) arc(180:90:.5 and 1) node[midway,left] {$\bm{r}_n$};
        \draw[ultra thick,mid arrow] (5,-2) arc(-90:-180:.5 and 1) node[midway,left] {$\bm{r'}_n$};
    \end{tikzpicture}
    \right\rbrack\,.
\end{equation}
Thus, the matrix element is given by \eqref{eq:matrix_element_6j}.

\subsection{$\grSU(3)$ and $\grSU(4)$ 6$j$-symbols}\label{app:su3_6j}

To evaluate the expressions in Section~\ref{sec:strong_coupling_bw}, we need the values of $D_{jk}$ for various representations. Here we will give the explicit values for $\grSU(3)$ and $\grSU(4)$ and give examples of how they are computed.

As illustrative examples, we will work out the $\grSU(3)$ $D_{jk}(\bm{1},\bm{1};\bm{8},\bm{8})$ and $D_{jk}(\bm{3},\bm{3};\bm{3},\bm{3})$, where we have suppressed multiplicity labels because they are all 1 (and note that $D_{jk}(\bm{8},\bm{8};\bm{1},\bm{1}) = 0$). In $D_{jk}(\bm{1},\bm{1};\bm{8},\bm{8})$, the 6$j$-symbols appearing are
\begin{equation}
    \begin{tikzpicture}[baseline=0cm,scale=1.2]
        \draw[dashed,ultra thick,mid arrow] (90:.7) node[fill,circle,minimum size=5,inner sep=0] {} arc(90:-30:.7) node[midway,right] {$\bm{8}$};
        \draw[dashed,ultra thick,mid arrow] (-30:.7) node[fill,circle,minimum size=5,inner sep=0,label=-30:{$l$}] {} arc(-30:-150:.7) node[midway,below] {$\bm{8}$};
        \draw[dotted,ultra thick,mid arrow] (210:.7) node[fill,circle,minimum size=5,inner sep=0] {} arc(210:90:.7) node[midway,left] {$\bm{1}$};
        \draw[ultra thick,dashed,mid arrow] (0,0) -- (90:.7);
        \draw[ultra thick,dashed,mid arrow] (0,0) -- (-30:.7);
        \draw[ultra thick,dashed,mid arrow] (210:.7) -- (0,0);
        \node[fill,minimum size=5,inner sep=0,label={[shift={(.1cm,-.1cm)}]150:$j$}] at (0,0) {};
    \end{tikzpicture} = \begin{tikzpicture}[baseline=0cm]
        \node[fill,minimum size=5,inner sep=0,label=180:{$j$}] at (0,0) {};
        \node[fill,circle,minimum size=5,inner sep=0,label=0:{$l$}] at (1,0) {};
        \draw[dashed,ultra thick,mid arrow] (0,0) arc(180:0:.5);
        \draw[dashed,ultra thick,mid arrow] (1,0) arc(0:-180:.5);
        \draw[dashed,ultra thick,mid arrow] (0,0) -- (1,0);
    \end{tikzpicture}, \qquad \begin{tikzpicture}[baseline=0cm,scale=1.2]
        \draw[dotted,ultra thick,mid arrow] (90:.7) node[fill,circle,minimum size=5,inner sep=0] {} arc(90:-30:.7) node[midway,right] {$\bm{1}$};
        \draw[dashed,ultra thick,mid arrow] (-30:.7) node[fill,circle,minimum size=5,inner sep=0] {} arc(-30:-150:.7) node[midway,below] {$\bm{8}$};
        \draw[dashed,ultra thick,mid arrow] (210:.7) node[fill,circle,minimum size=5,inner sep=0,label=210:{$l$}] {} arc(210:90:.7) node[midway,left] {$\bm{8}$};
        \draw[ultra thick,dashed,mid arrow] (0,0) -- (90:.7);
        \draw[ultra thick,dashed,mid arrow] (0,0) -- (-30:.7);
        \draw[ultra thick,dashed,mid arrow] (210:.7) -- (0,0);
        \node[fill,minimum size=5,inner sep=0,label={[shift={(-.1cm,-.1cm)}]30:$k$}] at (0,0) {};
    \end{tikzpicture} = \frac{1}{8}\begin{tikzpicture}[baseline=0cm]
        \node[fill,circle,minimum size=5,inner sep=0,label=180:{$l$}] at (0,0) {};
        \node[fill,minimum size=5,inner sep=0,label=0:{$k$}] at (1,0) {};
        \draw[dashed,ultra thick,mid arrow] (0,0) arc(180:0:.5);
        \draw[dashed,ultra thick,mid arrow] (1,0) arc(0:-180:.5);
        \draw[dashed,ultra thick,mid arrow] (0,0) -- (1,0);
    \end{tikzpicture}
\end{equation}
and so it suffices to compute the diagrams
\begin{equation}
    d^0_{jl} = \begin{tikzpicture}[baseline=0cm]
        \node[fill,minimum size=5,inner sep=0,label=180:{$j$}] at (-.5,0) (left) {};
        \node[fill,circle,minimum size=5,inner sep=0,label=0:{$l$}] at (.5,0) (right) {};
        \draw[ultra thick,dashed] (0,0) circle (.5) (-.5,0) -- (.5,0);
    \end{tikzpicture}\,.
\end{equation}
We computed the invariants at the square vertex in Section~\ref{sec:suN_fermions}. The invariants at the circular vertex must satisfy \eqref{eq:clebsch_orthogonality}. From~\cite{Haber:2019sgz} we see that
\begin{equation}
    f^{ABC}f^{BCD} = 3\delta^{AD}\,, \qquad d^{ABC} d^{BCD} = \frac{5}{3}\delta^{AD}\,,
\end{equation}
and so we can take our normalized invariants to be $\frac{1}{\sqrt{3}}f^{ABC}$ and $\sqrt{\frac{3}{5}}d^{ABC}$. We then find
\begin{equation}
	d^0_{11} = 4\sqrt{6}\,, \qquad d^0_{22} = 4\sqrt{10}\,, \qquad d^0_{12} = d^0_{21} = 0
\end{equation}
which implies
\begin{equation}
    D_{jk}(\bm{1},\bm{1};\bm{8},\bm{8}) = \begin{pmatrix} \frac{5}{2} & 0 \\ 0 & \frac{3}{2}\end{pmatrix}\,.
\end{equation}

For the $p = 1$ universe, all the $\bm{r}_n$ representations are fundamentals, and so we need the 6$j$ symbols
\begin{equation}
    d^1_j = \begin{tikzpicture}[baseline=0cm,scale=1.2]
        \draw[ultra thick,mid arrow] (90:.7) node[fill,circle,minimum size=5,inner sep=0] {} arc(90:-30:.7) node[midway,right] {$\bm{3}$};
        \draw[ultra thick,mid arrow] (-30:.7) node[fill,circle,minimum size=5,inner sep=0] {} arc(-30:-150:.7) node[midway,below] {$\bm{3}$};
        \draw[ultra thick,mid arrow] (210:.7) node[fill,circle,minimum size=5,inner sep=0] {} arc(210:90:.7) node[midway,left] {$\bm{3}$};
        \draw[ultra thick,dashed,mid arrow] (0,0) -- (90:.7); 
        \draw[ultra thick,dashed,mid arrow] (0,0) -- (-30:.7); 
        \draw[ultra thick,dashed,mid arrow] (210:.7) -- (0,0);
        \node[fill,minimum size=5,inner sep=0,label={[shift={(0cm,0cm)}]-90:$j$}] at (0,0) {};
    \end{tikzpicture}\,.
\end{equation}
The invariants on the rim of this diagram are proportional to the fundamental generators $(T^A)_{ab}$. We normalize the generators by $\tr\left(T^A T^B\right) = \frac{1}{2}\delta^{AB}$, and so they satisfy
\begin{equation}
    T^a_{AB} T^a_{BC} = C_2(\textbf{fund}) \,\delta_{AC} = \frac{4}{3}\delta_{AC}\,.
\end{equation}
Thus, the normalization \eqref{eq:clebsch_orthogonality} requires that we use $\frac{\sqrt{3}}{2}(T^A)_{ab}$ for these vertices. Then, using results from~\cite{Haber:2019sgz}, we find
\begin{equation}
    d^1_1 = \frac{9\sqrt{3}}{4\sqrt{2}}\,, \qquad d^1_2 = \frac{15}{4\sqrt{2}}
\end{equation}
which implies
\begin{equation}
	D_{jk}(\bm{3},\bm{3};\bm{3},\bm{3}) = \frac{1}{32} \begin{pmatrix} 25 & 15\sqrt{3} \\ -15\sqrt{3} & -27 \end{pmatrix}\,.
\end{equation}

These results suffice to write down $h^{(1)}$ for the $p = 0$ and $p = 1$ universes in the $\grSU(3)$ theory. To go to second order for the $p = 0$ universe, we need
\begin{equation}
\begin{split}
	D^{11;11}_{jk}(\bm{8},\bm{8};\bm{1},\bm{8}) &= \begin{pmatrix} \frac{5}{4\sqrt{2}} & 0 \\ 0 & 0 \end{pmatrix}, \qquad D^{11;12}_{jk}(\bm{8},\bm{8};\bm{1},\bm{8}) = \begin{pmatrix} 0 & \frac{\sqrt{15}}{4\sqrt{2}} \\ 0 & 0 \end{pmatrix}\,, \\
	D^{11;21}_{jk}(\bm{8},\bm{8};\bm{1},\bm{8}) &= \begin{pmatrix} 0 & 0 \\ -\frac{\sqrt{15}}{4\sqrt{2}} & 0 \end{pmatrix}, \qquad D^{11;22}_{jk}(\bm{8},\bm{8};\bm{1},\bm{8}) = \begin{pmatrix} 0 & 0 \\ 0 & -\frac{3}{4\sqrt{2}} \end{pmatrix}\,.
\end{split}
\end{equation}
To go to second order for the $p = 1$ universe, we need
\begin{equation}
	D_{jk}(\bm{3},\bm{3};\bm{3},\bm{6}) = \frac{1}{16\sqrt{2}}\begin{pmatrix} 25 & -5\sqrt{3} \\ 5\sqrt{3} & -3 \end{pmatrix}\,, \qquad D_{jk}(\bm{3},\bm{3};\bm{3},\bm{15}) = \frac{1}{32}\begin{pmatrix} \sqrt{5} & -\sqrt{15} \\ \sqrt{15} & -3\sqrt{5} \end{pmatrix}\,.
\end{equation}

For the strong-coupling calculations in the $\grSU(4)$ theory, we need several more 6$j$-symbols. We compute these using GroupMath \cite{Fonseca:2020vke}. This entails first solving the conditions \eqref{eq:cond1} -- \eqref{eq:cond3} to find a basis of invariants on the irreps $(\bm{15}, \bm{64}, \bm{64})$. There is some arbitrariness in this process because the conditions are unaffected by $\grSO(3)$ rotations of the basis\footnote{We similarly had an $\grSO(2)$ freedom in the basis for these invariants in the $\grSU(3)$ case, but there it was natural to take one of the invariants to be proportional to the symmetric $d$-symbol and the other proportional to the antisymmetric $f$-symbol. In this case we have one symmetric and two antisymmetric invariants; we are working in a basis where the first invariant is symmetric and the second and third are antisymmetric.}, so the following results are all basis-dependent, but they can be used to derive the basis-independent strong coupling results given in Section \ref{sec:strong_coupling_su3}.

To compute the projection of the Hamiltonian to the strong-coupling ground state subspace in the $p = 0$ or $p = 2$ universes, we use
\begin{equation}
	D_{jk}\left(\bm{6},\bm{6};\bm{15},\bm{15}\right) = \begin{pmatrix}
		\frac{25}{12} & 0 & 0 \\
		0 & \frac{2}{5} & \frac{7}{10} \\
		0 & \frac{7}{10} & \frac{49}{40}
	\end{pmatrix}, \qquad D_{jk}\left(\bm{15},\bm{15};\bm{6},\bm{6}\right) = \begin{pmatrix}
		0 & 0 & 0 \\
		0 & \frac{121}{100} & -\frac{77}{100} \\
		0 & -\frac{77}{100} & \frac{49}{100}
	\end{pmatrix}\,.
\end{equation}
To compute this projection in the $p = 1$ universe, we use
\begin{equation}
\begin{aligned}
	D_{jk}\left(\bm{20},\bm{20};\bm{\overline{4}},\bm{\overline{4}}\right) &= \begin{pmatrix}
		\frac{5}{6} & \frac{2}{3} \sqrt{\frac{2}{5}} & \frac{7}{3 \sqrt{10}} \\
		\frac{2}{3} \sqrt{\frac{2}{5}} & \frac{16}{75} & \frac{28}{75} \\
 		\frac{7}{3 \sqrt{10}} & \frac{28}{75} & \frac{49}{75}
	\end{pmatrix}\,, \\
	D_{jk}\left(\bm{\overline{4}},\bm{\overline{4}};\bm{20},\bm{20}\right) &= \begin{pmatrix}
		\frac{325}{288} & \frac{181}{288}\sqrt{\frac{5}{2}} & \frac{133}{288}\sqrt{\frac{5}{2}} \\
 		\frac{181}{288}\sqrt{\frac{5}{2}} & \frac{6541}{2880} & -\frac{707}{2880} \\
 		\frac{133}{288}\sqrt{\frac{5}{2}} & -\frac{707}{2880} & \frac{2989}{2880}
	\end{pmatrix}\,.
\end{aligned}
\end{equation}
The results for the $p = 2$ and $p = 3$ universes can be obtained from the $p = 0,1$ results using the symmetries and their mixed anomalies.

To compute the second-order correction, we need to consider the many possible states that could be reached by acting with the Hamiltonian on one of the strong-coupling ground states. For example, in the $p = 0$ universe, the ground states have $\bm{15}$ on the even links and $\bm{6}$ on the odd links. When we act with the Hamiltonian, one of the even links could transition to the $\bm{20'}$, the $\bm{45}$, the $\bm{\overline{45}}$, or the $\bm{84}$. This means we need the following:
{\small
\begin{equation}
\begin{aligned}
	D_{jk}(\bm{6},\bm{6};\bm{15},\bm{20'}) &= \begin{pmatrix}
		0 & 0 & 0 \\
 		0 & \frac{15 \sqrt{3}}{16} & 0 \\
 		0 & 0 & 0
 	\end{pmatrix}\,, \qquad
 	D_{jk}(\bm{6},\bm{6};\bm{15},\bm{84}) = \begin{pmatrix}
 		0 & 0 & 0 \\
 		0 & \frac{\sqrt{\frac{7}{5}}}{48} & -\frac{\sqrt{\frac{7}{5}}}{24} \\
 		0 & -\frac{\sqrt{\frac{7}{5}}}{24} & \frac{\sqrt{\frac{7}{5}}}{12}
 	\end{pmatrix}\,, \\
 	D_{jk}(\bm{6},\bm{6};\bm{15},\bm{45}) &= D_{kj}(\bm{6},\bm{6};\bm{15},\bm{\overline{45}}) = \begin{pmatrix}
 		\frac{5}{24 \sqrt{3}} & \frac{13}{12 \sqrt{30}} & -\frac{7}{24 \sqrt{30}} \\
 		-\frac{13}{12 \sqrt{30}} & -\frac{169}{300 \sqrt{3}} & \frac{91}{600 \sqrt{3}} \\
 		\frac{7}{24 \sqrt{30}} & \frac{91}{600 \sqrt{3}} & -\frac{49}{1200 \sqrt{3}}
 	\end{pmatrix}\,.
\end{aligned}
\end{equation}
}%
Similarly, one of the odd links could transition to the $\bm{10}$, the $\bm{\overline{10}}$, or the $\bm{64}$. For the $\bm{10}$ and $\bm{\overline{10}}$, the relevant $D_{jk}$ values are
{\small
\begin{equation}
\begin{aligned}
	D_{jk}(\bm{15},\bm{15};\bm{6},\bm{10}) &= \begin{pmatrix}
		-\frac{5}{12}\sqrt{\frac{5}{3}} & \frac{19}{12 \sqrt{6}} & \frac{7}{12 \sqrt{6}} \\
 		-\frac{19}{12 \sqrt{6}} & \frac{361}{120 \sqrt{15}} & \frac{133}{120 \sqrt{15}} \\
 		-\frac{7}{12 \sqrt{6}} & \frac{133}{120 \sqrt{15}} & \frac{49}{120 \sqrt{15}}
	\end{pmatrix}\,, &
	D_{jk}(\bm{15},\bm{15};\bm{6},\bm{\overline{10}}) &= \begin{pmatrix}
		-\frac{5 \sqrt{\frac{5}{3}}}{12} & -\frac{19}{12 \sqrt{6}} & -\frac{7}{12 \sqrt{6}} \\
 		\frac{19}{12 \sqrt{6}} & \frac{361}{120 \sqrt{15}} & \frac{133}{120 \sqrt{15}} \\
 		\frac{7}{12 \sqrt{6}} & \frac{133}{120 \sqrt{15}} & \frac{49}{120 \sqrt{15}} 
	\end{pmatrix}\,.
\end{aligned}
\end{equation}
}%
When one of the odd links becomes a $\bm{64}$, we have to account for the fact that in $\bm{15}\otimes\bm{64}$ there are three copies of the $\bm{64}$. There are thus many symbols to compute; we find
{\small
\begin{equation}
\begin{aligned}
	D_{jk}^{11;11}(\bm{15},\bm{15};\bm{6},\bm{64}) &= \begin{pmatrix}
		\frac{5}{6\sqrt{6}} & 0 & 0 \\
		0 & 0 & 0 \\
		0 & 0 & 0
	\end{pmatrix}\,, \\
	D_{jk}^{11;22}(\bm{15},\bm{15};\bm{6},\bm{64}) &= \begin{pmatrix}
		0 & 0 & 0 \\
 		0 & -\frac{11}{200 \sqrt{6}} & \frac{7}{200 \sqrt{6}} \\
 		0 & \frac{7}{200 \sqrt{6}} & -\frac{49}{2200 \sqrt{6}}
 	\end{pmatrix}\,, \\
 	D_{jk}^{11;33}(\bm{15},\bm{15};\bm{6},\bm{64}) &= \begin{pmatrix}
		0 & 0 & 0 \\
 		0 & -\frac{77}{600 \sqrt{6}} & -\frac{161}{600 \sqrt{6}} \\
 		0 & -\frac{161}{600 \sqrt{6}} & -\frac{3703}{6600 \sqrt{6}}
 	\end{pmatrix}\,, \\
 	D_{jk}^{11;12}(\bm{15},\bm{15};\bm{6},\bm{64}) = -D_{kj}^{11;21}(\bm{15},\bm{15};\bm{6},\bm{64}) &= \begin{pmatrix}
 		0 & -\frac{\sqrt{\frac{11}{10}}}{12} & \frac{7}{12 \sqrt{110}} \\
 		0 & 0 & 0 \\
 		0 & 0 & 0
 	\end{pmatrix}\,,\\
 	D_{jk}^{11;13}(\bm{15},\bm{15};\bm{6},\bm{64}) = -D_{kj}^{11;31}(\bm{15},\bm{15};\bm{6},\bm{64}) &= \begin{pmatrix}
 		0 & \frac{\sqrt{\frac{77}{30}}}{12} & \frac{23 \sqrt{\frac{7}{330}}}{12} \\
 		0 & 0 & 0 \\
 		0 & 0 & 0
 	\end{pmatrix}\,,\\
 	D_{jk}^{11;23}(\bm{15},\bm{15};\bm{6},\bm{64}) = D_{kj}^{11;32}(\bm{15},\bm{15};\bm{6},\bm{64}) &= \begin{pmatrix}
 		0 & 0 & 0 \\
 		0 & \frac{11 \sqrt{\frac{7}{2}}}{600} & \frac{23 \sqrt{\frac{7}{2}}}{600} \\
 		0 & -\frac{7 \sqrt{\frac{7}{2}}}{600} & -\frac{161 \sqrt{\frac{7}{2}}}{6600}
 	\end{pmatrix}\,.
\end{aligned}
\end{equation}
}

For the strong coupling ground states with $\bm{r}_0 = \bm{20}$ and $\bm{r}_1 = \bm{\overline{4}}$, the even links could transition to the $\bm{36}$, the $\bm{60}$, or the $\bm{140}$:
\begin{equation}
\begin{aligned}
	D_{jk}(\bm{\overline{4}},\bm{\overline{4}};\bm{20},\bm{36}) = D_{kj}(\bm{4},\bm{4};\bm{\overline{20}},\bm{\overline{36}}) &= \begin{pmatrix}
		-\frac{25 \sqrt{5}}{96} & \frac{95}{96 \sqrt{2}} & \frac{35}{96 \sqrt{2}} \\
 		-\frac{95}{96 \sqrt{2}} & \frac{361}{192 \sqrt{5}} & \frac{133}{192 \sqrt{5}} \\
 		-\frac{35}{96 \sqrt{2}} & \frac{133}{192 \sqrt{5}} & \frac{49}{192 \sqrt{5}}
 	\end{pmatrix}\,,\\
 	D_{jk}(\bm{\overline{4}},\bm{\overline{4}};\bm{20},\bm{60}) = D_{kj}(\bm{4},\bm{4};\bm{\overline{20}},\bm{\overline{60}})&= \begin{pmatrix}
		\frac{5 \sqrt{3}}{32} & \frac{\sqrt{\frac{3}{10}}}{32} & -\frac{7 \sqrt{\frac{3}{10}}}{32} \\
 		-\frac{\sqrt{\frac{3}{10}}}{32} & -\frac{\sqrt{3}}{1600} & \frac{7 \sqrt{3}}{1600} \\
 		\frac{7 \sqrt{\frac{3}{10}}}{32} & \frac{7 \sqrt{3}}{1600} & -\frac{49 \sqrt{3}}{1600} 
 	\end{pmatrix}\,,\\
 	D_{jk}(\bm{\overline{4}},\bm{\overline{4}};\bm{20},\bm{140}) = D_{kj}(\bm{4},\bm{4};\bm{\overline{20}},\bm{\overline{140}}) &= \begin{pmatrix}
		\frac{5 \sqrt{7}}{288} & -\frac{11 \sqrt{\frac{7}{10}}}{288} & -\frac{23 \sqrt{\frac{7}{10}}}{288} \\
 		\frac{11 \sqrt{\frac{7}{10}}}{288} & -\frac{121 \sqrt{7}}{14400} & -\frac{253 \sqrt{7}}{14400} \\
 		\frac{23 \sqrt{\frac{7}{10}}}{288} & -\frac{253 \sqrt{7}}{14400} & -\frac{529 \sqrt{7}}{14400}
 	\end{pmatrix}.
\end{aligned}
\end{equation}
The odd links could transition to the $\bm{\overline{20}}$ or the $\bm{\overline{36}}$, both of which have multiplicity labels on neighboring vertices. For the $\bm{\overline{20}}$ transition we have
\begin{equation}
\begin{aligned}
	D_{jk}^{11;11}(\bm{20},\bm{20};\bm{\overline{4}},\bm{\overline{20}}) = D_{kj}^{11;11}(\bm{\overline{20}},\bm{\overline{20}};\bm{4},\bm{20}) &= \begin{pmatrix}
		\frac{\sqrt{5}}{24} & -\frac{49}{120 \sqrt{2}} & -\frac{7}{120 \sqrt{2}} \\
 		\frac{49}{120 \sqrt{2}} & -\frac{2401}{1200 \sqrt{5}} & -\frac{343}{1200 \sqrt{5}} \\
 		\frac{7}{120 \sqrt{2}} & -\frac{343}{1200 \sqrt{5}} & -\frac{49}{1200 \sqrt{5}}
 	\end{pmatrix}\,,\\
 	D_{jk}^{11;12}(\bm{20},\bm{20};\bm{\overline{4}},\bm{\overline{20}}) = D_{kj}^{11;21}(\bm{\overline{20}},\bm{\overline{20}};\bm{4},\bm{20}) &= \begin{pmatrix}
		-\frac{5}{24} & -\frac{11}{24 \sqrt{10}} & \frac{7}{24 \sqrt{10}} \\
 		-\frac{49}{24 \sqrt{10}} & -\frac{539}{1200} & \frac{343}{1200} \\
 		-\frac{7}{24 \sqrt{10}} & -\frac{77}{1200} & \frac{49}{1200} 
 	\end{pmatrix}\,,\\
 	D_{jk}^{11;21}(\bm{20},\bm{20};\bm{\overline{4}},\bm{\overline{20}}) = D_{kj}^{11;21}(\bm{\overline{20}},\bm{\overline{20}};\bm{4},\bm{20}) &= \begin{pmatrix}
		\frac{5}{24} & -\frac{49}{24 \sqrt{10}} & -\frac{7}{24 \sqrt{10}} \\
 		-\frac{11}{24 \sqrt{10}} & \frac{539}{1200} & \frac{77}{1200} \\
 		\frac{7}{24 \sqrt{10}} & -\frac{343}{1200} & -\frac{49}{1200}
 	\end{pmatrix}\,,\\
 	D_{jk}^{11;22}(\bm{20},\bm{20};\bm{\overline{4}},\bm{\overline{20}}) = D_{kj}^{11;22}(\bm{\overline{20}},\bm{\overline{20}};\bm{4},\bm{20}) &= \begin{pmatrix}
		-\frac{5 \sqrt{5}}{24} & -\frac{11}{24 \sqrt{2}} & \frac{7}{24 \sqrt{2}} \\
 		\frac{11}{24 \sqrt{2}} & \frac{121}{240 \sqrt{5}} & -\frac{77}{240 \sqrt{5}} \\
 		-\frac{7}{24 \sqrt{2}} & -\frac{77}{240 \sqrt{5}} & \frac{49}{240 \sqrt{5}}
 	\end{pmatrix}\,.
\end{aligned}
\end{equation}
For the $\bm{\overline{36}}$ transition we have
\begin{equation}
\begin{aligned}
	D_{jk}^{11;11}(\bm{20},\bm{20};\bm{\overline{4}},\bm{\overline{36}}) = D_{kj}^{11;11}(\bm{\overline{20}},\bm{\overline{20}};\bm{4},\bm{36}) &= \begin{pmatrix}
		-\frac{5}{2 \sqrt{646}} & \frac{71}{12 \sqrt{1615}} & -\frac{7}{12 \sqrt{1615}} \\
 		-\frac{3}{\sqrt{1615}} & \frac{71}{25 \sqrt{646}} & -\frac{7}{25 \sqrt{646}} \\
 		-\frac{21}{4 \sqrt{1615}} & \frac{497}{100 \sqrt{646}} & -\frac{49}{100 \sqrt{646}}
 	\end{pmatrix}\,,\\
 	D_{jk}^{11;12}(\bm{20},\bm{20};\bm{\overline{4}},\bm{\overline{36}}) = D_{kj}^{11;21}(\bm{\overline{20}},\bm{\overline{20}};\bm{4},\bm{36}) &= \begin{pmatrix}
		\frac{5 \sqrt{\frac{7}{1938}}}{6} & -\frac{\sqrt{\frac{7}{4845}}}{12} & -\frac{43 \sqrt{\frac{7}{4845}}}{12} \\
 		\sqrt{\frac{7}{4845}} & -\frac{\sqrt{\frac{7}{1938}}}{25} & -\frac{43 \sqrt{\frac{7}{1938}}}{25} \\
 		\frac{7 \sqrt{\frac{7}{4845}}}{4} & -\frac{7 \sqrt{\frac{7}{1938}}}{100} & -\frac{301 \sqrt{\frac{7}{1938}}}{100} 
 	\end{pmatrix}\,,\\
 	D_{jk}^{11;21}(\bm{20},\bm{20};\bm{\overline{4}},\bm{\overline{36}}) = D_{kj}^{11;12}(\bm{\overline{20}},\bm{\overline{20}};\bm{4},\bm{36}) &= \begin{pmatrix}
		-\frac{5 \sqrt{\frac{7}{1938}}}{4} & \frac{71 \sqrt{\frac{7}{4845}}}{24} & -\frac{7 \sqrt{\frac{7}{4845}}}{24} \\
 		-\frac{31 \sqrt{\frac{7}{4845}}}{8} & \frac{2201 \sqrt{\frac{7}{1938}}}{600} & -\frac{217 \sqrt{\frac{7}{1938}}}{600} \\
 		\frac{\sqrt{\frac{119}{285}}}{8} & -\frac{71 \sqrt{\frac{119}{114}}}{600} & \frac{7 \sqrt{\frac{119}{114}}}{600}
 	\end{pmatrix}\,,\\
 	D_{jk}^{11;22}(\bm{20},\bm{20};\bm{\overline{4}},\bm{\overline{36}}) = D_{kj}^{11;22}(\bm{\overline{20}},\bm{\overline{20}};\bm{4},\bm{36}) &= \begin{pmatrix}
		\frac{35}{36 \sqrt{646}} & -\frac{7}{72 \sqrt{1615}} & -\frac{301}{72 \sqrt{1615}} \\
 		\frac{217}{72 \sqrt{1615}} & -\frac{217}{1800 \sqrt{646}} & -\frac{9331}{1800 \sqrt{646}} \\
		 -\frac{7 \sqrt{\frac{17}{95}}}{72} & \frac{7 \sqrt{\frac{17}{38}}}{1800} & \frac{301 \sqrt{\frac{17}{38}}}{1800}
 	\end{pmatrix}\,.
\end{aligned}
\end{equation}
The matrix elements for transitions of the other $p = 1$ strong coupling ground states are related, and also given in the equations above. Likewise, the matrix elements for transitions of the $p = 2$ and $p = 3$ strong coupling ground states are also given in terms of those above.

\subsection{Properties of $C^A_i$}\label{app:c_properties}

Here we discuss the three properties of the $(\textbf{adj}, \bm{R}, \bm{\bar{R}})$ invariants needed in Section~\ref{sec:suN_fermions}. We will formulate them all in terms of 6$j$-symbols. We do not have proofs that these properties of 6$j$-symbols hold, but we have checked explicitly for $\grSU(2)$, $\grSU(3)$, $\grSU(4)$, $\grSO(5)$, and $G_2$.

First we address
\begin{equation}\label{eq:c_property_1_app}
    \left\lbrack C^A_i, C^B_j\right\rbrack + \left\lbrack C^B_i, C^A_j\right\rbrack = 0\,.
\end{equation}
Let us define a four-point invariant
\begin{equation}
    (T_{ij})^{AB}_{ab} = \frac{1}{2}\left(\begin{tikzpicture}[scale=.9,baseline=-.5cm]
        \draw[ultra thick,dashed,mid arrow] (1,-2) -- (0,-1) node[fill,minimum size=5,inner sep=0,label=-90:{$i$}] {};
        \draw[ultra thick,dashed,mid arrow] (0,-1) -- (0,0) node[fill,minimum size=5,inner sep=0,label=90:{$j$}] {};
        \draw[ultra thick,dashed,mid arrow] (0,0) -- (1,1);
        \draw[ultra thick,dotted,blue,mid arrow] (0,-1) -- (-1,-2);
        \draw[ultra thick,dotted,blue,mid arrow] (0,0) -- (-1,1);
    \end{tikzpicture} + \begin{tikzpicture}[scale=.9,baseline=-.5cm]
        \draw[ultra thick,dashed,mid arrow] (1,-2) -- (0,-1) node[fill,minimum size=5,inner sep=0,label=-90:{$i$}] {};
        \draw[ultra thick,dashed,mid arrow] (0,-1) -- (0,0) node[fill,minimum size=5,inner sep=0,label=90:{$j$}] {};
        \draw[ultra thick,dashed,mid arrow] (0,0) -- (1,1);
        \draw[ultra thick,dotted,blue,mid arrow] (0,0) -- (-1,-2);
        \draw[ultra thick,dotted,blue,mid arrow] (0,-1) -- (-1,1);
    \end{tikzpicture}\right) = {{(C_i)^{A}}_a}^c {{(C_j)^{B}}_c}^b + (A\leftrightarrow B)\,.
\end{equation}
Note that we have made the $\bm{R}$ indices $a,b$ explicit. The statement \eqref{eq:c_property_1_app} is equivalent to $T_{ij} = T_{ji}$.

We can expand $T_{ij}$ into four-point invariants in a crossed channel, with some representation $\lambda$ in the symmetric square of the adjoint being exchanged:
\begin{equation}
    \left(S^\lambda_{\alpha\beta}\right)^{AB}_{ab} = \begin{tikzpicture}[scale=.9,baseline=0cm]
        \draw[ultra thick,dashed,mid arrow] (2,0) -- (3,1); 
        \draw[ultra thick,dashed,mid arrow] (3,-1) -- (2,0);
        \draw[ultra thick,dotted,blue,mid arrow] (0,0) -- (-1,-1);
        \draw[ultra thick,dotted,blue,mid arrow] (0,0) -- (-1,1);
        \draw[ultra thick,mid arrow] (0,0) node[fill,circle,minimum size=5,inner sep=0,label=180:{$e$}] {} -- node[above] {$\lambda$} (2,0) node[fill,circle,minimum size=5,inner sep=0,label=0:{$f$}] {};
    \end{tikzpicture} = {(M_e)^{AB}}_\sigma {{(N_{f})^\sigma}_a}^b\,.
\end{equation}
Let the expansion be given by
\begin{equation}
    T_{ij} = \sum_{\lambda,e,f} \alpha(i,j; \lambda,e,f) S^\lambda_{e,f}\,.
\end{equation}
The $s$-channel invariants satisfy the orthogonality relation
\begin{equation}
    (S^\lambda_{ef})^{AB}_{ab} (S^{\lambda'}_{e'f'})^{AB}_{ab} = (\dim\lambda) \delta_{\lambda\lambda'}\delta_{ee'}\delta_{ff'}\,,
\end{equation}
so the expansion coefficients are
\begin{equation}\label{eq:fij_contraction}
\begin{split}
    \alpha(i,j;\lambda,e,f) &= \frac{1}{\dim\lambda} {{(C_i)^{(A}}_a}^c {{(C_j)^{B)}}_c}^b{(M_e)_{AB}}^\sigma {{(N_f)_\sigma}^a}_b \\
    &= \frac{1}{\dim\lambda}\left(\begin{tikzpicture}[baseline=0cm]
		\draw[ultra thick,dashed,mid arrow] (0,0) node[fill,circle,minimum size=5,inner sep=0,label=-90:{$j$}] {} -- (90:1) node[fill,circle,minimum size=5,inner sep=0,label=90:{$i$}] {};
		\draw[ultra thick,dotted,blue,mid arrow] (0,0) -- (-150:1) node[fill,circle,black,minimum size=5,inner sep=0,label=-150:{\color{black} $e$}] {};
		\draw[ultra thick,dotted,blue,mid arrow] (90:1) arc(90:210:1);
		\draw[ultra thick,dashed,mid arrow] (90:1) arc(90:-30:1);
		\draw[ultra thick,dashed,mid arrow] (-30:1) -- (0,0);
		\draw[ultra thick,mid arrow] (-150:1) arc(-150:-30:1) node[midway,below] {$\lambda$} node[fill,circle,minimum size=5,inner sep=0,label=-30:{$f$}] {};
	\end{tikzpicture}+\begin{tikzpicture}[baseline=0cm]
		\draw[ultra thick,dashed,mid arrow] (0,0) node[fill,circle,minimum size=5,inner sep=0,label=-90:{$j$}] {} -- (90:1) node[fill,circle,minimum size=5,inner sep=0,label=90:{$i$}] {};
		\draw[ultra thick,dotted,blue,mid arrow] (0,0) ..controls (150:1).. (180:1) arc(180:210:1) node[fill,circle,black,minimum size=5,inner sep=0,label=-150:{\color{black} $e$}] {};
		\draw[ultra thick,dotted,blue,mid arrow] (90:1) -- (210:1);
		\draw[ultra thick,dashed,mid arrow] (90:1) arc(90:-30:1);
		\draw[ultra thick,dashed,mid arrow] (-30:1) -- (0,0);
		\draw[ultra thick,mid arrow] (-150:1) arc(-150:-30:1) node[midway,below] {$\lambda$} node[fill,circle,minimum size=5,inner sep=0,label=-30:{$f$}] {};
	\end{tikzpicture}\right)\,.
\end{split}
\end{equation}
The property \eqref{eq:c_property_1_app} is equivalent to the claim that $\alpha(i,j;\lambda,e,f) = \alpha(j,i;\lambda,e,f)$.

We also need our invariants to satisfy
\begin{equation}\label{eq:c_property_2_app}
    f^{ABC} C^B_{i} C^C_{j} - f^{ABC} C^B_{j} C^C_{i} = 0\,.
\end{equation}
The first term on the left is a $(\mathbf{adj},\bm R,\bm{\bar{R}})$ invariant, and so we can expand it in terms of the $C^A_i$'s themselves:
\begin{equation}
	f^{ABC} C^B_{i} C^C_{j} = \sum \beta_{ij,k} C^A_k\,.
\end{equation}
The $\beta_{ij,k}$ coefficients are 6$j$-symbols:
\begin{equation}
	\beta_{ij,k} = \frac{1}{\dim\bm{R}}\begin{tikzpicture}[baseline=0cm]
		\draw[ultra thick,dashed,mid arrow] (0,0) node[fill,circle,minimum size=5,inner sep=0,label=-90:{$k$}] {} -- (90:1) node[fill,circle,minimum size=5,inner sep=0,label=90:{$i$}] {};
		\draw[ultra thick,dotted,blue,mid arrow] (0,0) -- (-150:1) node[fill,circle,black,minimum size=5,inner sep=0,label=-150:{\color{black} $f$}] {};
		\draw[ultra thick,dotted,blue,mid arrow] (90:1) arc(90:210:1);
		\draw[ultra thick,dashed,mid arrow] (90:1) arc(90:-30:1);
		\draw[ultra thick,dashed,mid arrow] (-30:1) -- (0,0);
		\draw[ultra thick,blue,dotted,mid arrow] (-150:1) arc(-150:-30:1) node[fill=black,circle,minimum size=5,inner sep=0,label=-30:{\color{black} $j$}] {};
	\end{tikzpicture}\,,
\end{equation}
where by $f$ we mean that the invariant on three adjoints is the $f$-symbol.  The condition \eqref{eq:c_property_2_app} is equivalent to the claim that $\beta_{ij,k} = \beta_{ji,k}$.

In addition, the derivation in Section~\ref{sec:suN_fermions} relied upon having a basis in which
\begin{equation}\label{eq:c_property_3_app}
	\sum_{j=1}^{\rk G} \left\lbrace C^A_j, C^B_j\right\rbrace = \delta^{AB}\,.
\end{equation}
If we change the basis of invariants by $C^A_j = Q_{jk} C^A_k$, then this condition becomes
\begin{equation}
	\left(Q_{jk} Q_{jk'}\right)\left\lbrace C^A_k, C^B_{k'}\right\rbrace = \delta^{AB}\,,
\end{equation}
where all repeated indices are summed. Using a Cholesky decomposition, any symmetric $\rk G\times \rk G$ matrix can be written in the form $Q^T Q$, which appears on the left-hand side of this equation, and so we just need to show that the identity $s$-channel invariant on the right can be written as a sum of the symmetrized $t$- and $u$-channel invariants on the left. That is, diagrammatically,
\begin{equation}
	\sum_{k,k'=1}^{\rk G} M_{(kk')} \left(\begin{tikzpicture}[scale=.9,baseline=-.5cm]
        \draw[ultra thick,dashed,mid arrow] (1,-2) -- (0,-1) node[fill,circle,minimum size=5,inner sep=0,label=-90:{$k$}] {};
        \draw[ultra thick,dashed,mid arrow] (0,-1) -- (0,0) node[fill,circle,minimum size=5,inner sep=0,label=90:{$k'$}] {};
        \draw[ultra thick,dashed,mid arrow] (0,0) -- (1,1);
        \draw[ultra thick,dotted,blue,mid arrow] (0,-1) -- (-1,-2);
        \draw[ultra thick,dotted,blue,mid arrow] (0,0) -- (-1,1);
    \end{tikzpicture} + \begin{tikzpicture}[scale=.9,baseline=-.5cm]
        \draw[ultra thick,dashed,mid arrow] (1,-2) -- (0,-1) node[fill,circle,minimum size=5,inner sep=0,label=-90:{$k$}] {};
        \draw[ultra thick,dashed,mid arrow] (0,-1) -- (0,0) node[fill,circle,minimum size=5,inner sep=0,label=90:{$k'$}] {};
        \draw[ultra thick,dashed,mid arrow] (0,0) -- (1,1);
        \draw[ultra thick,dotted,blue,mid arrow] (0,-1) -- (-1,1);
        \draw[ultra thick,dotted,blue,mid arrow] (0,0) -- (-1,-2);
    \end{tikzpicture}\right) = \sqrt{\dim G}\times \begin{tikzpicture}[scale=.9,baseline=0cm]
        \draw[ultra thick,dashed,mid arrow] (3,-1) -- (2,0); 
        \draw[ultra thick,dashed,mid arrow] (2,0) -- (3,1);
        \draw[ultra thick,dotted,blue,mid arrow] (0,0) -- (-1,-1);
        \draw[ultra thick,dotted,blue,mid arrow] (0,0) -- (-1,1);
        \draw[ultra thick,gray,dotted,mid arrow] (2,0) node[black,fill,circle,minimum size=5,inner sep=0] {} -- node[black,above] {$1$} (0,0) node[black,fill,circle,minimum size=5,inner sep=0] {};
    \end{tikzpicture}\,,
\end{equation}
where as a matrix $M = Q^T Q$.

We can expand this equation into $s$-channel invariants with representation $\lambda$ exchanged, where $\lambda$ is in the symmetric square of the adjoint. We find the linear system
\begin{equation}\label{eq:c_property_2_linear}
	\sum_{k,k'=1}^{\rk G} A_{\lbrace \lambda,e,f\rbrace, (kk')} M_{(kk')} = b_{\lbrace \lambda, e, f\rbrace} \equiv \begin{cases} \sqrt{\dim G} & \lambda = 1 \\ 0 & \text{otherwise}, \end{cases}
\end{equation}
where the matrix $A$ has $6j$-symbols as entries:
\begin{equation}\label{eq:constraint_6j}
	A_{\lbrace \lambda,e,f\rbrace, (kk')} = 2\times \begin{tikzpicture}[baseline=0cm]
		\draw[ultra thick,dashed,mid arrow] (0,0) node[fill,circle,minimum size=5,inner sep=0,label=-90:{$k$}] {} -- (90:1) node[fill,circle,minimum size=5,inner sep=0,label=90:{$k'$}] {};
		\draw[ultra thick,dotted,blue,mid arrow] (0,0) -- (-150:1) node[fill,circle,black,minimum size=5,inner sep=0,label=-150:{\color{black} $e$}] {};
		\draw[ultra thick,dotted,blue,mid arrow] (90:1) arc(90:210:1);
		\draw[ultra thick,dashed,mid arrow] (90:1) arc(90:-30:1);
		\draw[ultra thick,dashed,mid arrow] (-30:1) -- (0,0);
		\draw[ultra thick,mid arrow] (-150:1) arc(-150:-30:1) node[midway,below] {$\lambda$} node[fill,circle,minimum size=5,inner sep=0,label=-30:{$f$}] {};
	\end{tikzpicture}\,.
\end{equation}
This is generically an overconstrained system for the $\binom{\rk G}{2}$ variables $M_{(kk')}$. For instance, when $G = \grSU(N_c)$, there are $(N_c-1)^2 = (\rk G)^2$ invariants of two adjoint representations and two copies of $\bm{R}$ that are symmetric in the adjoint indices, so we have this many equations in \eqref{eq:c_property_2_linear}. The satisfiability of \eqref{eq:c_property_3_app} is equivalent to the claim $b$ is in the span of the columns of $A$.

We do not yet have a proofs of these claims, but we have checked them in the cases $\grSU(2)$, $\grSU(3)$, $\grSO(5)$, $\grSU(4)$, $\grSO(5)$, and $G_2$. Proving these statements in general will likely require some more detailed knowledge of $6j$-symbols appearing in the conditions above, which are all special cases of the 6$j$-symbols that appear in our lattice Hamiltonian.

\section{Solving a Majorana chain}\label{app:majorana_chain}

In this appendix, we will explain how to solve the Majorana chain that appears in the strong coupling expansion in Section~\ref{sec:strong_coupling_bw}.

We consider a chain with $N\in 2\mathbb{Z}_+$ sites and Majorana fermions $\lambda_{n,j}$, with $j=1,\ldots,\rk G$, on each site. The Hamiltonian is of the form
\begin{align}
	H = -\frac{i}{2}\sum_{n=0}^{\frac{N}{2}-1} \sum_{j,j'=1}^{\rk G} \left(A_{jj'}\lambda_{2n,j} \lambda_{2n+1,j'}+B_{jj'}\lambda_{2n+1,j}\lambda_{2n+2,j'}\right)\,.
\end{align}
The Majorana fermions satisfy $\{\lambda_{n,j},\lambda_{m,j'}\} = 2\delta_{nm}\delta_{jj'}$. We can define complex fermions
\begin{align}
	c_{n,j} = \frac{1}{2}(\lambda_{2n,j}+i\lambda_{2n+1,j})\,,\qquad n=0,\ldots,\frac{N}{2}-1\,,\qquad j=1,\ldots,\rk G\,.
\end{align}
Writing the Hamiltonian in terms of these operators yields a chain of complex fermions on $\frac{N}{2}$ sites:
\begin{align}
	H=  \frac{1}{2}\sum_{n=0}^{\frac{N}{2}-1}\sum_{j,j'=1}^{\rk G}\left(A_{jj'}(c_{n,j}+c_{n,j}^{\dagger})(c_{n,j'}-c_{n,j'}^{\dagger})+B_{jj'}(c_{n,j}-c_{n,j}^{\dagger})(c_{n+1,j'}+c_{n+1,j'}^{\dagger})\right)\,.
\end{align}
By introducing Fourier modes
\begin{align}
	c_{n,j} = \sqrt{\frac{2}{N}}\sum_k e^{-ink} \tilde c_{k,j}\,,\qquad k=\frac{4\pi p}{N}\qquad\text{for}\qquad p=0,\ldots,\frac{N}{2} - 1\,,
\end{align}
the Hamiltonian can be put into the form
\begin{align}
	 H =\frac{1}{2}\sum_k \sum_{j,j'=1}^{\rk G}\begin{pmatrix}
	(\vec{\tilde{c}}_k)^\dagger & \vec{\tilde{c}}_{-k}
\end{pmatrix}
M_{k}
\begin{pmatrix}
	\vec{\tilde{c}}_{k}\\
(	\vec{\tilde{c}}_{-k})^\dagger
\end{pmatrix}\,,
\end{align}
where $\vec{\tilde{c}}_k = \left(\tilde{c}_{k,1},\ldots,\tilde{c}_{k,\rk G}\right)$ and $M_k$ is a $2\rk G\times 2\rk G$ matrix given by
\begin{align}\label{eq:m_block}
	M_{k} = \frac{1}{2}\begin{pmatrix}
		A+A^T -e^{-ik}B-e^{ik}B^T & -A+A^T-e^{-ik}B+e^{ik}B^T \\
		A - A^T+e^{-ik}B-e^{ik}B^T & -A - A^T +e^{-ik}B +e^{ik}B^T
	\end{pmatrix}\,.
\end{align}
The matrix $M_k$ is Hermitian and satisfies the relations
\begin{align}\label{eq:m_relation}
	\{\sigma_1,M_k\} = 0\,,\qquad  -\sigma_1 M_{-k}^*\sigma_1 = M_k\,,
\end{align}
where the Pauli matrix $\sigma_1$ acts on the $2\times 2$ block structure of \eqref{eq:m_block}. The first relation implies that the spectrum of $M_k$ consists of pairs $\pm \epsilon_r$, while the second implies that the spectrum of $M_{-k}$ is minus that of $M_k$. Combining these statements, we see that $M_k$ and $M_{-k}$ have the same eigenvalues and $\rk G$ of them are nonnegative. Let $\Lambda_k=\operatorname{diag}(\epsilon_{k,1},\ldots, \epsilon_{k,\rk G})$ be a diagonal matrix with $\epsilon_{k,j}\ge 0$ being the $\rk G$ nonnegative eigenvalues of $M_k$. Let $u_k$ and $v_k$ be $\rk G\times \rk G$ matrices such that
\begin{align}
	M_k\begin{pmatrix}
		u_k\\
		v_k
	\end{pmatrix} = \begin{pmatrix}
	u_k\\
	v_k
	\end{pmatrix}\Lambda_k\,;
\end{align}
that is, when we concatenate the $j$th column of $u_k$ with the $j$th column of $v_k$, we find the eigenvector of $M_k$ with eigenvalue $\epsilon_{k,j}$.

The second relation in \eqref{eq:m_relation} also implies that $M_k$ can be diagonalized as
\begin{align}\label{eq:diagonal}
	U_k ^\dagger M_k U_k = \begin{pmatrix}
		\Lambda_k&0\\
		0& -\Lambda_{-k}
	\end{pmatrix},\qquad U_k = \begin{pmatrix}
		u_k & v_{-k}^*\\
		v_k & u^*_{-k}
	\end{pmatrix}\,.
\end{align}
Using $U_k$, we can define fermionic annihilation operators $\xi_k$ by
\begin{align}
	\begin{pmatrix}
		\vec{\xi}_k\\
		\vec{\xi}_{-k}
	\end{pmatrix} = U_k^\dagger \begin{pmatrix}
	\vec{c}_k\\
	(\vec{c}_{-k})^\dagger
	\end{pmatrix}\,,
\end{align}
which satisfy the canonical relations $\{\xi_k,\xi_{k'}\}=0$ and $\{\xi_k,\xi_{k'}^\dagger\}=\delta_{kk'}$ due to the block structure of $U_k$. In terms of the new operators $\xi_k$ the Hamiltonian takes the form
\begin{align}
	H=\frac 12\sum_k  \begin{pmatrix}
		(\vec{\xi}_k)^\dagger & \vec{\xi}_{-k}
	\end{pmatrix}\begin{pmatrix}
		\Lambda_k&0\\
		0& - \Lambda_{-k}
	\end{pmatrix}\begin{pmatrix}
		\vec{\xi}_k \\
		(\vec{\xi}_{-k})^\dagger
	\end{pmatrix} = \sum _{k}\sum_{j=1}^{\rk G} \epsilon_{k,j}\xi^\dagger_{k,j}\xi_{k,j}-\frac{1}{2}\sum _{k} \sum_{j=1}^{\rk G} \epsilon_{k,j}\, .
\end{align}
As all excitations $\epsilon_{k,i}$ are positive the ground state is characterized by $\xi_{k}^a\ket{0}=0$ and has energy
\begin{align}
	\braket{0|H|0} = -\frac{1}{2}\sum_{k,i}\epsilon_{k,i}\, .
\end{align}
In the limit of an infinite Majorana chain $N\rightarrow \infty$, the sum over the Brillouin zone can be evaluated using an integral
\begin{align}
	\frac{\braket{0|H|0}}{N} = -\frac{1}{8\pi}\int^{2\pi}_0 dk\, \tr [\Lambda_k]\,.
\end{align}

Finally, the expectation value of another quadratic operator
\begin{align}
	H' = \frac{1}{2}\sum_k \begin{pmatrix}
		c^\dagger_k & c_{-k}
	\end{pmatrix} M'_k
	\begin{pmatrix}
		c_k\\
		c^\dagger_{-k}
	\end{pmatrix}
\end{align}
is given by
\begin{align}
	\braket{0| H'|0}=-\frac{1}{2}\sum_k \tr[\tilde{ A}'_k]\,,
\end{align}
where
\begin{align}
	\begin{pmatrix}
		\tilde{A}'_k & \tilde{B}'_k\\
		\tilde{B}'^\dagger _k &  -\tilde{A}'^{ T}_{-k}
	\end{pmatrix}=U_k^\dagger \begin{pmatrix}
		A'_k & B'_k\\
		B'^\dagger _k &  -A'^{ T}_{-k}
	\end{pmatrix}U_k \,.
\end{align}

\section{Instanton action for a small circle}\label{app:instanton}
In this appendix, we review the derivation of the instanton action responsible for the exponential suppression of the energy splitting on a small circle with anti-periodic fermions with a small mass $m$ found in Section~\ref{sec:numerics_small_circle}. We will follow the setup and notation of~\cite{Dempsey:2024ofo}.

On a small circle with circumference $L\ll g^{-1}$, the dynamics of adjoint QCD$_2$ is well-approximated by integrating out all non-zero momentum modes, since they acquire a mass of order $L^{-1}$. For anti-periodic fermions, the only zero mode is given by the gauge holonomy around the compact direction~\cite{Lenz:1994du,Dempsey:2024ofo}. To leading order one can treat the holonomy as time-independent and integrate out the fermions in such a background. To that end, let us pick a gauge in which the holonomy is diagonal:
\begin{align}
	\exp\left[i\int_0^\beta d\tau \, A_\tau \right] = \operatorname{diag}\left(e^{ia_1},e^{ia_2},\ldots,e^{ia_{N_c}}\right)\, ,\qquad a_1+a_2+\ldots+ a_{N_c} =0\,,
\end{align}
which corresponds to a gauge potential
\begin{align}\label{eq:small_a}
	A_\tau  = \frac{1}{L}\operatorname{diag} (a_1,a_2,\ldots,a_{N_c})\,, \qquad A_x =0\, .
\end{align}
For this parametrization to be in one-to-one correspondence with physically inequivalent configurations, one has to identify the values of $\bm a =(a_1,a_2,\ldots,a_{N_c})$ under permutations and translations of the form $(a_j,a_k)\rightarrow(a_j+2\pi,a_k-2\pi)$ for $j\neq k$. Equivalently, we can restrict the range of $\bm a$. One way to do this is by restricting $\bm a$ to values in a fundamental domain given by the simplex with vertices
\begin{align}
	\bm v_k = \bigg (\underbrace{\frac{2\pi k}{N_c},\ldots,\frac{2\pi k}{N_c}}_{N_c-k},\underbrace{\frac{2\pi (k-N_c)}{N_c},\ldots,\frac{2\pi (k-N_c)}{N_c}}_{k}\bigg)\,,
\end{align}
with $k=0,\ldots,N_c-1$, within the hyperplane $a_1+a_2+\ldots+ a_{N_c} = 0$.

We then want to integrate out the fermions in the background \eqref{eq:small_a}. Despite the fermions having a small non-zero mass $m$, the mass can be set to zero in a first approximation of instanton action. Thus the relevant effective potential is
\begin{align}\label{eq:pot}
	V_\text{eff}(\bm a) = \frac{1}{2\pi L }\sum_{i<j}\min(a_i-a_j,2\pi -(a_i-a_j))^2\,.
\end{align}
The effective potential has minima at each of the $N_c$ corners of the fundamental domain. Naively, the $N_c$ minima give rise to $N_c$ degenerate ground states, but for a non-zero fermion mass this degeneracy is lifted by instanton effects. To compute the instanton action, we also need the kinetic term in the parametrization \eqref{eq:small_a}:
\begin{align}\label{eq:kin}
	T_\text{eff}=-\frac{1}{2g^2}\int dx \, \tr_\text{fund}(F_{\mu\nu}F^{\mu\nu}) = \frac{1}{2g^2L}\dot{\bm a}^2\, .
\end{align}
Consider an instanton that interpolates between $\bm v_0$ and $\bm v_k$ as\footnote{Note that we do not lose any generality by starting at $\bm{v}_0$ since the minima are all related by center symmetry.}
\begin{align}\label{eq:ansatz}
	\bm a(\theta) = \frac{\theta}{2\pi} \bm v_k \, , \qquad 0\le \theta\le 2\pi\,.
\end{align}
The Euclidean Lagrangian along the ansatz \eqref{eq:ansatz} is given by
\begin{equation}\label{eq:efflag}
	\begin{aligned}
	L_E = T_\text{eff}+V_\text{eff} &= \frac{\bm v_k^2}{4\pi^2g^2}\dot \theta^2+\frac{1}{2\pi L}\sum_{\substack{1\le i\le j\\k+1\le j\le N_c}}\min(\theta,2\pi-\theta)^2\\
	&= \frac{1}{2}\frac{2k(N_c-k)}{N_cg^2L}\dot\theta^2+\frac{k(N_c-k)}{2\pi L}\min(\theta,2\pi-\theta)^2\, .
\end{aligned}
\end{equation}
Strictly speaking this effective theory should not be trusted for the trajectory in question because the potential barrier between minima is of order $L^{-1}$, but as noted in \cite{Smilga:1994hc} this ``flawed'' computation nevertheless yields the correct instanton action.

In general, for a one-dimensional system with Euclidean Lagrangian $L_E =\frac{1}{2m}\dot q^2+V(q)$, the classical action for an instanton interpolating between $q_1$ and $q_2$ is
\begin{align}
	S_I = \int_{q_1}^{q_2}dq\, \sqrt{2m[V(q)-V(q_1)]}\,.
\end{align}
Thus, for the effective Lagrangian \eqref{eq:efflag}, we find that the instanton action for the transition from $\bm v_0$ to $\bm v_k$ is
\begin{align}\label{eq:int_act}
	S_I^{(k)} = \frac{k(N_c-k)}{\sqrt{N_c}}\frac{\sqrt{2}\pi^{3/2}}{gL}\,.
\end{align}
The instanton action is minimized for the ``nearest-neighbor'' transition $k=1$ or $k=N_c-1$. This gives the exponential suppression in \eqref{eq:su2_abc_gap} and \eqref{eq:su3_abc_gap}.
Note that the instanton action we find has an extra factor of $\sqrt 2$ compared to what is reported in~\cite{Cherman:2019hbq,Bergner:2024ttq}. 
Our result including this factor is supported by the numerical calculations in Section~\ref{sec:numerics_small_circle}. 

 The leading-order dependence on the fermion mass $m$ comes from fermion zero modes about the instanton trajectory $\eqref{eq:ansatz}$. There are $N_c-1$ such zero modes~\cite{Smilga:1994hc}, and so the expected energy splitting is given by
\begin{align}
	\Delta E \sim m^{N_c-1}\exp\left[-\frac{(N_c-1)}{\sqrt{N_c}}\frac{\sqrt{2}\pi^{3/2}}{gL}\right]\,.
\end{align}

\bibliographystyle{ssg}
\bibliography{su3}

\end{document}